\documentclass[prx,aps,twocolumn,nolongbibliography,superscriptaddress]{revtex4-1}

\usepackage[colorlinks,bookmarks=true,citecolor=blue,linkcolor=red,urlcolor=blue,citecolor=blue]{hyperref}
\usepackage{bm}
\usepackage{amsmath}
\usepackage{cancel}
\usepackage{amssymb}
\usepackage{amsfonts}
\usepackage{bbm}
\usepackage{braket}
\usepackage{xcolor}
\usepackage{appendix}
\usepackage{enumitem}
\usepackage{graphicx}
\usepackage{tikz}
\usepackage{float}
\usepackage{algorithm}
\usepackage{algpseudocode}
\usepackage{bbm}
\usepackage{subfigure}
\usepackage{color}
\usepackage{nicematrix}
\usepackage{xspace}
\usepackage{refcount}

\usepackage{tikz}
\usetikzlibrary{shapes.geometric, arrows.meta, positioning, decorations.pathreplacing, calc}

\begin{document}

\title{Efficient simulation of low-entanglement bosonic Gaussian states in polynomial time}

\author{Tong Liu}
\affiliation{Institute of Physics, Chinese Academy of Sciences, Beijing 100190, China}
\affiliation{School of Physical Sciences, University of Chinese Academy of Sciences, Beijing 100049, China}

\author{Hui-Ke Jin}
\email[Corresponding author: ]{jinhk@shanghaitech.edu.cn}
\affiliation{School of Physical Science and Technology, ShanghaiTech University, Shanghai 201210, China}

\author{Tao Xiang}
\affiliation{Institute of Physics, Chinese Academy of Sciences, Beijing 100190, China}
\affiliation{School of Physical Sciences, University of Chinese Academy of Sciences, Beijing 100049, China}

\author{Hong-Hao Tu}
\email[Corresponding author: ]{h.tu@lmu.de}
\affiliation{Faculty of Physics and Arnold Sommerfeld Center for Theoretical Physics, Ludwig-Maximilians-Universit\"at M\"unchen, 80333 Munich, Germany}

\begin{abstract}
Bosonic Gaussian states are ubiquitous in quantum optics and condensed matter physics. While they are efficiently handled within the Gaussian formalism, sampling requires calculating amplitudes in the boson occupation basis. This step, however, is hindered by a significant bottleneck due to the hafnian. We present an efficient algorithm that converts pure bosonic Gaussian states into matrix product states (MPSs), thereby establishing a versatile tool for probing bosonic Gaussian systems in settings where direct Gaussian-formalism-based calculations become inefficient. Our method combines a Gaussian singular value decomposition with a projected-creation-operator mapping that constructs local MPS tensors without computing hafnians. Benchmarking on covariance matrices from the Jiuzhang~2.0 and Jiuzhang~4.0 Gaussian boson sampling experiments demonstrates substantial speedups over previous tensor-network approaches in the low-entanglement regime relevant to lossy devices. The method provides a scalable classical simulation framework for bosonic Gaussian states with limited entanglement. In this regime, a target accuracy can be achieved with a bond dimension that remains computationally tractable, thereby extending the applicability of MPS-based methods to a broad range of bosonic systems.
\end{abstract}

\maketitle

\section{Introduction}
\label{sec:introduction}

The quest for quantum advantage---the demonstration of a quantum device performing a task provably intractable for any classical machine---has become a central objective in quantum science and technology. Among the platforms proposed for this pursuit, the boson sampling problem~\cite{Aaronson2011,Broome2013,Tillmann2013,Spring2013,Crespi2013,WangH2017,HeY2017,Loredo2017} has emerged as a leading candidate. A particularly prominent variant is Gaussian boson sampling (GBS)~\cite{Lund2014}, in which squeezed vacuum states are injected into a linear-optical circuit and subsequently measured by photon-number-resolving detectors~\cite{ZhongHS2019,ZhongHS2020,Zhong2021,Madsen2022,Deng2023}. The output probabilities are proportional to matrix hafnians~\cite{Hamilton2017}, and since hafnian computing is \#P-hard~\cite{Valiant1979,Aaronson2011}, the exact classical simulation of GBS is widely believed to require exponential time. Crucially, this hardness arises not from the Gaussian nature of the input states, but from the non-Gaussianity of the photon-counting basis, where amplitudes are governed by the hafnian.

This theoretical hardness, however, is softened by two important considerations. First, realistic photonic systems inevitably experience noise and photon loss. When the effective loss is sufficiently large, the computational hardness of GBS can break down, making the problem efficiently simulable on classical
hardware~\cite{Kalai2014,Rahimi-Keshari2015,Oszmaniec2018,Qi2020,Aharonov2023,Oh2023,Oh2024}. Second, and more fundamentally, \#P-hardness of the exact problem does not preclude the existence of efficient \emph{approximate} classical algorithms. This leads to a central dichotomy: quantum advantage is determined not by the idealized complexity of GBS, but by whether the best classical approximation algorithms can reproduce experimental data with high fidelity and efficiency. Pushing the limits of classical simulation is therefore essential for charting the true frontier of quantum computation.

Beyond its role in assessing quantum advantage, progress in classical algorithms for bosonic Gaussian systems has broad implications across theoretical physics. For instance, efficient classical algorithms for bosonic Gaussian systems are central to a wide range of applications, including Schwinger-boson mean-field theories~\cite{Auerbach-Book,Arovas1988,Sachdev1992,WangF2006}, the simulation of open quantum systems with Gaussian environments~\cite{Keeling2025,Link2024,ChenC2025}, and lattice gauge theories with bosonic matter fields~\cite{Chanda2020}. Existing classical approaches fall largely into two categories~\cite{Jerrum2004,Aaronson2012,LimY2025}: tensor-network-based methods that rely on brute-force hafnian calculations~\cite{Oh2024}, and Monte Carlo sampling algorithms~\cite{Jerrum2004,Aaronson2012,LimY2025,Quesada2022,Blumer2022,Dodd2025}. The former still suffer from exponential scaling as the matrices involved in hafnian computations grow, while the latter typically lack controlled error bounds, leaving the practical limits of classical simulation unclear.

In this work, we introduce a simulation framework that sidesteps the hafnian bottleneck by leveraging the matrix product state (MPS) formalism~\cite{Ostlund1995,Vidal2003b,Perez-Garcia2007,Verstraete2008,Schollwoeck2011,Cirac2021,XiangT-Book} together with an efficient procedure for constructing its tensors. The efficiency of the approach is enabled by two key components. First, we apply a Gaussian singular value decomposition (GSVD) that serves as an analytical pre-processing step, identifying a compressed basis for the system and substantially reducing the number of bosonic modes that must be treated explicitly. Second, operating on this reduced system, we propose a projected-creation-operator (PCO) mapping algorithm that constructs the MPS representation. In parallel with recent methodological advances for converting fermionic Gaussian states into MPSs~\cite{Fishman2015,Schuch2019,WuYH2020,JinHK2020,Aghaei2020,Petrica2021,Nuesseler2021,JinHK2022a,LiuT2025a,LiKL2025,JinHK2025,Hille2025}, this algorithm enables an efficient conversion from bosonic Gaussian states to MPSs. Crucially, the whole procedure avoids computational steps whose cost scales exponentially with the number of bosonic modes, allowing simulations to access larger boson-number regimes with substantially reduced computational resources compared to previous MPS-based approaches. After the MPS conversion, the sampling procedure can be performed by standard MPS-sampling routines as in existing workflows~\cite{Oh2024}. We emphasize that the present work focuses on the conversion algorithm and does not aim to improve the sampling procedure itself.

This paper is structured as follows. In Sec.~\ref{sec:method}, we present our algorithm for converting bosonic Gaussian states to MPSs, including the GSVD decomposition and the PCO mapping procedure. In Sec.~\ref{sec:results}, we benchmark the method on data from recent Gaussian boson sampling experiments, including both Jiuzhang 2.0 and Jiuzhang 4.0. Section~\ref{sec:summary} concludes with a summary and outlook. The main text is supplemented with three Appendices: Appendix~\ref{app:paired_wf} derives the paired-form wave function from the Bogoliubov transformation; Appendix~\ref{app:fast_search} provides implementation details of the PCO mapping; Appendix~\ref{app:opt_Gamma} describes the optimization procedure for extracting the pure-state covariance matrix from experimental data.

\section{Method}
\label{sec:method}

\begin{figure*}[hbtp]
    \centering
    \includegraphics[width=0.94\linewidth]{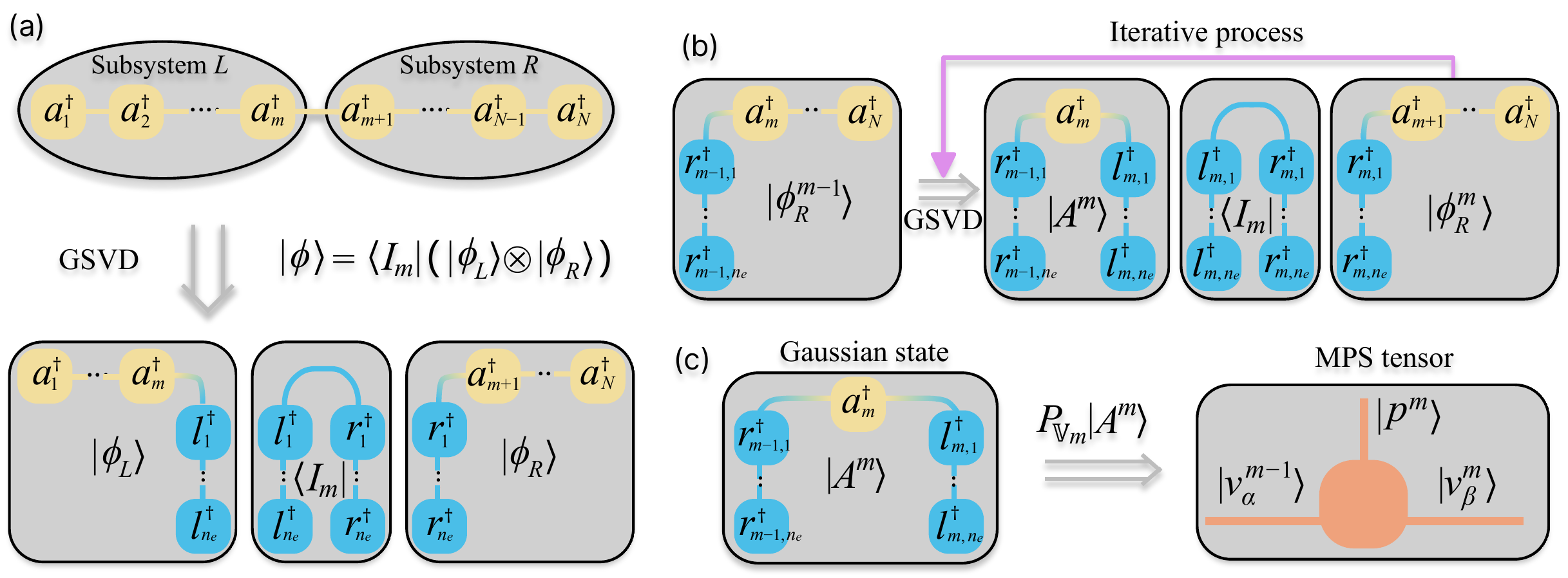}
    \caption{(a) Schematic of GSVD on the $m$-th bond. A pure BGS $|\phi\rangle$ is decomposed as $|\phi\rangle=\langle I_m|(|\phi_L\rangle\otimes|\phi_R\rangle)$, where $|\phi_L\rangle$ ($|\phi_R\rangle$) is a BGS of physical and virtual modes in the subsystem $L$ ($R$) and $|I_m\rangle$ is a maximally entangled BGS of virtual modes.  (b) Iterative GSVD steps. At the $m$-th step, the GSVD gives $|\phi^{m-1}_R\rangle=\langle I_m|(|A^m\rangle\otimes|\phi^m_R\rangle)$, where $|\phi^{m}_R\rangle$ becomes the input in the $(m+1)$-th GSVD step. For notational simplicity, we display the situation with the number of left and right virtual modes to be equal, i.e., $n^{m-1}_e=n^m_e=n_e$. (c) Conversion of a local BGS $|A^m\rangle$ to an MPS tensor, involving truncation on the physical and virtual boson Fock basis states. This truncation is performed using a projection operator $P_{\mathbbm{V}_m}$, where $\mathbbm{V}_m$ represents the local kept subspace. }
    \label{fig:GSVD}
\end{figure*}

In this section, we present an efficient classical algorithm for converting bosonic Gaussian states (BGSs) into MPSs. Our method includes two major stages: (i) Using the convariance matrix formalism, we apply an iterative Gaussian singular value decomposition. This macroscopic procedure factorizes an $N$-mode BGS into a sequence of smaller Gaussian states ${|A^m\rangle}$ that naturally form the backbone of an MPS. (ii) At the microscopic level, we introduce a projected-creation-operator mapping algorithm that computes the MPS tensor entries from each $|A^m\rangle$ while completely avoiding operations that scale exponentially with the number of bosonic modes. The workflow is summarized in Fig.~\ref{fig:GSVD}.

\subsection{Brief review of bosonic Gaussian states}
\label{sec:methodA}

We begin by reviewing the formalism of BGSs and introducing the notations used in this work. Consider a system of $N$ bosonic modes with creation and annihilation operators $a^{\dagger}_j$ and $a_j$ ($j=1,\ldots,N$). Each mode can be divided into canonical position and momentum operators:
\begin{equation}
a^\dagger_j=\frac{1}{\sqrt{2}}(\hat{x}_j - i\hat{p}_j),\quad
a_j=\frac{1}{\sqrt{2}}(\hat{x}_j + i\hat{p}_j),
\end{equation}
where $\hat{x}_j$ and $\hat{p}_k$ satisfy the canonical commutation relations, $[\hat{x}_j, \hat{p}_k] = i\delta_{jk}$. It is convenient to collect these operators into vector form:
\begin{align}
\hat{\boldsymbol{x}}=(\hat{x}_1, \ldots, \hat{x}_N)^T,
\quad \hat{\boldsymbol{p}}=(\hat{p}_1, \ldots, \hat{p}_N)^T,
\end{align}
Throughout this work, boldface symbols denote vectors and matrices, while their components appear in normal font.

Generally, BGSs include both pure and mixed states. Here, we restrict ourselves to pure BGSs. An $N$-mode pure BGS $|\phi\rangle$ is fully characterized by its $2N \times 2N$ covariance matrix $\boldsymbol{\Gamma}$~\footnote{Local displacements are omitted because they can always be removed locally and do not affect entanglement properties.}:
\begin{equation}
\boldsymbol{\Gamma} =
\frac{1}{2}\begin{pmatrix}
\langle\langle \hat{x}_j ,\hat{x}_k\rangle\rangle &
\langle\langle\hat{x}_j, \hat{p}_k\rangle\rangle  \\
\langle\langle\hat{p}_j ,\hat{x}_k\rangle\rangle &
\langle\langle\hat{p}_j, \hat{p}_k\rangle\rangle
\end{pmatrix}_{1\leq{}j,k\leq{}N}.
\label{eq:GammaMat}
\end{equation}
The double braket is defined by
\begin{equation}
\langle\langle{}O_i,O_j\rangle\rangle\equiv\langle\{O_i,O_j\}\rangle-\langle O_i\rangle\langle O_j\rangle,
\end{equation}
where $\{\cdot,\cdot\}$ denotes anticommutator and the expectation value $\langle \cdot \rangle$ is taken with respect to $|\phi\rangle$. The matrix $\boldsymbol{\Gamma}$ is real, symmetric, and positive-definite. Because $|\phi\rangle$ is Gaussian, all higher-order correlators of $\hat{\boldsymbol{x}}$ and $\hat{\boldsymbol{p}}$ follow from $\boldsymbol{\Gamma}$ via Wick's theorem.

The covariance matrix formalism provides a convenient route to the reduced density operators of a BGS. Consider a bipartition of the $N$-mode system into subsystems $L$ and $R$, containing $m$ and $\bar{m} \equiv N-m$ modes, respectively. Tracing out either subsystem produces a mixed Gaussian state, which is fully characterized by the principal submatrix of the full covariance matrix $\boldsymbol{\Gamma}$. Specifically, the covariance matrix for the subsystem $L$, denoted as $\boldsymbol{\Gamma}_{LL}$, is a $2m \times 2m$ matrix obtained by restricting the indices of $\boldsymbol{\Gamma}$ in Eq.~\eqref{eq:GammaMat} to the modes in $L$ ($1 \leq j,k \leq m$). Similarly, the $2\bar{m} \times 2\bar{m}$ covariance matrix $\boldsymbol{\Gamma}_{RR}$ for the subsystem $R$ is obtained by restricting the indices to the modes in $R$ ($m+1 \leq j,k \leq N$).

Williamson's theorem ensures that the positive-definite covariance matrices $\boldsymbol{\Gamma}_{LL}$ and $\boldsymbol{\Gamma}_{RR}$ can be brought into a diagonal form by symplectic transformations. There exist symplectic matrices $\boldsymbol{M}_L$ and $\boldsymbol{M}_R$ such that
\begin{align}
\boldsymbol{\Gamma}_{AA} = \frac{1}{2}\boldsymbol{M}_A\begin{pmatrix}
\boldsymbol{D}_A & \boldsymbol{0} \\
\boldsymbol{0} & \boldsymbol{D}_A
\end{pmatrix}\boldsymbol{M}_A^T, \quad A \in \{L, R\},
\label{eq:symDiag}
\end{align}
where the diagonal matrices $\boldsymbol{D}_L$ and $\boldsymbol{D}_R$ take the form
\begin{equation}
    \boldsymbol{D}_L = \begin{pmatrix}
      \oplus_{q}D_q   &  \\
         & \boldsymbol{\mathbbm{1}}_{m-n_e}
    \end{pmatrix},~
    \boldsymbol{D}_R = \begin{pmatrix}
      \oplus_{q}D_q   &  \\
         & \boldsymbol{\mathbbm{1}}_{\bar{m}-n_e}
    \end{pmatrix}
\end{equation}
with $n_e \leq \min(m, \bar{m})$. As we shall see below, $n_e$ is the number of entangled mode pairs contributing to the bipartite entanglement. The symplectic eigenvalues satisfy $D_q > 1$ and are arranged in descending order. Each $D_q$ appears twice in Eq.~\eqref{eq:symDiag}, reflecting the intrinsic twofold degeneracy of symplectic eigenvalues. This degeneracy introduces a gauge freedom in the choice of $\boldsymbol{M}_L$ and $\boldsymbol{M}_R$, which can be fixed by the off-diagonal block $\boldsymbol{\Gamma}_{LR}$:
\begin{equation}
\boldsymbol{\Gamma}_{LR} = \frac{1}{2}\boldsymbol{M}_L\begin{pmatrix}
\boldsymbol{P}_{LR}&\boldsymbol{0}  \\
  \boldsymbol{0} &-\boldsymbol{P}_{LR}\end{pmatrix}\boldsymbol{M}_R^T \, .
\end{equation}
Note that $\boldsymbol{\Gamma}_{LR}$ is the $2m\times{}2\bar{m}$ submatrix of $\boldsymbol{\Gamma}$ by restricting Eq.~\eqref{eq:GammaMat} to $1\leq{}j\leq{}m$ and $m+1\leq{}k\leq{}N$. The matrix $\boldsymbol{P}_{LR}$ is an $m \times \bar{m}$ matrix defined by
\begin{equation}
\boldsymbol{P}_{LR}=\begin{cases}
\begin{pmatrix}
    \sqrt{\boldsymbol{D}^2_L-\boldsymbol{\mathbbm{1}}_m}& \boldsymbol{0}
\end{pmatrix}, & m\le\bar{m} \, , \\[2mm]
\begin{pmatrix}
    \sqrt{\boldsymbol{D}^2_R-\boldsymbol{\mathbbm{1}}_{\bar{m}}} \\
\boldsymbol{0}
\end{pmatrix}, & m>\bar{m} \, .
\end{cases}
\end{equation}

The symplectic matrices $\boldsymbol{M}_L$ and $\boldsymbol{M}_R$ define Bogoliubov transformations in two subsystems. They can be written in terms of complex matrices $\boldsymbol{U}_A$ and $\boldsymbol{V}_A$ (with $A \in \{L, R\}$) as
\begin{equation}
\boldsymbol{M}_A = \begin{pmatrix}
-\mathrm{Im}{(\boldsymbol{U}_A+\boldsymbol{V}_A)}  &
-\mathrm{Re}{(\boldsymbol{U}_A+\boldsymbol{V}_A)} \\
~~\mathrm{Re}{(\boldsymbol{U}_A-\boldsymbol{V}_A)} &
-\mathrm{Im}{(\boldsymbol{U}_A-\boldsymbol{V}_A)}
\end{pmatrix}.
\label{eq:MUV}
\end{equation}
These transformations define Bogoliubov modes:
\begin{equation}
b^{\dagger}_{A,q} = \sum_{j\in{}A} [ (U_{A})_{j,q}a^{\dagger}_j-(V_{A})_{j,q} a_j ]
\label{eq:bmode}
\end{equation}
with $q = 1,\ldots,m$ ($q = 1,\ldots,\bar{m}$) for the subsystem $L$ ($R$). Below, we refer to these as $b$-modes; similar notation (e.g., $a$-modes and $d$-modes) will be used later to distinguish different mode bases.

In the basis of the $b$-modes, the BGS $|\phi\rangle$ admits a mode-wise Schmidt decomposition
\begin{equation}
    |\phi\rangle = |\text{vac}\rangle_{L} \otimes \left( \bigotimes_{q=1}^{n_e} |\phi_q\rangle \right)  \otimes |\text{vac}\rangle_{R},
\label{eq:SchmidtD}
\end{equation}
where $|\text{vac}\rangle_{L}$ and $|\text{vac}\rangle_{R}$ are the product vacua of the unentangled modes, which therefore do not contribute to the bipartite entanglement. The $b$-modes associated with them will be referred to as \emph{frozen modes} hereafter. Each $|\phi_q\rangle$ in Eq.~\eqref{eq:SchmidtD} is a two-mode Gaussian state,
\begin{equation}
|\phi_q\rangle = \sqrt{1-\Lambda_q^2} \sum_{n=0}^{\infty} \Lambda_q^n |n\rangle_{L,q} \otimes |n\rangle_{R,q}
\label{eq:SVD1}
\end{equation}
with squeezing parameter $\Lambda_q=\sqrt{\frac{D_q-1}{D_q+1}}$, where $|n\rangle_{A,q}$ ($A \in \{L,R\}$) are the Fock states of the Bogoliubov modes, defined by $|n\rangle_{A,q}= \frac{1}{\sqrt{n!}} (b^\dagger_{A,q})^n |0\rangle_{A,q}$ ($q = 1,\ldots,n_e)$. These $b$-modes contribute to the entanglement between two subsystems; they will be referred to as \emph{active modes}.

Using the Schmidt decomposition in Eq.~\eqref{eq:SchmidtD}, the entanglement entropy of the BGS $|\phi\rangle$ with respect to the bipartition is
\begin{equation}
 S = \sum_{q=1}^{n_e} S_q ,
\end{equation}
where each entangled mode pair contributes
\begin{equation}
S_q = -\log(1-\Lambda_q^2) - \frac{\Lambda_q^2}{1-\Lambda_q^2}\log(\Lambda_q^2).
\end{equation}
The entanglement is fully controlled by the squeezing parameter $\Lambda_q$ (equivalently by the symplectic eigenvalue $D_q$). The entropy $S_q$ increases monotonically with $\Lambda_q$: it vanishes when $\Lambda_q \rightarrow 0$ (i.e., $D_q \rightarrow 1$), corresponding to an unentangled mode pair for which $|\phi_q\rangle$ reduces to the vacuum, and it diverges in the maximally entangled limit $\Lambda_q \rightarrow 1$ (i.e., $D_q \rightarrow \infty$).

In numerical implementations, it is convenient to introduce a tolerance parameter $D^*$ slightly above unity. Modes with symplectic eigenvalues $D_q$ in the range $1 < D_q < D^*$ are effectively unentangled and can be treated as frozen modes in the numerical implementation.

\subsection{Iterative MPS construction via Gaussian singular value decomposition}

Having established the necessary formalism in Sec.~\ref{sec:methodA}, we now introduce an iterative GSVD procedure. This macro-level step decomposes the full $N$-mode BGS $|\phi\rangle$ into a sequence of interconnected, smaller local Gaussian states $\{|A^m\rangle\}$, naturally establishing the MPS structure.

\subsubsection{Gaussian Singular Value Decomposition}
\label{sec:GSVD}

To formulate the GSVD in a way that closely parallels the MPS construction, it is useful to discuss a Gaussian analogue of the SVD. We refer to this construction as the GSVD. Conceptually, the GSVD factorizes a pure BGS across a bipartition into three pure BGSs by introducing a set of virtual modes.

We begin by rewriting the Schmidt decomposition in Eq.~\eqref{eq:SchmidtD} as
\begin{equation}
|\phi\rangle \propto \prod_{q=1}^{n_e} \mathrm{exp}\left[\Lambda_q b^{\dagger}_{L,q}b^{\dagger}_{R,q} \right] |0\rangle_{b_L,b_R} \, ,
\label{eq:SVD2}
\end{equation}
where $|0\rangle_{b_L,b_R} $ denotes the joint vacuum of all $b$-modes [Eq.~\eqref{eq:bmode}], including both active and frozen ones.

To obtain a factorized representation, we introduce, for each active mode pair, two auxiliary or \emph{virtual} modes $l_{q}$ and $r_{q}$ ($q=1,\ldots,n_e$), which play the role of Schmidt legs in the Gaussian setting. Using these virtual modes, we define two Gaussian states:
\begin{align}
|\phi_L\rangle &= \prod_{q=1}^{n_e}\mathrm{exp}\left[b^{\dagger}_{L,q}l^{\dagger}_{q}\right]|0\rangle_{b_L,l},\nonumber \\
|\phi_R\rangle &= \prod_{q=1}^{n_e}\mathrm{exp}\left[ \Lambda_q b^{\dagger}_{R,q}r^{\dagger}_{q}\right]|0\rangle_{b_R,r} ,
\label{eq:SVD3}
\end{align}
which may be viewed as Gaussian maps from the virtual Hilbert spaces to the physical ones. The original state is then recovered by contracting the virtual modes through a maximally entangled Gaussian state,
\begin{align}
|\phi\rangle \propto \langle I_m|\left(|\phi_L\rangle \otimes |\phi_R\rangle\right),
\label{eq:SVD3_factorization}
\end{align}
where
\begin{align}
|I_m\rangle &= \prod_{q=1}^{n_e}\mathrm{exp}\left[l^{\dagger}_{q}r^{\dagger}_{q}\right]|0\rangle_{l,r}.
\label{eq:bond}
\end{align}
is the product of $n_e$ identical two-mode maximally entangled states of virtual modes. The structure in Eqs.~\eqref{eq:SVD3_factorization} and \eqref{eq:bond} is the exact Gaussian analogue of the factorization step in the MPS constructions, with ${l_q}$ and ${r_q}$ playing the role of virtual indices.

To make contact with our GSVD implementation, we rewrite $|\phi_L\rangle$ in terms of the original physical modes $a_{L,j}$. As shown in Appendix~\ref{app:paired_wf}, the Gaussian nature ensures that $|\phi_L\rangle$ can be rewritten as
\begin{align}
|\phi_L\rangle \propto \mathrm{exp}\left[\frac{1}{2}\begin{pmatrix}\boldsymbol{a}^{\dagger}_L&\boldsymbol{l}^{\dagger}\end{pmatrix}\boldsymbol{Q}_L\begin{pmatrix}
    \boldsymbol{a}^{\dagger}_L \\[2mm]
    \boldsymbol{l}^{\dagger}
\end{pmatrix}\right]|0\rangle_{a_L,l} \, ,
\label{eq:Q_matrix}
\end{align}
where the vector
\begin{align}
\begin{pmatrix}
\boldsymbol{a}^{\dagger}_L & \boldsymbol{l}^{\dagger}
\end{pmatrix}
= (a_1^\dagger,\ldots,a_m^\dagger,\;
 l_1^\dagger,\ldots,l_{n_e}^\dagger),
\end{align}
collects the creation operators of physical and virtual modes. The $(m+n_e)\times(m+n_e)$ matrix $\boldsymbol{Q}_L$ is determined entirely by the Bogoliubov matrices $\boldsymbol{U}_L$ and $\boldsymbol{V}_L$ in Eq.~\eqref{eq:MUV}, and its explicit form is presented in Appendix~\ref{app:paired_wf}. An analogous expression holds for $|\phi_R\rangle$. The arrangement of physical and virtual modes in this GSVD is illustrated in Fig.~\ref{fig:GSVD}(a).

The GSVD thus provides a Gaussian factorization of the original bipartite state into two Gaussian states, each supported on one subsystem, which are contracted through a maximally entangled Gaussian state of virtual modes~\footnote{Strictly speaking, the GSVD of a pure BGS takes the form $|\phi\rangle = \langle I_m|\left(|\phi_L\rangle \otimes |\phi_R\rangle\right)$, where $|\phi_L\rangle = \prod_{q=1}^{n_e}\mathrm{exp}\left[b^{\dagger}_{L,q}l^{\dagger}_{q}\right]|0\rangle_{b_L,l}$ and $|\phi_R\rangle = \prod_{q=1}^{n_e}\mathrm{exp}\left[b^{\dagger}_{R,q}r^{\dagger}_{q}\right]|0\rangle_{b_R,r}$ are Gaussian ``isometries'', and the bond state $|I_m\rangle = \prod_{q=1}^{n_e}\mathrm{exp}\left[\Lambda_q l^{\dagger}_{q}r^{\dagger}_{q}\right]|0\rangle_{l,r}$ carries the Gaussian ``singular values'' $\Lambda_q$. For convenience of the ensuing algorithm, we absorb the factors $\Lambda_q$ into $|\phi_R\rangle$ and take $|I_m\rangle$ to be a maximally entangled state of virtual modes mediating the entanglement across the cut.}. These virtual modes mediate the entanglement across the cut and supply the structure underpinning the iterative decomposition algorithm introduced below.

\subsubsection{Iterative Decomposition Algorithm}

Having established the GSVD for a single bipartition, we now build the full MPS representation by applying the decomposition iteratively for the whole system. At each step we factor out one physical mode together with its associated virtual modes, producing one local Gaussian state.  The algorithm proceeds as follows:

\begin{enumerate}
    \item \textbf{First mode $m=1$:} We start from the full state $|\phi\rangle$ and partition the system into left part $L_1$, containing the first mode $\{a^\dagger_1\}$, and right part $R_1$, containing the remaining physical modes $\{a^\dagger_2, \ldots, a^\dagger_N\}$. Applying the GSVD produces $|\phi\rangle \rightarrow \langle I_1|(|A^1\rangle \otimes |\phi_{R}^1\rangle)$. The state $|A^{1}\rangle$ is a BGS defined on the first physical mode $a^\dagger_1$ and a single \emph{outgoing} virtual mode $l^\dagger_{1}$, which carries all entanglement between site $1$ and the rest of the chain. The residual state $|\phi_{R}^1\rangle$ lives on physical modes $\{a^\dagger_2, \ldots, a^\dagger_N\}$ and contains the corresponding \emph{incoming} virtual mode $\{r^\dagger_{1}\}$. In the paired form [cf. Eq.~\eqref{eq:Q_matrix}], $|A^1\rangle$ and $|\phi_{R}^1\rangle$ read
    \begin{align}
    |A^1\rangle &= \mathrm{exp}\!\left[\frac{1}{2}\sum_{\mu,\nu}(Q^{1})_{\mu\nu}c^{\dagger}_{1,\mu}c^{\dagger}_{1,\nu}\right]|0\rangle_{c_1},\nonumber\\
    |\phi_{R}^1\rangle &= \mathrm{exp}\!\left[\frac{1}{2}\sum_{\mu,\nu}(Q^{1}_{R})_{\mu\nu}d^{\dagger}_{1,\mu}d^{\dagger}_{1,\nu}\right]|0\rangle_{d_1},
    \end{align}
    with $\boldsymbol{c}^\dagger_1=(a^\dagger_1,l^\dagger_1)$ and $\boldsymbol{d}^\dagger_1=(r^\dagger_1,a^\dagger_2,\ldots,a^\dagger_N)$. The entanglement mediator $|I_1\rangle$ is the maximally entangled Gaussian state, $|I_1\rangle = \mathrm{exp}(l^{\dagger}_{1}r^{\dagger}_{1})|0\rangle_{l_1,r_1}$.
    Note that the first cut can have at most one entangled mode. If the first physical mode is unentangled with the rest, then $|A^1\rangle$ reduces to a trivial single-mode BGS and one may simply begin the algorithm at $m=2$.

    \item \textbf{Bulk modes $m=2, \ldots, N-1$:} At step $m$, the input is the residual state $|\phi_{R}^{m-1}\rangle$, which is supported on physical modes $\{a^\dagger_m, \ldots, a^\dagger_N\}$ and the $n^{m-1}_e$ incoming virtual modes $\{r^\dagger_{m-1,q}\}$ ($q = 1,\ldots,n^{m-1}_{e}$) produced at the previous step. Here the notation $n^{m-1}_e$ denotes the number of entangled mode pairs at the $(m-1)$-th cut; the superscript is an index and should not be interpreted as an exponent. We now define a new partition: left part $L_m$, consisting of $n^{m-1}_{e}$ incoming virtual modes $\{r^\dagger_{m-1,q}\}$ and one physical mode $a^\dagger_m$; right part $R_m$, consisting of the remaining physical modes $\{a^\dagger_{m+1}, \ldots, a^\dagger_N\}$. Performing the GSVD on $|\phi_{R}^{m-1}\rangle$ across this partition yields the local state $|A^m\rangle$ and the next residual state $|\phi_{R}^{m}\rangle$. The state $|A^m\rangle$ is a BGS supported on three sets of modes: the $n^{m-1}_{e}$ incoming virtual modes $\{r^\dagger_{m-1,q}\}$, the physical mode $a^\dagger_m$, and a new set of $n^m_{e}$ outgoing virtual modes $\{l^\dagger_{m,q}\}$ ($q=1,\ldots,n^m_{e}$). In the paired form, $|A^m\rangle$ and $|\phi_{R}^m\rangle$ are given by
    \begin{align}
    |A^m\rangle &= \mathrm{exp}\!\left[\frac{1}{2}\sum_{\mu,\nu}(Q^{m})_{\mu\nu}c^{\dagger}_{m,\mu}c^{\dagger}_{m,\nu}\right]|0\rangle_{c_m},\nonumber\\
    |\phi_{R}^m\rangle &= \mathrm{exp}\!\left[\frac{1}{2}\sum_{\mu,\nu}(Q^{m}_{R})_{\mu\nu}d^{\dagger}_{m,\mu}d^{\dagger}_{m,\nu}\right]|0\rangle_{d_m},
    \label{eq:Am_ket}
    \end{align}
    with $\boldsymbol{d}^\dagger_m = (r^\dagger_{m,1},\ldots,r^\dagger_{m,n_e^m},a^\dagger_{m+1},\ldots,a^\dagger_N)$ and $\boldsymbol{c}^\dagger_m = (r^\dagger_{m-1,1},\ldots,r^\dagger_{m-1,n_e^{m-1}},a^\dagger_m,l^\dagger_{m,1},\ldots,l^\dagger_{m,n_e^m})$.
    This three-leg structure (the incoming virtual leg, the local physical leg, and the outgoing virtual leg) is precisely the structure of an MPS local tensor. The entanglement mediator $|I_m\rangle$ connects $\{l^\dagger_{m,q}\}$ to the incoming virtual modes $\{r^\dagger_{m,q}\}$ of the next step.

    \item \textbf{Last mode $m=N$:} The final residual state $|\phi_{R}^{N-1}\rangle$ contains the last physical mode $a^\dagger_N$ and the incoming virtual modes $\{r^\dagger_{N-1,q}\}$. Since there is no right subsystem to further partition, the GSVD is stopped. The local state $|A^N\rangle$ is identified with $|\phi_{R}^{N-1}\rangle$ and terminates the MPS. Its paired form reads
    \begin{align}
    |A^N\rangle = \mathrm{exp}\!\left[\frac{1}{2}\sum_{\mu,\nu}(Q^{N})_{\mu\nu}c^{\dagger}_{N,\mu}c^{\dagger}_{N,\nu}\right]|0\rangle_{c_N},
    \end{align}
    with $\boldsymbol{c}^\dagger_N=(r^\dagger_{N-1,1},\ldots,r^\dagger_{N-1,n_e^{N-1}},a^\dagger_N)$.

\end{enumerate}

After this iterative procedure, the BGS $|\phi\rangle$ is rewritten as
\begin{align}
    |\phi\rangle \propto \left( \otimes_{m=1}^{N-1} \langle I_m| \right) \left( \otimes_{m=1}^{N} |A^m\rangle\right),
    \label{eq:GMPS}
\end{align}
where $\{|A^m\rangle\}$ is a set of BGSs associated with physical modes, each playing the role of an MPS local map, as shown in Fig.~\ref{fig:GSVD}(b). The form in Eq.~\eqref{eq:GMPS} is equivalent to the Gaussian MPS introduced in Ref.~\cite{Schuch2008}. However, this form is not sufficient for our purposes, as practical computations require a concrete tensor with finite bond dimension. In the next subsection, we show how to \emph{optimally} truncate $|A^m\rangle$ into a concrete, finite-dimensional tensor using the truncation procedure described there.

\subsection{Generating MPS local tensor}
\label{sec:methodC}

The GSVD-based iterative decomposition produces a set of BGSs $\{|A^m\rangle\}$, where each $|A^m\rangle$ is supported on a local physical mode and its incoming and outgoing virtual modes. Each local state can be expanded in the Fock basis as
\begin{align}
|A^m\rangle = \sum_{\alpha,p,\beta} A^m_{\alpha p \beta}
|v^{m-1}_\alpha\rangle\otimes|p^m\rangle\otimes|v^{m}_\beta\rangle,
\label{eq:Am_definition}
\end{align}
where $|p^m\rangle$ labels the physical Fock states ($p=0,1,\ldots$) at site $m$, and $|v^{m-1}_\alpha\rangle$ ($|v_\beta^{m}\rangle$) labels the virtual Fock states associated with the virtual mode set $\{r^{\dagger}_{m-1,1},\cdots,r^{\dagger}_{m-1,n_e^{m-1}}\}$ ($\{l^{\dagger}_{m,1},\cdots,l^{\dagger}_{m,n_e^m}\}$) [see Eq.~\eqref{eq:Am_ket} and Fig.~\ref{fig:GSVD}(c)]. At this stage, both the physical and virtual indices are, in general, infinite-dimensional. Later in this subsection, we truncate both physical and virtual Fock basis states, yielding a finite-dimensional tensor $A^m_{\alpha p \beta}$ with $p=0,\ldots,d-1$ and $\alpha,\beta=0,\ldots,D-1$. In the standard MPS formalism, $D$ corresponds to the MPS bond dimension $\chi$~\cite{Vidal2003b,Perez-Garcia2007,Verstraete2008,Schollwoeck2011}.

In principle, obtaining the MPS tensor elements amounts to computing overlaps between $|A^m\rangle$ and multi-mode Fock states. A brute-force approach would require evaluating matrix hafnians of rapidly growing size and is therefore computationally prohibitive.

In this subsection, we introduce an efficient algorithm to convert $|A^m\rangle$ into its tensor representation $A^m_{\alpha p \beta}$ \emph{without} computing any hafnian. Our approach proceeds in two stages. First, in Sec.~\ref{sec:projected_subspace}, we identify an optimal finite-dimensional subspace of the local bosonic Hilbert space and project $|A^m\rangle$ into this subspace. Second, in Sec.~\ref{sec:pco_algorithm}, we develop a fast algorithm to evaluate all expansion coefficients in this subspace.

\subsubsection{The Kept Subspace and Projection}
\label{sec:projected_subspace}

To convert each $|A^m\rangle$ into a finite-dimensional MPS tensor, $A^m_{\alpha p \beta}$, we must project $|A^m\rangle$, which lives in an infinite-dimensional bosonic Hilbert space, onto a \emph{truncated} subspace of fixed dimension. Our first task is therefore to identify, for a given bond dimension $D$, the subspace that retains the maximal amount of physically relevant information.

For a bulk tensor $|A^m\rangle$ ($m = 2,\ldots, N-1$), the basis is spanned by three sets of modes: the $n^{m-1}_e$ incoming virtual modes $\{r^\dagger_{m, q}\}$ from the left bond, the local physical mode $a^\dagger_m$, and the $n^{m}_e$ outgoing virtual modes $\{l^\dagger_{m, q}\}$ for the right bond. The physical mode is naturally expanded by the Fock states $\{|p\rangle\}$ satisfying $a^{\dagger}_m a_m |p\rangle = p |p\rangle$ ($p=0,1,\ldots$), truncated by keeping a threshold occupation number $p \leq d-1$. For obtaining the optimal bases of the left and right virtual modes, a natural idea is to use the Schmidt decomposition. If one traces out all physical modes $\{a^\dagger_{m+1},\ldots,a^{\dagger}_{N}\}$ in $|\phi^{m}_R\rangle$, the reduced density operator for the virtual modes $\{r^\dagger_{m, q}\}$ reads
\begin{align}
\rho_{m} = \frac{1}{Z} e^{-H_{m}},
\label{eq:rho_m}
\end{align}
where the entanglement Hamiltonian $H_m$ is given by
\begin{align}
H_{m}=-\sum^{n^{m}_e}_{q=1}2\mathrm{log}(\Lambda_q)r^{\dagger}_{m,q}r_{m,q} \, ,
\label{eq:HE}
\end{align}
and the factor
\begin{align}
Z = \prod_{q=1}^{n^m_e} \frac{1}{1-\Lambda^2_q}
\label{eq:Z}
\end{align}
ensures the normalization $\mathrm{Tr}(\rho_m)=1$. This form directly follows from Eq.~\eqref{eq:SVD3}. Consequently, the eigenstates of $\rho_m$ are simply the Fock states of $\{r^\dagger_{m, q}\}$, and the eigenvalues (Schmidt coefficients) follow from the entanglement energies defined by $H_m$. The most relevant states (i.e., those we should keep in the kept subspace) are precisely the eigenstates with the \textit{lowest} entanglement energies. Since the modes ${l^\dagger_{m,q}}$ and ${r^\dagger_{m,q}}$ form a maximally entangled pair in $|I_{m}\rangle$, the kept virtual spaces for $\{r^{\dagger}_{m,q}\}$ and $\{l^{\dagger}_{m,q}\}$ are identical.

Concretely, the optimal $D$-dimensional basis for the virtual modes $\{l^\dagger_{m, q}\}$ (or equivalently $\{r^\dagger_{m, q}\}$) is as follows:
\begin{enumerate}
    \item Enumerate virtual Fock states $|v^{m}_\beta\rangle = |n^{m}_{\beta,1}, \dots, n^{m}_{\beta,n^m_e}\rangle$, where $n^{m}_{\beta,q}$ denotes the occupation number of the virtual mode $r^\dagger_{m,q}$.
    \item Compute their entanglement energies $E^m_\beta = -\sum_q 2n^{m}_{\beta,q}\mathrm{log}(\Lambda_q)$.
    \item Sort states by increasing $E^m_\beta$ and retain the first $D$ states.
\end{enumerate}
The basis for the incoming virtual modes ${r^\dagger_{m-1,q}}$ can be obtained in a similar manner.

The above procedure allows us to construct the optimal virtual bases $\{|v^{m-1}_\alpha\rangle\}$ and $\{|v^{m}_{\beta}\rangle\}$ for the incoming and outgoing virtual modes $\{r^\dagger_{m-1, q}\}$ and $\{l^\dagger_{m, q}\}$. Together with the truncated physical basis, they define a computationally manageable subspace:
\begin{align}
\mathbbm{V}_m = \{|v^{m-1}_\alpha\rangle \otimes |p^m\rangle \otimes |v^{m}_\beta\rangle\}.
\end{align}
The finite-dimensional MPS tensor is then obtained by projecting $|A^m\rangle$ into this subspace [see Fig.~\ref{fig:GSVD}(c)]:
\begin{equation}
A^m_{\alpha p \beta} = \left( \langle v^{m-1}_{\alpha} |\otimes \langle p^m |\otimes \langle v^{m}_{\beta} | \right) P_{\mathbbm{V}_m}|A^m\rangle,
\label{eq:Am_apb}
\end{equation}
where $P_{\mathbbm{V}_m}$ denotes the projector from the infinite-dimensional Hilbert space (before truncation) onto the subspace $\mathbbm{V}_m$. Importantly, the projected state, defined as $|\tilde{A}^m\rangle \equiv P_{\mathbbm{V}_m}|A^m\rangle$, is no longer Gaussian because of the projection. Although this sacrifices the analytic simplicity of the Gaussian formalism, the truncation is optimal at the level of each bond, in the sense that it is performed with respect to the reduced density operator (i.e., a Schmidt/SVD truncation). We emphasize that this procedure yields a locally optimal approximation within the full MPS manifold, but does not guarantee a globally optimal approximation at a fixed bond dimension. Nevertheless, such locally optimal truncation schemes are standard in MPS approaches~\cite{Vidal2003b,Verstraete2008} and are known to provide highly accurate approximations in practice. As we shall see, this choice enables us to construct a computationally efficient, finite-dimensional MPS representation.

\subsubsection{The Core Algorithm: Sequential Application of Projected Creation Operators}
\label{sec:pco_algorithm}

A direct calculation of the MPS tensor entries in Eq.~\eqref{eq:Am_apb} would require calculating hafnians, a computationally expensive task we circumvent with an alternative algorithmic approach. Recall that the local state $|A^m\rangle$ is a BGS defined on three sets of modes: the incoming virtual modes from the left bond, $\{r^\dagger_{m-1,q}\}$ ($q = 1,\ldots,n^{m-1}_{e}$); the local physical mode, $a^\dagger_m$; and the outgoing virtual modes for the right bond, $\{l^\dagger_{m,q}\}$ ($q = 1,\ldots,n^{m}_{e}$).
As a BGS, $|A^m\rangle$ has a paired-form representation as in Eq.~\eqref{eq:Q_matrix}. For notational simplicity, we define a vector $\boldsymbol{d}^\dagger$,
\begin{align}
\boldsymbol{d}^{\dagger} = (r^{\dagger}_{m-1,1}, \dots, r^{\dagger}_{m-1,n^{m-1}_e}, a^{\dagger}_m, l^{\dagger}_{m,1}, \dots, l^{\dagger}_{m,n^{m}_e}),
\label{eq:combine_modes}
\end{align}
which contains $n_{\text{modes}} = n_e^{m-1} + n_e^m + 1$ modes. Up to an unimportant normalization factor, the local BGS $|A^m\rangle$ can be written as
\begin{align}
|A^m\rangle=\mathrm{exp}\left[\frac{1}{2}\sum_{\zeta,\zeta^{\prime} = 1}^{n_{\text{modes}}} Q^m_{\zeta, \zeta^{\prime}}d^{\dagger}_{\zeta} d^{\dagger}_{\zeta^{\prime}}\right]|0\rangle_d,
\label{eq:local_Am_state}
\end{align}
where $|0\rangle_d$ is the vacuum of all $d$-modes, and $\boldsymbol{Q}^m$ is a symmetric matrix.

The key to our algorithm lies in a property of the truncated subspace $\mathbbm{V}_m$. By construction, this basis is ordered by entanglement energy, and the creation operators $d^\dagger_{\zeta}$ always map states to others with equal or higher entanglement energy. Consequently, a creation operator cannot map a state with higher entanglement energy outside $\mathbbm{V}_m$ to a state with lower entanglement energy inside $\mathbbm{V}_m$. This ensures that the subspace is closed under the action of creation operators, leading to an important identity:
\begin{align}
\label{eq:pass_projector_restated}
P_{\mathbbm{V}_m} d^{\dagger}_{\zeta} = P_{\mathbbm{V}_m} d^{\dagger}_{\zeta} P_{\mathbbm{V}_m}, \; \forall \zeta.
\end{align}
Using this identity, we can expand the exponential in Eq.~\eqref{eq:local_Am_state} and write the projected state $P_{\mathbbm{V}_m}|A^m\rangle$ as
\begin{align}
|\tilde{A}^m\rangle = P_{\mathbbm{V}_m}|A^m\rangle =
\prod_{1 \leq \zeta \leq \zeta^{\prime} \leq n_{\text{modes}}} \mathbbm{P}^m_{\zeta,\zeta^{\prime}} |0\rangle_d \, ,
\end{align}
where $\mathbbm{P}^m_{\zeta,\zeta^{\prime}}$ are the PCOs, defined as
\begin{equation}
\mathbbm{P}^m_{\zeta,\zeta^{\prime}} = P_{\mathbbm{V}_m}\sum^{\tilde{n}_{\zeta,\zeta^{\prime}}}_{\kappa=0}\frac{1}{\kappa!}\left[\left(1-\frac{\delta_{\zeta,\zeta^{\prime}}}{2}\right)Q^m_{\zeta,\zeta^{\prime}}d^{\dagger}_{\zeta} d^{\dagger}_{\zeta^{\prime}}\right]^{\kappa}.
\end{equation}
Here, the polynomial series for each PCO is finite, truncated at the maximum allowed joint occupation $\tilde{n}_{\zeta,\zeta^{\prime}}$ for modes $\zeta$ and $\zeta^{\prime}$ within $\mathbbm{V}_m$ ($\tilde{n}_{\zeta,\zeta}=\lfloor{\tilde{n}_{\zeta}/2}\rfloor$, where $\tilde{n}_{\zeta}$ is the maximum allowed occupation number of the $d_{\zeta}$-mode within $\mathbbm{V}_m$).

The construction of $|\tilde{A}^m\rangle$ is an iterative procedure. It begins with the vacuum state $|0\rangle_d$, which is one of the kept states in $\mathbbm{V}_m$. The state $|\tilde{A}^m\rangle$ is then built by sequentially applying each PCO one by one. Note that the order of application of the PCOs is irrelevant due to bosonic commutation relations. The practical application of a single $\mathbbm{P}^m_{\zeta,\zeta^{\prime}}$ does not involve writing down its explicit $dD^2 \times dD^2$ matrix form. Instead, we compute its action directly on the state vector, which is stored as a rank-3 tensor of size $D \times d \times D$. The ``search-and-update'' algorithm proceeds as follows:
\begin{enumerate}
    \item Iterate through each of the $dD^2$ basis states $|v_k\rangle \in \mathbbm{V}_m$ and its coefficient $c_k$ in the current state vector.
    \item For each non-zero $c_k$, apply the PCO $\mathbbm{P}^m_{\zeta,\zeta^{\prime}}$ to the basis state $|v_k\rangle$. This generates a linear combination of distinct target basis states:
    \begin{align}   \mathbbm{P}^m_{\zeta,\zeta^{\prime}}|v_k\rangle&=\sum_{\kappa}g^{\zeta,\zeta^{\prime}}_{\kappa}|v^{\zeta,\zeta^{\prime}}_{k,\kappa}\rangle, \nonumber
    \end{align}
    where $|v^{\zeta,\zeta^{\prime}}_{k,\kappa}\rangle$ is the unit basis vector in $\mathbbm{V}_m$ such that
    \begin{align}
    |v^{\zeta,\zeta^{\prime}}_{k,\kappa}\rangle \propto (d^{\dagger}_{\zeta} d^{\dagger}_{\zeta^{\prime}})^{\kappa}|v_k\rangle,
    \label{eq:target}
    \end{align}
    and $g^{\zeta,\zeta^{\prime}}_{\kappa}$ is the corresponding expansion coefficient, whose explicit value is derived in Appendix~\ref{app:fast_search}.
    \item For each resulting target state $|v^{\zeta,\zeta^{\prime}}_{k, \kappa}\rangle$, we identify its index in the basis of $\mathbbm{V}_m$ and add the corresponding contribution ($c_k \times g^{\zeta,\zeta^{\prime}}_{\kappa}$) to the resulting vector.
\end{enumerate}

We now analyze the computational complexity. The cost of applying one PCO is dominated by the loop over the $dD^2$ coefficients, with each step involving $\mathcal{O}(\tilde{n}_{\zeta,\zeta^{\prime}})$ operations. The cost of a single application is therefore $\mathcal{O}(\tilde{n}_{\zeta,\zeta^{\prime}} dD^2)$. To obtain $|\tilde{A}^m\rangle$, we must apply a PCO for each unique pair of modes. The total number of PCOs scales as $\mathcal{O}(n_{\text{modes}}^2)$. Denoting the average occupation number in $|A^m\rangle$ as $\tilde{n}=\frac{1}{n_{\text{modes}}}\sum_{\zeta=1}^{n_{\text{modes}}}\langle A^m|d^{\dagger}_{\zeta}d_{\zeta}|A^m\rangle$ and assuming $n_e^{m-1} \approx n_e^m \approx n_e$, the computational cost for obtaining the MPS tensor from $|\tilde{A}^m\rangle$ is $\mathcal{O}(n_e^2 \cdot \tilde{n} \cdot dD^2)$. This polynomial scaling with the number of modes indicates that the notorious \#P problem of calculating matrix hafnians is circumvented via our PCO mapping algorithm.

Before proceeding, we emphasize that while the formal complexity of applying a single PCO $\mathbbm{P}^m_{\zeta,\zeta^{\prime}}$ is $\mathcal{O}(\tilde{n}_{\zeta,\zeta^{\prime}} dD^2)$, this procedure involves substantial computational subtleties. Although the multiplication operation itself is mathematically straightforward, the challenge lies in efficiently determining the index of the resulting target basis state. For the details of the implementation, we refer interested readers to Appendix~\ref{app:fast_search}, where an optimized state-indexing algorithm is presented. The key procedure is to establish a mapping from initial states to target states for each mode pair $d^{\dagger}_{\zeta} d^{\dagger}_{\zeta^{\prime}}$. The construction of mappings can be performed separately for the incoming and outgoing virtual spaces as $r^{\dagger}_{m-1,q}$ and $l^{\dagger}_{m,q}$ do not influence the states in each other's virtual space. This separation enables the construction of all mode mappings with $\mathcal{O}(\tilde{n}\cdot n^2_e \cdot D)$ complexity---negligible compared to the overall $\mathcal{O}(\tilde{n} \cdot n_e^2 \cdot dD^2)$ computational cost. A key consequence of these mappings is that the action of a PCO  $\mathbbm{P}^m_{\zeta,\zeta^{\prime}}$ on the subspace $\mathbbm{V}_m$ is highly sparse. The coefficients of only a small fraction of the $dD^2$ basis vectors need to be updated, indicating that the actual complexity is even lower than $\mathcal{O}(\tilde{n} \cdot n_e^2 \cdot dD^2)$. This represents a crucial improvement over methods relying on direct hafnian calculations~\cite{Oh2024}, which become computationally intractable when the number of entangled modes becomes large, leading to an exponential scaling complexity of $\mathcal{O}( 2^{\tilde{n}} \cdot dD^2)$.

\begin{figure}[tp]
    \centering
    \includegraphics[width=1.0\linewidth]{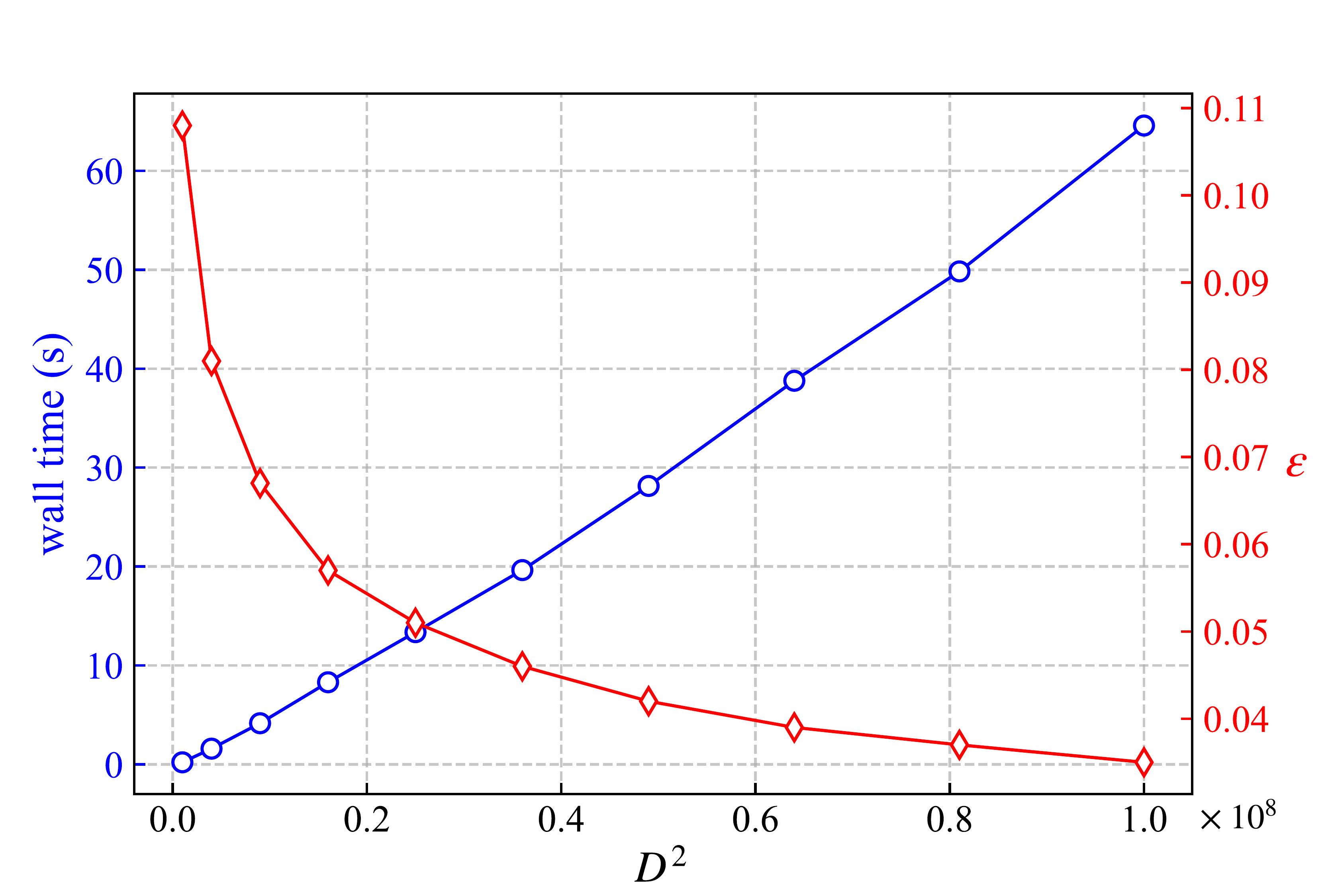}
    \caption{Performance of the PCO mapping algorithm for the Jiuzhang~2.0 J2-P65-5  system. The wall timed (blue line) and truncation error (red line) as a function of $D^2$ for constructing the MPS tensor on the $72$th mode. The error bars for the wall time are contained within the markers. All computations were performed on on a laptop~\cite{laptop}. }
    \label{fig:Jiuzhang2}
\end{figure}

We further explore how the theoretical complexity, $\mathcal{O}(\tilde{n} n_e^2 d D^2)$, scales with the entanglement properties of the system. The bond dimension $D$ required to achieve a target discarded weight $\varepsilon$ is determined by the entanglement energies $E^m_\beta$ [see Eqs.~\eqref{eq:rho_m}, \eqref{eq:HE} and the discussion below Eq.~\eqref{eq:Z}]:
\begin{equation}
D(\varepsilon) = \min \left\{ D ~\Big|~ 1 - \frac{1}{Z} \sum_{\beta=1}^{D} e^{-E^m_\beta} \le \varepsilon \right\}.
\end{equation}
Physically, a flatter entanglement spectrum, corresponding to stronger entanglement, leads to a slower decay of the Schmidt coefficients $e^{-E^m_\beta}$ and thus requires a larger bond dimension $D$. As shown in Figs.~\ref{fig:Jiuzhang2} and \ref{fig:S64}, our numerical results demonstrate that $n_e$ and $\tilde{n}$ remain relatively stable as $D$ increases to suppress truncation errors. This indicates that increasing entanglement is predominantly reflected in the growth of the required bond dimension $D$.

\begin{figure*}[t]
    \centering
    \includegraphics[width=1\linewidth]{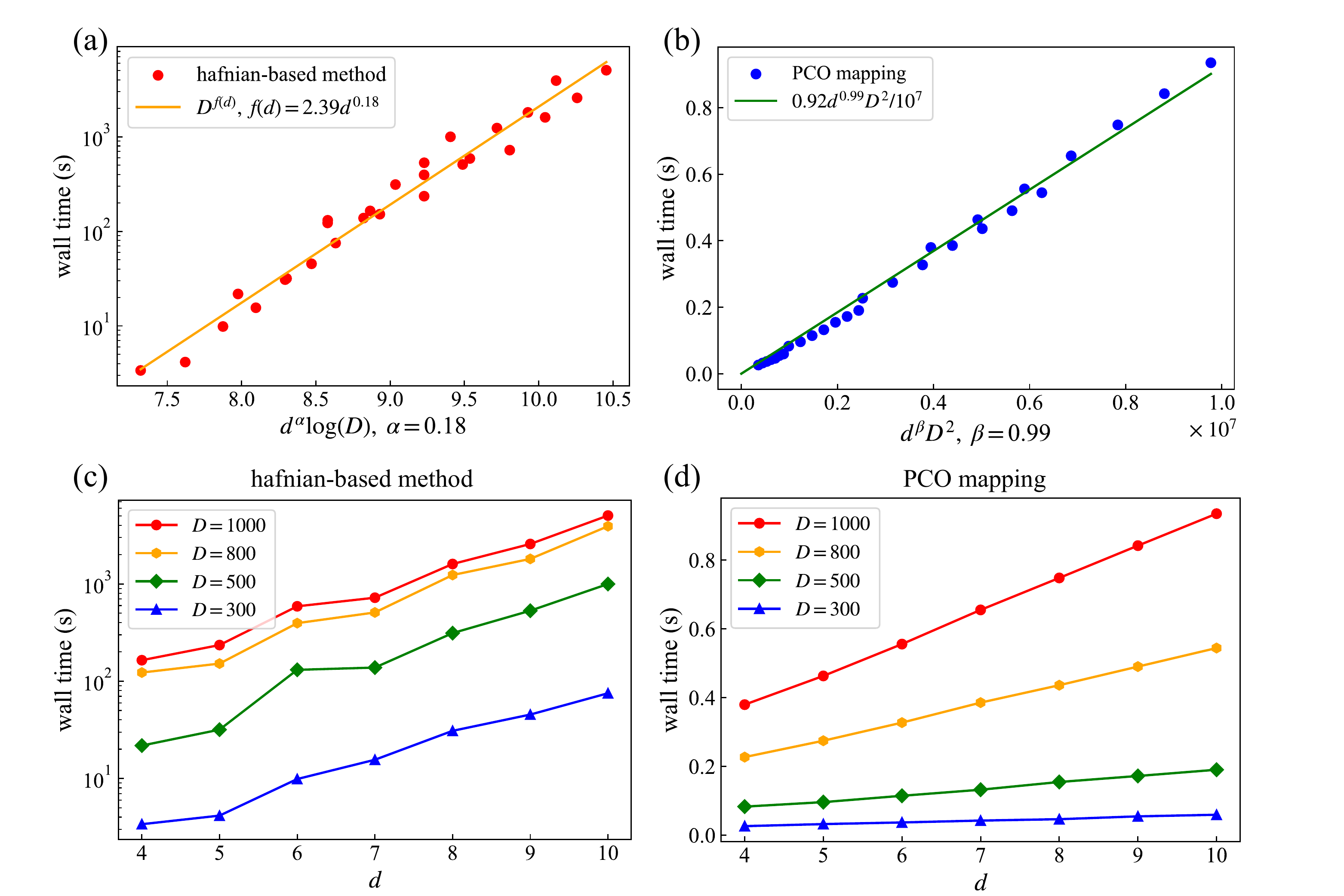}
    \caption{Comparison between the hafnian-based and the PCO mapping methods. The performance is benchmarked by constructing an MPS tensor for the 7th mode in Jiuzhang~2.0 J2-P65-5 system.  (a, b) Data collapse of the measured wall times for (a) the hafnian-based method, shown as a function of $d^{\alpha}\log(D)$, and (b) the PCO mapping method, as a function of $d^{\beta}D^2$. (c, d) The same wall time results replotted to show their dependence on the local dimension $d$, with separate curves for each value of $D$. The wall times for the hafnian-based method (a, c) are presented on a logarithmic scale.
    The error bars for the wall time are smaller than the marker size.}
    \label{fig:time_compare}
\end{figure*}

\section{numerical results}
\label{sec:results}

To benchmark our algorithm, we apply it to simulate data from recent large-scale GBS experiments (Jiuzhang~2.0~\cite{Zhong2021} and Jiuzhang~4.0~\cite{LiuHL2025}) and compare the results with those obtained using the hafnian-based approach~\cite{Oh2024}. Note that the raw experimental data give covariance matrices describing \emph{mixed} Gaussian states, which arise from decoherence processes such as photon loss. In practice, this decoherence highly suppresses the overall entanglement. Consequently, our task is divided into two steps. First, before implementing the MPS conversion, we extract a pure BGS from the experimentally produced mixed-state covariance matrix. Second, we construct the MPS representation for the pure BGS using our method described in Sec.~\ref{sec:method}. The resulting MPS provides a direct route to efficient classical simulations, e.g., generating boson samples~\cite{Oh2024}. We emphasize that this section benchmarks the MPS conversion stage rather than introducing a new sampling algorithm.

To prepare the input for our MPS conversion algorithm, we first isolate the pure-state component from the experimentally produced mixed-state covariance matrix $\boldsymbol{\Gamma}_{\text{th}}$:
\begin{align}
    \boldsymbol{\Gamma}_{\text{th}} = \boldsymbol{\Gamma}_{p} + \boldsymbol{W},
    \label{eq:decompose}
\end{align}
where $\boldsymbol{\Gamma}_{p}$ corresponds to the pure BGS and $\boldsymbol{W}$ is a positive semi-definite matrix accounting for the thermal noise~\cite{Quesada2022}. This decomposition interprets the mixed state as a pure quantum state subjected to classical noise which does not introduce any quantum entanglement. Note that the entanglement is characterized by the total number of squeezed photons:
\begin{align}
    N_{\text{eff}} = \frac{1}{2}  \text{tr}(\boldsymbol{\Gamma}_{p}-\boldsymbol{\mathbbm{1}}) \, .
\end{align}
Therefore, the optimal decomposition is obtained by minimizing the trace of $\boldsymbol{\Gamma}_{p}$ subject to the constraint that  $\boldsymbol{W}$ remains positive semidefinite~\cite{Oh2023}. This ensures that the input state of our algorithm possesses a minimal photon number and, consequently, minimal overall entanglement.

For systems with a small number of photon modes, the optimal decomposition can be found using brute-force nonlinear optimization~\cite{Oh2024}. However, the computational cost of this approach becomes prohibitive at the scale of recent experiments generating thousands of output photon modes~\cite{LiuHL2025}. To address this scaling challenge, we derive a controllable optimization algorithm for large-scale systems; see details in Appendix.~\ref{app:opt_Gamma}.

We benchmark our algorithm against data from two recent GBS experiments: Jiuzhang~2.0~\cite{Zhong2021} and Jiuzhang~4.0~\cite{LiuHL2025}. The pure-state covariance matrices $\boldsymbol{\Gamma}_{p}$, serving as the input states for MPS conversion, are extracted from the experimental data using our optimization algorithm in Appendix~\ref{app:opt_Gamma}. In Sec.~\ref{sec:J2}, we focus on the smaller-scale Jiuzhang~2.0 experiment. For comparison, the hafnian-based approach in Ref.~\cite{Oh2024}, which successfully simulated this experiment, provides a baseline. We shall see that our method achieves an order-of-magnitude speedup in the MPS preparation time. We then proceed to the significantly larger-scale Jiuzhang~4.0 experiment in Sec.~\ref{sec:J4}. For this scale, the hafnian-based approach becomes infeasible due to the large computational cost. Consequently, we exclusively employ our optimization algorithm to evaluate its performance and estimate the MPS preparation time for S64, the smallest-scale configuration for the Jiuzhang~4.0 experiment.

\subsection{Jiuzhang~2.0}
\label{sec:J2}

The first benchmark utilizes data from the Jiuzhang~2.0 experiment~\cite{Zhong2021}, one of the most challenging GBS platforms for classical simulation~\cite{Oh2024}. We focus on its most computationally demanding configuration, J2-P65-5, which has an effective squeezed-photon number $N_{\text{eff}} \approx 4.96$ and therefore represents the hardest instance to simulate.

We now demonstrate the performance and accuracy of our algorithm on the J2-P65-5 data~\cite{Zhong2021}. A key advantage of our GSVD-based PCO mapping algorithm is its inherent parallelism: each MPS tensor can be generated independently. We therefore focus on constructing the tensor at the central (72nd) mode of the 144-mode system. This bipartition carries one of the largest entanglement entropies and typically constitutes the computational bottleneck; all other tensors can be produced in comparable or less time.

For this benchmark, we evaluate two quantities: the wall time (computational performance) and the truncation error $\varepsilon$ (simulation fidelity). As discussed in Sec.~\ref{sec:projected_subspace}, converting a Gaussian state into an MPS tensor requires projecting the infinite-dimensional Gaussian basis onto a finite $D$-dimensional Fock subspace. This projection introduces a local truncation error $\varepsilon_m$ associated with bond $m$, defined by the discarded spectral weight of the entanglement Hamiltonian [Eq.~\eqref{eq:HE}],
\begin{equation}
\varepsilon_m = 1 - \frac{1}{Z} \sum_{\beta=1}^{D} e^{-E^{m}_{\beta}}.
\end{equation}
The global truncation error of the MPS is taken as the maximum over all bonds,
\begin{equation}
\varepsilon = \max_{m \in \{1, \dots, N-1\}} \varepsilon_m \, .
\end{equation}
In practice, the largest error typically occurs at the central bonds, $m \sim N/2$.

Figure~\ref{fig:Jiuzhang2} shows the wall time and the corresponding truncation error $\varepsilon$ for this task, both plotted as a function of the squared bond dimension $D^2$. The maximum local photon number is set to three $(d=4)$. At the largest bond dimension considered, $D=10^4$, our algorithm generates the central MPS tensor in approximately one minute on a standard laptop~\cite{laptop}, while achieving a small truncation error of $\varepsilon\approx0.03$. For comparison, the hafnian-based method in Ref.~\cite{Oh2024} required 9.5 minutes for the same task on a high-performance A100 GPU. Achieving such a low error together with a substantial speed-up highlights both the accuracy and efficiency of our approach for this demanding task.

A key advantage of our PCO mapping algorithm lies in its computational scaling with the local physical dimension $d$. In hafnian-based methods, the wall time grows exponentially with $d$. This occurs because these methods compute the exact amplitude of each Fock state, and the computational cost of evaluating a hafnian increases exponentially with the \textit{total photon number} in that Fock state. As either the physical dimension $d$ or bond dimension $D$ is increased, Fock states with larger photon numbers appear, leading to exponentially more expensive amplitude evaluations. In contrast, our algorithm completely avoids this bottleneck.

To quantify this effect, we benchmark the wall time for generating a single MPS tensor. For hafnian methods, the computation becomes very expensive for central modes at high $d$. We therefore select the 7th tensor as a representative case that remains tractable for comparison. The benchmark is performed for $d = 4$ to $10$ and bond dimensions $D \in \{300, 500, 800, 1000\}$. The PCO mapping runs on a laptop~\cite{laptop}, while the hafnian-based computations~\cite{LiuMZ2023-Data} are carried out on a workstation equipped with AMD 7H12@2.6GHz CPUs and 512 GB of RAM. As shown in Fig.~\ref{fig:time_compare}, the results reveal a stark performance gap. The runtime of the hafnian-based method grows exponentially with $d$, fitting approximately to $\sim D^{2.39d^{0.18}}$. In contrast, our method exhibits a much more favorable polynomial scaling, with runtime $\sim d^{0.99}D^2$. This shift from exponential to polynomial scaling represents a fundamental improvement in computational efficiency, enabling practical simulations at large local dimensions and/or large bond dimensions.

To validate our method, we verified that the MPS generated by our algorithm is essentially identical to that obtained from the hafnian-based approach, achieving a wavefunction fidelity of 0.9999. Such high fidelity guarantees that any downstream sampling would reproduce the same statistical outcomes. Since our contribution lies in accelerating the MPS construction itself, we do not repeat the computationally intensive sampling procedure of Ref.~\cite{Oh2024}, which is orthogonal to the algorithmic improvement presented here.

\subsection{Jiuzhang 4.0}
\label{sec:J4}

The advanced Jiuzhang~4.0 experiment~\cite{LiuHL2025} serves as our next and more challenging benchmark. It represents a major step forward in scalability and fidelity, producing configurations with substantially larger effective squeezed-photon numbers $N_{\text{eff}}$, a key indicator of classical simulation complexity. We examine three representative instances reported in the experiment: S64 (4336 output modes), M256 (5104 output modes), and the largest L1024 (8176 output modes). The squeezed-photon numbers --- approximately $N_{\text{eff}}\approx7$ for S64 and $N_{\text{eff}}\approx29$ for M256, obtained using the gradient method in Appendix~\ref{app:opt_Gamma}, are consistent with those reported in the original study.

\begin{figure}[tp]
    \centering
    \includegraphics[width=1.0\linewidth]{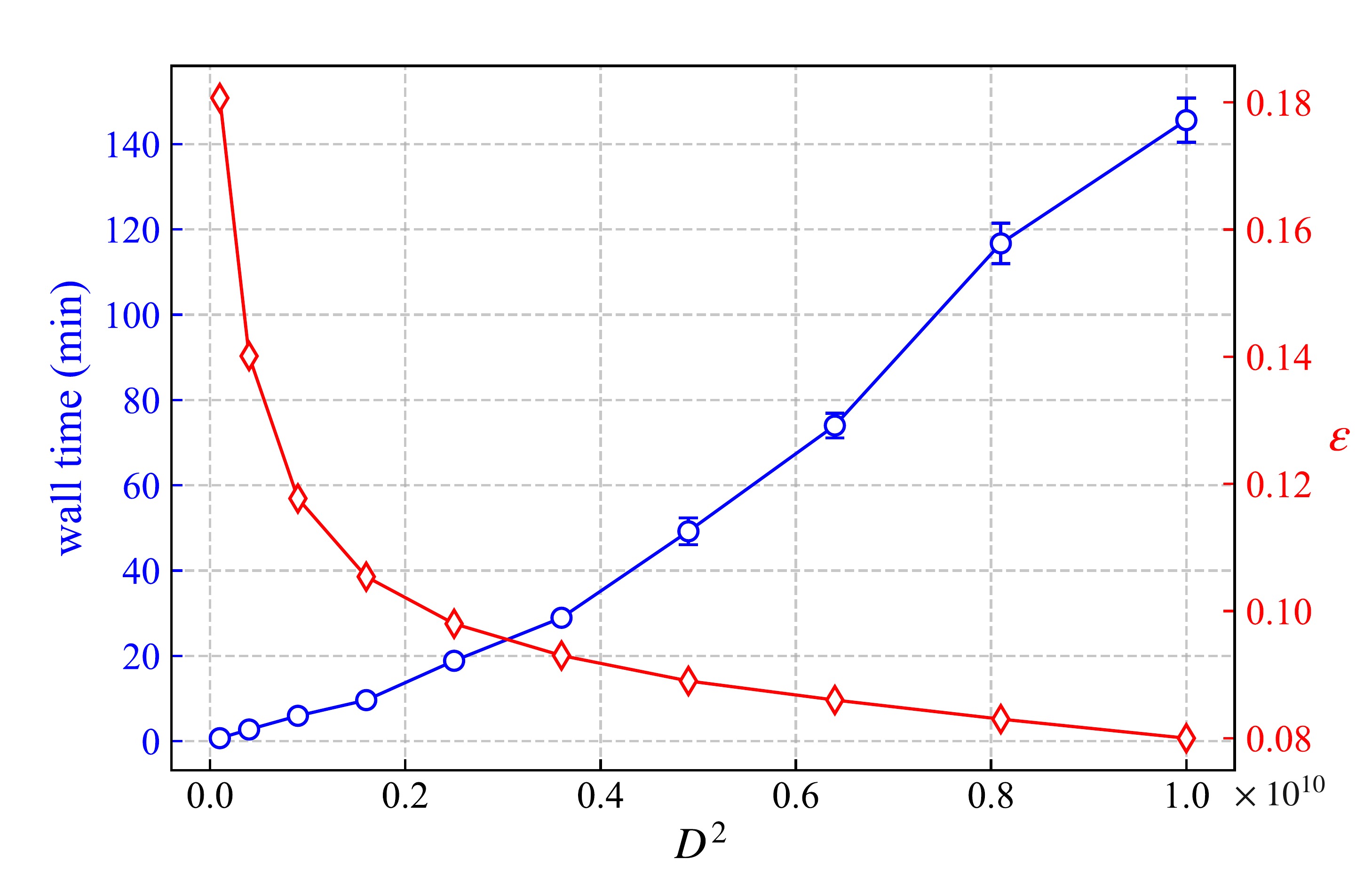}
    \caption{Performance of the PCO mapping algorithm for the Jiuzhang~4.0 S64 system. The wall time (blue line) and the truncation error (red line) as a function of $D^2$ for constructing the MPS tensor on the $2168$th mode. All computations were performed on a workstation with 32 Intel(R) Xeon(R) Gold 6326 CPUs and 512 GB of RAM.}
    \label{fig:S64}
\end{figure}

Because the large $N_{\text{eff}}$ values in M256 and L1024 generate very high entanglement, these two datasets are beyond the reach of our current MPS-based classical simulation. We therefore focus our performance evaluation on the more tractable S64 configuration. To benchmark the algorithm, we analyze a single central tensor (the 2168th mode of the 4336-mode system) and measure both the wall time and the associated truncation error, as shown in Fig.~\ref{fig:S64}. All simulations are performed on a workstation equipped with 32 Intel(R) Xeon(R) Gold 6326 CPUs and 512 GB of RAM. Although the results shown use a local dimension $d=1$, the procedure parallelizes naturally to larger $d$ across multiple nodes, as detailed in Appendix~\ref{app:fast_search}.

For the S64 system, a local tensor with bond dimension $D = 10^5$ can be generated in under 2.5 hours. At this scale, the resulting truncation error is $\varepsilon = 0.08$, which is sufficient for the practical purposes of this study. While the wall time indicates that even larger bond dimensions may be computationally accessible, the primary limitation becomes memory consumption, which scales as $\mathcal{O}(D^2)$. Sampling poses an even greater challenge: the memory required to store the full MPS exceeds the capacity of a single high-performance GPU such as an NVIDIA A100. We estimate that generating $10^7$ samples would require a highly parallelized workflow distributed over $4336d$ nodes, roughly taking on the order of ten days.

\section{Summary and Outlook}
\label{sec:summary}

In summary, we have developed an algorithm for efficiently converting low-entanglement bosonic Gaussian states into MPSs and incorporated it into a classical simulation framework for Gaussian boson sampling. This approach yields a substantial computational advantage over hafnian-based methods. For example, in the MPS preparation task for Jiuzhang~2.0, our algorithm completes in roughly one minute on a standard laptop, compared with 9.5 minutes for a hafnian-based implementation running on an A100 GPU.

We have further introduced a decomposition procedure that extracts the pure Gaussian component of mixed states arising in lossy GBS experiments. This method scales to systems with thousands of modes, enabling efficient preprocessing for large configurations such as S64 and M256 in Jiuzhang~4.0, and it may also assist recently proposed simulation strategies~\cite{Dodd2025}.

While MPS-based classical simulation becomes infeasible for extremely large-scale GBS experiments (such as Jiuzhang~4.0) where the required bond dimension grows prohibitively, our method remains highly effective for systems obeying area-law entanglement. This regime includes many condensed matter systems with short-range interactions. Moreover, unlike hafnian-based approaches, which are strongly affected by large single-mode occupation numbers (typical in condensed matter settings where entanglement is concentrated in a few modes), our algorithm is largely insensitive to such distributions. This robustness constitutes an additional advantage over conventional simulation schemes.

These features suggest that the method may have significant potential in quantum many-body computations. Possible applications include improving the spin-wave theory by incorporating Holstein-Primakoff boson-number constraints~\cite{LiuK2026} or, in combination with automatic differentiation techniques~\cite{LiaoHJ2019}, enabling efficient simulations of Gutzwiller-projected Schwinger boson mean-field states~\cite{Tay2011}.

\vspace{0.35cm}

\textbf{Data availability:}
The code for the article is available from the authors upon reasonable request. The data that support the findings of this article are openly available at \href{https://zenodo.org/records/18045305}{https://zenodo.org/records/18045305}.

\section*{Acknowledgments}
We thank Chao-Yang Lu for correspondence regarding the Jiuzhang 4.0 data. T.L. and T.X. are supported by the National Natural Science Foundation of China under Grant No. 12488201 and the Innovation Program for Quantum Science and Technology (Grant No. 2021ZD0301800). H.-K.J. acknowledges the support from National Natural Science Foundation of China under Grant No. 12504180 and the start-up funding from ShanghaiTech University.

\textbf{Contributions:}
H.-H. Tu and H.-K. Jin jointly conceived and supervised the project. T. Liu and H.-K. Jin developed the algorithm and performed the numerical simulations. T. Xiang provided scientific guidance and critically revised the manuscript. All authors contributed to data analysis and manuscript preparation.

\textbf{Competing interests:}
The authors declare no competing interests.

\appendix
\section{Derivation of the paired-form wave function}
\label{app:paired_wf}

In this Appendix, we derive the paired-form wave function from the Bogoliubov transformation. To this end, we adopt the setup from Sec.~\ref{sec:methodA}, namely, a free bosonic system with creation (annihilation) operators $a^\dagger_j$ ($a_j$) for $j = 1, \ldots, N$. The Bogoliubov modes $b_k$ that annihilate the BGS $|\phi\rangle$ (i.e., $b_k|\phi\rangle=0 \; \forall k$) are defined by
\begin{align}
b_{k} = \sum_{j=1}^N (U^*_{jk} a_{j} - V^{*}_{jk} a^{\dagger}_j),
\end{align}
or, more compactly, in the vector form:
\begin{align}
\boldsymbol{b}=\boldsymbol{U}^{\dagger}\boldsymbol{a}-\boldsymbol{V}^{\dagger}\boldsymbol{a}^{\dagger}.
\end{align}
To preserve the bosonic commutation relation, the Bogoliubov matrices must satisfy
\begin{align}
\boldsymbol{U}^*\boldsymbol{U}^T - \boldsymbol{V}\boldsymbol{V}^{\dagger} &= \boldsymbol{\mathbbm{1}}, & \boldsymbol{V}^*\boldsymbol{U}^T &= \boldsymbol{U}\boldsymbol{V}^{\dagger}, \nonumber \\
\boldsymbol{U}^{\dagger}\boldsymbol{U} - \boldsymbol{V}^{\dagger}\boldsymbol{V} &= \boldsymbol{\mathbbm{1}}, & \boldsymbol{U}^{T}\boldsymbol{V} &= \boldsymbol{V}^{T}\boldsymbol{U}.
\label{eq:UV}
\end{align}

Using $b_k|\phi\rangle = 0 \; \forall k$ together with the constraints in Eq.~\eqref{eq:UV}, the covariance matrix can be derived as
\begin{align}
\boldsymbol{\Gamma}=\frac{1}{2}\boldsymbol{M}\boldsymbol{M}^T,
\end{align}
where $\boldsymbol{M}$ is the symplectic matrix related to Bogoliubov matrices as
\begin{align}
\label{eq:M2UV}
\boldsymbol{M} &= \begin{pmatrix}
\frac{i\boldsymbol{U}+i\boldsymbol{V}-i\boldsymbol{U}^*-i\boldsymbol{V}^*}{2}  & -\frac{\boldsymbol{U}+\boldsymbol{V}+\boldsymbol{U}^*+\boldsymbol{V}^*}{2} \nonumber \\[2mm]
\frac{\boldsymbol{U}-\boldsymbol{V}+\boldsymbol{U}^*-\boldsymbol{V}^*}{2} & \frac{i\boldsymbol{U}-i\boldsymbol{V}-i\boldsymbol{U}^*+i\boldsymbol{V}^*}{2}
\end{pmatrix} \\
&= \begin{pmatrix}
-\text{Im}(\boldsymbol{U}+\boldsymbol{V})  &
-\text{Re}(\boldsymbol{U}+\boldsymbol{V}) \\
~~\text{Re}(\boldsymbol{U}-\boldsymbol{V}) &
-\text{Im}(\boldsymbol{U}-\boldsymbol{V})
\end{pmatrix}.
\end{align}
The symplectic property of $\boldsymbol{M}$ can be verified using Eq.~\eqref{eq:UV}.

If $\boldsymbol{U}$ is invertible, the BGS $|\phi\rangle$ admits a paired-form expression:
\begin{align}
|\phi\rangle \propto \exp\left(\frac{1}{2} \sum_{i,j=1}^N Q_{ij} a^{\dagger}_i a^{\dagger}_j\right)|0\rangle_a,
\label{eq:paired-form}
\end{align}
where
\begin{align}
\label{eq:matrix_Q}
\boldsymbol{Q}=(\boldsymbol{U}^{\dagger})^{-1}\boldsymbol{V}^{\dagger}
\end{align}
and $|0\rangle_a$ denotes the vacuum of $a$-modes.
The fact that the matrix $\boldsymbol{Q}$ is symmetric follows directly from the identity $\boldsymbol{U}^{\dagger}\boldsymbol{V}^* = \boldsymbol{V}^{\dagger}\boldsymbol{U}^*$.

For normalizable BGSs, the covariance matrix formalism is directly applicable. By diagonalizing the covariance matrix through Williamson decomposition
and then utilizing Eqs.~\eqref{eq:M2UV} and~\eqref{eq:matrix_Q}, the paired-form wave function can be readily obtained. However, this framework fails for non-normalizable states like $|\phi^L\rangle$ in Eq.~\eqref{eq:SVD3}.

To address this issue, we devise a method to handle the divergent norm. To demonstrate it, we consider the state $|\phi^L\rangle$ in Eq.~\eqref{eq:SVD3}. To obtain a well-defined norm throughout the derivation, we introduce a regularized state,
\begin{align}
|\tilde{\phi}^L\rangle = (1-\tilde{\beta}^2)^{m/2} \exp\left[\tilde{\beta}\sum_{q=1}^{m} b^{\dagger}_{L,q} l^{\dagger}_{q}\right](|0\rangle_{b,l}\otimes|0\rangle_{\bar{l}}).
\end{align}
Here, the parameter $\tilde{\beta} < 1$ is included to ensure the convergence of the norm. This
construction also incorporates auxiliary modes $l^{\dagger}_{q>n_e}$, which are maximally entangled with the unoccupied Bogoliubov modes (denoted by $b^{\dagger}_{L,q>n_e}$) of the left subsystem. $|0\rangle_{b,l}$ denotes the vacuum annihilated by $b_{L,q}$ with $q\in[1,m]$ and $l_q$ with $q\in[1,n_e]$, and $|0\rangle_{\bar{l}}$ represents the vacuum of the auxiliary space, which comprises the $m - n_e$ bosonic modes associated with $l^{\dagger}_{q>n_e}$.   Then, the target state $|\phi^L\rangle$ is recovered via the limit
\begin{align}
|\phi^L\rangle = \lim_{\tilde{\beta} \to 1} {}_{\bar{l}} \langle 0 |\tilde{\phi}^L\rangle.
\end{align}
Although these extra virtual modes are not strictly necessary, their inclusion significantly simplifies the subsequent derivation.

It is easy to see that $|\tilde{\phi}^L\rangle$ is annihilated by the following normalized modes:
\begin{align}
d_{+,q}=\frac{l_q-\tilde{\beta}b^{\dagger}_{L,q}}{\sqrt{1-\tilde{\beta}^2}}, \quad
d_{-,q}=\frac{b_{L,q}-\tilde{\beta}l^{\dagger}_q}{\sqrt{1-\tilde{\beta}^2}}.
\end{align}
Clearly, these modes become ill-defined in the limit $\tilde{\beta}\to1$, further justifying the need to introduce $\tilde{\beta}$ as a regulator. By defining vectors of creation and annihilation operators as
\begin{align}
\boldsymbol{O}^{\dagger}=(O^{\dagger}_1,O^{\dagger}_2,\cdots,O^{\dagger}_m)^T, \quad
\boldsymbol{O}=(O_1,O_2,\cdots,O_m)^T
\end{align}
with $O\in\{b,f,d_{+},d_{-}\}$, $d_{+,q}$ and $d_{-,q}$ can be written in a compact form:
\begin{align}
\begin{pmatrix}
\boldsymbol{d}_{+} \\
\boldsymbol{d}_{-} \\
\boldsymbol{d}^{\dagger}_{+} \\
\boldsymbol{d}^{\dagger}_{-} \\
\end{pmatrix}&=\frac{1}{\sqrt{1-\tilde{\beta}^2}}
\begin{pmatrix}
0 &\boldsymbol{\mathbbm{1}} &-\boldsymbol{\tilde{\beta}}  &0 \\
\boldsymbol{\mathbbm{1}} &0&0&-\boldsymbol{\tilde{\beta}} \\
-\boldsymbol{\tilde{\beta}}  & 0 &0 &\boldsymbol{\boldsymbol{\mathbbm{1}}}\\
0 &-\boldsymbol{\tilde{\beta}}  &\boldsymbol{\mathbbm{1}} & 0
\end{pmatrix}\begin{pmatrix}
\boldsymbol{b}_L\\
\boldsymbol{l}\\
\boldsymbol{b}^{\dagger}_L\\
\boldsymbol{l}^{\dagger}
\end{pmatrix}\nonumber\\[4mm]
&=\frac{1}{\sqrt{1-\tilde{\beta}^2}}\begin{pmatrix}
\boldsymbol{\tilde{\beta}}\boldsymbol{V}_L^T &\boldsymbol{\mathbbm{1}} &-\boldsymbol{\tilde{\beta}}\boldsymbol{U}_L^T &0 \\
\boldsymbol{U}_L^{\dagger} &0&-\boldsymbol{V}_L^{\dagger}&-\boldsymbol{\tilde{\beta}}\\
-\boldsymbol{\tilde{\beta}}{\boldsymbol{U}_L}^{\dagger} & 0 &\boldsymbol{\tilde{\beta}}{\boldsymbol{V}_L}^{\dagger} &\boldsymbol{\mathbbm{1}}\\
-\boldsymbol{V}_L^T &-\boldsymbol{\tilde{\beta}} &\boldsymbol{U}_L^T & 0
\end{pmatrix}\begin{pmatrix}
\boldsymbol{a}_L\\
\boldsymbol{l}\\
\boldsymbol{a}^{\dagger}_L\\
\boldsymbol{l}^{\dagger}
\end{pmatrix},
\end{align}
where $\boldsymbol{\mathbbm{1}}$ is the $m\times m$ identity matrix and $\boldsymbol{\tilde{\beta}}\equiv\tilde{\beta}\boldsymbol{\mathbbm{1}}$. The result in the last line follows from an application of Eq.~\eqref{eq:bmode}.
From Eq.~\eqref{eq:matrix_Q}, one can derive that (the factor $1/\sqrt{1-\tilde{\beta}^2}$ of two involved matrices cancel each other)
\begin{align}
\boldsymbol{Q}_L&=\lim_{\tilde{\beta}\to1}\begin{pmatrix}
\boldsymbol{\tilde{\beta}}\boldsymbol{V}_L^T &\boldsymbol{\mathbbm{1}}\\
\boldsymbol{U}_L^{\dagger} &0\\
\end{pmatrix}^{-1} \begin{pmatrix}
\boldsymbol{\tilde{\beta}}\boldsymbol{U}_L^T &0\\
\boldsymbol{V}_L^{\dagger} & \boldsymbol{\tilde{\beta}}\\
\end{pmatrix} \nonumber \\[3mm]
&=\begin{pmatrix}
(\boldsymbol{U}_L^{\dagger})^{-1}\boldsymbol{V}_L^{\dagger} & (\boldsymbol{U_L}^{\dagger})^{-1} \\
(\boldsymbol{U_L}^{*})^{-1} & -\boldsymbol{V}_L^T(\boldsymbol{U}_L^{\dagger})^{-1}
\end{pmatrix}.
\end{align}
Having taken the limit $\tilde{\beta}\to1$, the next step is to perform the projection onto $|0\rangle_{\bar{l}}$. This can be done by restricting $\boldsymbol{Q}_L$ to the first $m+n_e$ columns and rows, which eliminates all auxiliary virtual modes.

\section{Fast search of the target state}
\label{app:fast_search}

This appendix details the algorithm for identifying the index of the state $|v^{\zeta,\zeta^{\prime}}_{k,\kappa}\rangle$ in Eq.~\eqref{eq:target} and for determining the corresponding coefficient $g^{\zeta,\zeta^{\prime}}_{\kappa}$. To elucidate the fast index searching algorithm, it is more practical to analyze the incoming virtual ($\boldsymbol{r}^{\dagger}_{m-1}$), outgoing virtual ($\boldsymbol{l}^{\dagger}_{m}$), and physical modes ($a^{\dagger}_m$) independently, rather than adopting the unified mode representation presented in Eq.~\eqref{eq:combine_modes}. Our methodology is first illustrated through a base case with $|p\rangle=|0\rangle$, corresponding to the calculation of $D^2$ elements. The extension to higher local physical occupation numbers, involving the remaining $(d-1)D^2$ elements, follows naturally from this foundation and will be systematically addressed in subsequent discussion.

First of all, we should define three fundamental vector quantities associated with each mode: the input vector $v^{\text{in}}$, output vector $v^{\text{out}}$, and order vector $v^{\text{ord}}$.
As a concrete example, for the mode $l^{\dagger}_{m,q}$, we construct the corresponding vector set:
\begin{equation}
\bigl\{
v^{\text{in}}_{m,q},\
v^{\text{out}}_{m,q},\
v^{\text{ord}}_{m,q}
\bigr\},
\label{eq:vectorset}
\end{equation}
through the following algorithmic procedure:
\begin{algorithm}[H]
\caption{Construction of vector set~\eqref{eq:vectorset}}\label{alg:vecset}
\begin{algorithmic}[1]
\Require Virtual states $|v^{m}_{\beta}\rangle=|n^{m}_{\beta,1},\cdots,n^{m}_{\beta,n_e}\rangle$, for $\beta=0,\cdots,D-1$
\Ensure Vectors $v^{\mathrm{in}}_{{m},q}$, $v^{\mathrm{out}}_{{m},q}$, $v^{\mathrm{ord}}_{{m},q}$

\State Initialize empty vectors:
\State \quad $v^{\mathrm{in}}_{{m},q} \gets [\,]$
\State \quad $v^{\mathrm{out}}_{{m},q} \gets [\,]$
\State \quad $v^{\mathrm{ord}}_{{m},q} \gets [\,]$

\For{$\beta \gets D-1$ \textbf{down} \textbf{to} $0$}\label{line:reverse} \Comment{Reverse iteration}
    \State $s_{\mathrm{out}} \gets |v^{m}_{\beta}\rangle = |n^{m}_{\beta,1},\ldots,n^{m}_{\beta,q},\ldots,n^{m}_{\beta,m}\rangle$

    \If{$n^{m}_{\beta,q} \geq 1$}
        \For{$\kappa \gets 1$ \textbf{to} $n^{m}_{\beta,q}$}
            \State $s_{\mathrm{in}} \gets |n^{m}_{\beta,1},\ldots,n^{m}_{\beta,q}-\kappa,\ldots,n^{m}_{\beta,n_e}\rangle$
            \State Find index $\alpha$ such that $|v^{m}_{\alpha}\rangle = s_{\mathrm{in}}$

            \State $v^{\mathrm{in}}_{{m},q}.\mathsf{append}(\alpha)$
            \State $v^{\mathrm{out}}_{{m},q}.\mathsf{append}(\beta)$
            \State $v^{\mathrm{ord}}_{{m},q}.\mathsf{append}(\kappa)$
        \EndFor
    \EndIf
\EndFor
\State \Return $\{v^{\mathrm{in}}_{{m},q}, v^{\mathrm{out}}_{{m},q}, v^{\mathrm{ord}}_{{m},q}\}$ \Comment{Return constructed vectors}
\end{algorithmic}
\end{algorithm}
The information encoded in the vector set~\eqref{eq:vectorset} can be interpreted through the following physical relationship. For any index $l$, let $(\alpha, \beta, \kappa) = (v^{\mathrm{in}}_{m,q}[l], v^{\mathrm{out}}_{m,q}[l], v^{\mathrm{ord}}_{m,q}[l])$. These components should satisfy
\begin{equation}
\ket{v^m_{\beta}} = \sqrt{\frac{n^m_{\alpha,q}!}{n^m_{\beta,q}!}} \left(l^{\dagger}_{m,q}\right)^{\kappa} \ket{v^m_{\alpha}}.
\end{equation}
It becomes evident that the vector set~\eqref{eq:vectorset} provides an explicit mapping between initial and final states under operator applications. Specifically, for any selected initial state indexed in $v^{\mathrm{in}}_{m,q}$, the corresponding entry in $v^{\mathrm{out}}_{m,q}$ uniquely determines the resultant state after applying the operator $l^{\dagger}_{m,q}$ exactly $\kappa$ times, where $\kappa$ is specified in $v^{\mathrm{ord}}_{m,q}$.
The vector sets or mappings for all modes can be established in $O(n^m_e\tilde{n}D)$ steps; this computationally inexpensive preparation, however, can make the fast search of target states possible when applying PCOs.
Moreover, the application of $l^{\dagger}_{m,q}$ no longer requires iteration through all $D$ states in the complete basis $\{|v^m_{\beta}\rangle\}$. Instead, it suffices to iterate exclusively over the input states indexed in $v^{\mathrm{in}}_{m,q}$, which further prompts the efficiency of our method. The vector set
\begin{align}
\{v^{\text{in}}_{m-1,q},v^{\text{out}}_{m-1,q},v^{\text{ord}}_{m-1,q}\}
\end{align}
can be constructed similarly in $O(n^{m-1}_e\tilde{n}D)$ steps.

The PCOs $\mathbbm{P}^{m}_{\zeta,\zeta^{\prime}}$ can be systematically classified into three distinct categories according to the ranges of indices $\zeta$ and $\zeta^{\prime}$:
\begin{enumerate}
    \item \textbf{cross-type} one left and one right virtual mode are paired together ($\zeta\in[1,n^{m-1}_e]$, $\zeta^{\prime}\in[n^{m-1}_e+2,n_{mode}]$):
    \begin{equation*}
        P_{\mathbb{V}_m} \exp\left[g r^{\dagger}_{m-1,q}l^{\dagger}_{m,q^{\prime}}\right]
    \end{equation*}

    \item \textbf{intra-type} for distinct modes within the same virtual space ($\zeta,\zeta^{\prime}\in[1,n^{m-1}_e]$ or $\zeta,\zeta^{\prime}\in[n^{m-1}_e+2,n_{mode}]$, $\zeta\neq\zeta^{\prime}$):
    \begin{equation*}
        P_{\mathbb{V}_m} \exp\left[g r^{\dagger}_{m-1,q}r^{\dagger}_{m-1,q^{\prime}}\right] \ \text{and} \ P_{\mathbb{V}_m} \exp\left[g l^{\dagger}_{m,q}l^{\dagger}_{m,q^{\prime}}\right]
    \end{equation*}
    with $q \neq q^{\prime}$.

    \item \textbf{duplicate-type} with identical mode indices ($\zeta=\zeta^{\prime}\neq n^{m-1}_e+1$):
    \begin{equation*}
    \label{eq:virtual_only_PPCO}
        P_{\mathbb{V}_m} \exp\left[\tfrac{g}{2} r^{\dagger}_{m-1,q}r^{\dagger}_{m-1,q}\right] \ \text{and} \ P_{\mathbb{V}_m} \exp\left[\tfrac{g}{2} l^{\dagger}_{m,q}l^{\dagger}_{m,q}\right]
    \end{equation*}
\end{enumerate}
For notational simplicity, we use a uniform notation $g$ for all PCO coupling strengths. This convention does not imply that all PCOs share identical pair strengths---the actual values remain determined by the matrix $\boldsymbol{Q}^m$.

The explicit implementation of the cross-type PCO is detailed in Alg.~\ref{alg:PPCO1}. The procedure involves iterating through states within $\{|v^{m-1}_{\alpha}\rangle\otimes|0\rangle\otimes|v^m_{\beta}\rangle)\}$ and adjusting the coefficients of the corresponding target states. A key aspect of the implementation is the reverse ordering of $v^{\text{out}}$ [see line~\ref{line:reverse} of Alg.~\ref{alg:vecset}], which serves an important purpose: it guarantees that modified elements won't interfere with subsequent updates during the PCO execution [see Alg.~\ref{alg:PPCO1}, line~\ref{line:update1}].  Without this ordering, an auxiliary matrix $vC$ would be required to track intermediate changes, ultimately necessitating an additional summation of $vC$ and $C$ to obtain $\tilde{C}$, which would incur extra computational overhead and memory cost.
\begin{algorithm}[H]
\caption{Implementation of cross-type PCO}\label{alg:PPCO1}
\begin{algorithmic}[1]
\Require initial state $C_{\alpha\beta}$ $|v^{m-1}_{\alpha}\rangle \otimes |0\rangle \otimes |v^{L}_{\beta}\rangle$, cross-type PCO $P_{\mathbbm{V}_m}\mathrm{exp}\left[gr^{\dagger}_{m-1,q}l^{\dagger}_{m,q^{\prime}}\right]$
\Ensure outcome state $\tilde{C}_{\alpha\beta}$ $|v^{m-1}_{\alpha}\rangle \otimes |0\rangle \otimes |v^{L}_{\beta}\rangle$
\For{$l \gets 1$ \textbf{to} length($v^{\text{in}}_{m-1,q}$)}
\For{$r \gets 1$ \textbf{to} length($v^{\text{in}}_{m,q^{\prime}}$)}
\State $(\kappa_l,\kappa_r)\gets(v^{\text{ord}}_{m-1,q}[l],v^{\text{ord}}_{m,q^{\prime}}[r])$
\If{$\kappa_l=\kappa_r$}
\State$\kappa\gets\kappa_l$
\State $(\mu,\nu)\gets(v^{\text{out}}_{m-1,q}[l],v^{\text{out}}_{m,q^{\prime}}[r]$)
\State $(\sigma,\gamma)\gets(v^{\text{in}}_{m-1,q}[l],v^{\text{in}}_{m,q^{\prime}}[r]$)
\State $s^l_{\text{out}}\gets|n^{m-1}_{\mu,1},\cdots,n^{m-1}_{\mu,q},\cdots,n^{m-1}_{\mu,m-1}\rangle$
\State
$s^r_{\text{out}}\gets|n^m_{\nu,1},\cdots,n^m_{\nu,q^{\prime}},\cdots,n^m_{\nu,m}\rangle$
\State $s^l_{\text{in}}\gets|n^{m-1}_{\sigma,1},\cdots,n^{m-1}_{\sigma,q},\cdots,n^{m-1}_{\sigma,m-1}\rangle$
\State
$s^r_{\text{in}}\gets|n^m_{\gamma,1},\cdots,n^m_{\gamma,q^{\prime}},\cdots,n^m_{\gamma,m}\rangle$
\State $C_{\mu\nu} \gets C_{\mu\nu} + \frac{g^{\kappa}C_{\sigma\gamma}}{\kappa!} \sqrt{ \frac{n^{m-1}_{\mu,q}!n^{m}_{\nu,q^{\prime}}!}{n^{m-1}_{\sigma,q}!n^{m}_{\gamma,q^{\prime}}!}}$
\label{line:update1}
\EndIf
\EndFor
\EndFor
\State \Return $\tilde{C}_{\alpha\beta}\gets C_{\alpha\beta}$
\end{algorithmic}
\end{algorithm}

To discuss the implementation of intra-type PCOs, we should define the composition of vector sets or mappings. For instance, the composition of $\{v^{\text{in}}_{m,q}, v^{\text{out}}_{m,q}, v^{\text{ord}}_{m,q}\}$ and $\{v^{\text{in}}_{m,q^{\prime}},v^{\text{out}}_{m,q^{\prime}},v^{\text{ord}}_{m,q^{\prime}}\}$ are denoted as
\begin{equation}
\bigl\{
v^{\text{in}}_{m,q\circ q^{\prime}},\
v^{\text{out}}_{m,q\circ q^{\prime}},\
v^{\text{ord}}_{m,q\circ q^{\prime}}
\bigr\},
\label{eq:vectorset_compostion}
\end{equation}
which can be determined using the following algorithm:

\begin{algorithm}[H]
\caption{Composition of vector sets}
\label{alg:ordered-vector-composition}
\begin{algorithmic}[1]
\Require
    \Statex Two vector sets $V_i = \{v^{\text{in}}_{m,q}, v^{\text{out}}_{m,q}, v^{\text{ord}}_{m,q}\}$
    \Statex and $V_j = \{v^{\text{in}}_{m,q^{\prime}}, v^{\text{out}}_{m,q^{\prime}}, v^{\text{ord}}_{m,q^{\prime}}\}$
\Ensure
    \Statex Composite vector set $V_{i \circ q^{\prime}} = \{v^{\text{in}}_{m,q\circ q^{\prime}}, v^{\text{out}}_{m,q\circ q^{\prime}}, v^{\text{ord}}_{m,q\circ q^{\prime}}\}$

\State Initialize empty vectors $v^{\text{in}}_{m,q\circ q^{\prime}}$, $v^{\text{out}}_{m,q\circ q^{\prime}}$, $v^{\text{ord}}_{m,q\circ q^{\prime}}$
\State Compute all matching index pairs:
    \State $M \gets \Big\{ (l, r) \;\Big|\; (v^{\text{out}}_{m,q}[l], v^{\text{ord}}_{m,q}[l]) = (v^{\text{in}}_{m,q^{\prime}}[r], v^{\text{ord}}_{m,q^{\prime}}[r]) \Big\}$

\State Sort $M$ by ascending $r$ value
\For{$(l, r) \in M$} \Comment{Process in order of increasing $r$}
    \State $v^{\text{in}}_{m,q\circ q^{\prime}}.\mathsf{append}(v^{\text{in}}_{m,q}[l])$
    \State $v^{\text{out}}_{m,q\circ q^{\prime}}.\mathsf{append}(v^{\text{out}}_{m,q^{\prime}}[r])$
    \State $v^{\text{ord}}_{m,q\circ q^{\prime}}.\mathsf{append}(v^{\text{ord}}_{m,q^{\prime}}[r])$
\EndFor
\State \Return $V_{q \circ q^{\prime}} \gets \{v^{\text{in}}_{m,q\circ q^{\prime}}, v^{\text{out}}_{m,q\circ q^{\prime}}, v^{\text{ord}}_{m,q\circ q^{\prime}}\}$
\end{algorithmic}
\end{algorithm}
Then the implementation of intra-type PCOs follows Alg.~\ref{alg:PPCO2}. The composite mapping must be evaluated for all mode pairs $(l^{\dagger}_{m,q},l^{\dagger}_{m,q^{\prime}})$, this operation incurs only an $\mathcal{O}(\tilde{n}D)$ computational overhead, therefore the total preparation can be finished in $O(\tilde{n}\cdot n^2_e \cdot D)$ steps, which is the dominant cost of mappings construction.
The duplicate-type PCO has the most straightforward implementation as given in Alg.~\ref{alg:PPCO3}.

At this stage, we can efficiently implement all types of PCOs. Crucially, as demonstrated in Algs~\ref{alg:PPCO1},~\ref{alg:PPCO2} and \ref{alg:PPCO3}, the mappings are defined on individual left and right virtual states (each of dimension $D$) rather than their tensor product space (of dimension $D^2$). This separate treatment is essential for maintaining computational efficiency.

The construction of local tensor with empty physical space begins with the vacuum initialization $|\mathrm{vac}\rangle = |v^{m-1}_0\rangle \otimes |0\rangle \otimes |v^m_0\rangle$, implemented by setting $C^{0}_{00} = 1$ while all other coefficients $C^{p}_{\alpha\beta}$ vanish. This vacuum state is then dressed by applying the complete set of virtual-only PCOs.
For local tensor with higher physical occupation number, the construction involves:
\begin{enumerate}
    \item Physical excitation through the physical modes involved PCOs:
    \begin{align}
        &P_{\mathbb{V}_m}\exp[gr^{\dagger}_{m-1,q}a^{\dagger}_m] & \quad (1 \leq q \leq n^{m-1}_e) \nonumber \\
        &P_{\mathbb{V}_m}\exp[\tfrac{g}{2}a^{\dagger}_ma^{\dagger}_m] \nonumber \\
        &P_{\mathbb{V}_m}\exp[gl^{\dagger}_{m,q^{\prime}}a^{\dagger}_m] & \quad (1 \leq q^{\prime} \leq n^m_e)
    \end{align}

    \item Independent processing of each occupation sector using virtual-mode PCOs, enabling parallel computation across different physical occupation numbers.
\end{enumerate}
Therefore, the computation of local tensor elements can be distributed across at least $d$ CPU/GPU cores. The parallelization of our algorithm can be further enhanced by incorporating some virtual modes as physical modes, but this approach ruin the clear separation between physical and virtual modes, potentially introducing unnecessary computations. To achieve optimal efficiency, it is crucial to consider the trade-off between enhanced parallelism and the wasted resources.
However, since our polynomial-time algorithm demonstrates sufficient efficiency, implementing only the basic parallelization scheme utilizing $d$ CPU cores is enough.

\begin{algorithm}[H]
\caption{Implementation of intra-type PCO}\label{alg:PPCO2}
\begin{algorithmic}[1]
\Require initial state $C_{\alpha\beta}$ $|v^{m-1}_{\alpha}\rangle \otimes |0\rangle \otimes |v^{m}_{\beta}\rangle$, intra-type PCO $P_{\mathbbm{V}_m}gl^{\dagger}_{m,q}l^{\dagger}_{m,q}$
\Ensure outcome state $\tilde{C}_{\alpha\beta}$ $|v^{m-1}_{\alpha}\rangle \otimes |0\rangle \otimes |v^{m}_{\beta}\rangle$
\For{$\mu \gets 0$ \textbf{to} $D-1$}
\For{$r \gets 1$ \textbf{to} length($v^{\text{in}}_{m,q\circ q^{\prime}}$)}
\State $(\nu,\gamma,\kappa)\gets(v^{\text{out}}_{m,q\circ q^{\prime}}[r],v^{\text{in}}_{m,q \circ q^{\prime}}[r],v^{\text{ord}}_{m,q\circ q^{\prime}}[r])$
\State $s^l_{\text{out}}=s^l_{\text{in}}=|v^{m-1}_{\mu}\rangle$
\State
$s^r_{\text{out}}\gets|v^{m}_{\nu}\rangle=|n^m_{\nu,1},\cdots,n^m_{\nu,q},\cdots,n^m_{\nu,q^{\prime}},\cdots,n^m_{\nu,m}\rangle$
\State
$s^r_{\text{in}}\gets|v^{m}_{\gamma}\rangle=|n^m_{\gamma,1},\cdots,n^m_{\gamma,q},\cdots,n^m_{\gamma,q^{\prime}},\cdots,n^m_{\gamma,m}\rangle$
\State $C_{\mu\nu}\gets C_{\mu\nu}+\frac{g^{\kappa}C_{\mu\gamma}}{\kappa!}\sqrt{\frac{ {n^{m}_{\nu,q}!n^{m}_{\nu,q^{\prime}}!}}{ {n^{m}_{\gamma,q}!n^{m}_{\gamma,q^{\prime}}!}}}$
\EndFor
\EndFor
\State \Return $\tilde{C}_{\alpha\beta}\gets C_{\alpha\beta}$
\end{algorithmic}
\end{algorithm}

\begin{algorithm}[H]
\caption{Implementation of duplicate-type PCO}\label{alg:PPCO3}
\begin{algorithmic}[1]
\Require initial state $C_{\alpha\beta}$ $|v^{m-1}_{\alpha}\rangle \otimes |0\rangle \otimes |v^{m}_{\beta}\rangle$, duplicate-type PCO $P_{\mathbbm{V}_m}gl^{\dagger}_{m,q}l^{\dagger}_{m,q^{\prime}}$
\Ensure outcome state $\tilde{C}_{\alpha\beta}$ $|v^{m-1}_{\alpha}\rangle \otimes |0\rangle \otimes |v^{m}_{\beta}\rangle$
\For{$\mu \gets 0$ \textbf{to} $D-1$}
\For{$r \gets 1$ \textbf{to} length($v^{\text{in}}_{m,q}$)}
\State $(\nu,\gamma,\kappa)=(v^{\text{out}}_{m,q}[r],v^{\text{in}}_{m,q}[r],v^{\text{ord}}_{m,q}[r])$
\If{$\mathrm{mod}(\kappa,2)=0$}
\State $s^l_{\text{out}}=s^l_{\text{in}}=|v^{m-1}_{\mu}\rangle$
\State
$s^r_{\text{out}}\gets|v^{m}_{\nu}\rangle=|n^m_{\nu,1},\cdots,n^m_{\nu,q},\cdots,n^m_{\nu,m}\rangle$
\State
$s^r_{\text{in}}\gets|v^{m}_{\gamma}\rangle=|n^m_{\gamma,1},\cdots,n^m_{\gamma,q},\cdots,n^m_{\gamma,m}\rangle$
\State $C_{\mu\nu}\gets C_{\mu\nu}+\frac{(g/2)^{\kappa/2}C_{\mu\gamma}}{(\kappa/2)!}\sqrt{\frac{ {n^{m}_{\nu,q}!}}{ {n^{m}_{\gamma,q}!}}}$
\EndIf
\EndFor
\EndFor
\State \Return $\tilde{C}_{\alpha\beta}\gets C_{\alpha\beta}$

\end{algorithmic}
\end{algorithm}

\section{Filtration of the quantum channel}
\label{app:opt_Gamma}

The covariance matrix of a pure BGS can be expressed as
\begin{align}
\boldsymbol{\Gamma}_{p} = \frac{1}{2} \boldsymbol{S} \boldsymbol{S}^T,
\end{align}
where $\boldsymbol{S}$ is a symplectic matrix.
Consequently, the optimal decomposition problem translates to minimizing the trace of $\boldsymbol{\Gamma}_p$ over the symplectic manifold, subjected to the constraint that the matrix $\boldsymbol{W} = \boldsymbol{\Gamma}_{\text{th}} - \boldsymbol{\Gamma}_p$ remains positive semi-definite ($\boldsymbol{W}\geq{}0$).

Directly handling this positive semi-definite constraint on the symplectic manifold is non-trivial. We therefore reframe the problem by adopting an indirect approach. We initialize the optimization with $\boldsymbol{S}$ as the identity matrix, which minimizes $\text{tr}(\boldsymbol{\Gamma}_p)$ but typically results in a $\boldsymbol{W}$ that is not positive semi-definite.
The optimization then proceeds by iteratively updating $\boldsymbol{S}$ in a direction that {\em penalizes the negative eigenvalues} of $\boldsymbol{W}$, effectively pushing the smallest eigenvalue up towards zero.
The final $\boldsymbol{\Gamma}_p$ is the result of this trade-off between minimizing the trace and satisfying the physical constraint.

To enforce the constraint $\boldsymbol{\Gamma}_{\text{th}} - \frac{1}{2} \boldsymbol{S} \boldsymbol{S}^T \geq 0$, we frame the problem as the minimization of a loss function designed to maximize the smallest eigenvalue of the constrained matrix. The loss is defined as:
\begin{equation}
    \mathcal{L}(\boldsymbol{S}) = - \lambda_{\min} \left( \boldsymbol{\Gamma}_{\text{th}} - \frac{1}{2} \boldsymbol{S} \boldsymbol{S}^T \right),
\label{eq:loss}
\end{equation}
where $\lambda_{\min}(\cdot)$ is the minimum eigenvalue. The optimization algorithm, a Riemannian gradient descent on the symplectic manifold, seeks to drive $\mathcal{L}(\boldsymbol{S})$ to a non-positive value.
The full algorithm proceeds as follows:
\begin{enumerate}
    \item \textbf{Initialization}: We begin with $\boldsymbol{S}^{(0)}=\mathbf{I}$. This choice minimizes the trace of the pure part but generally violates the positive semi-definite constraint, resulting in an initial loss $\mathcal{L}(\boldsymbol{S}^{(0)}) > 0$.

    \item \textbf{Gradient Computation}: At each iteration $k$, we compute the Riemannian gradient of $\mathcal{L}(\boldsymbol{S})$ with respect to $\boldsymbol{S}^{(k)}$. This gradient is a tangent vector $\boldsymbol{G}^{(k)} \in \mathfrak{sp}(2N)$, an element of the symplectic Lie algebra.

    \item \textbf{Retraction}: We update the matrix by moving along the gradient direction. To ensure the updated matrix remains on the symplectic manifold, we use the exponential map as a retraction:
    \[
    \boldsymbol{S}^{(k+1)} = \boldsymbol{S}^{(k)} \exp\left( -\alpha^{(k)} \boldsymbol{G}^{(k)} \right),
    \]
    where $\alpha^{(k)} > 0$ is a suitably chosen step size. This operation guarantees that $\boldsymbol{S}^{(k+1)}$ is also a symplectic matrix.

    \item \textbf{Convergence Criterion}: The iterations continue until the loss function becomes non-positive, $\mathcal{L}(\boldsymbol{S}) \leq \epsilon$, for a predefined small tolerance $\epsilon \ge 0$. At this point, the constraint is satisfied, and a valid decomposition has been found.
\end{enumerate}

\begin{figure}[ht]
    \centering
    \includegraphics[width=0.9\linewidth]{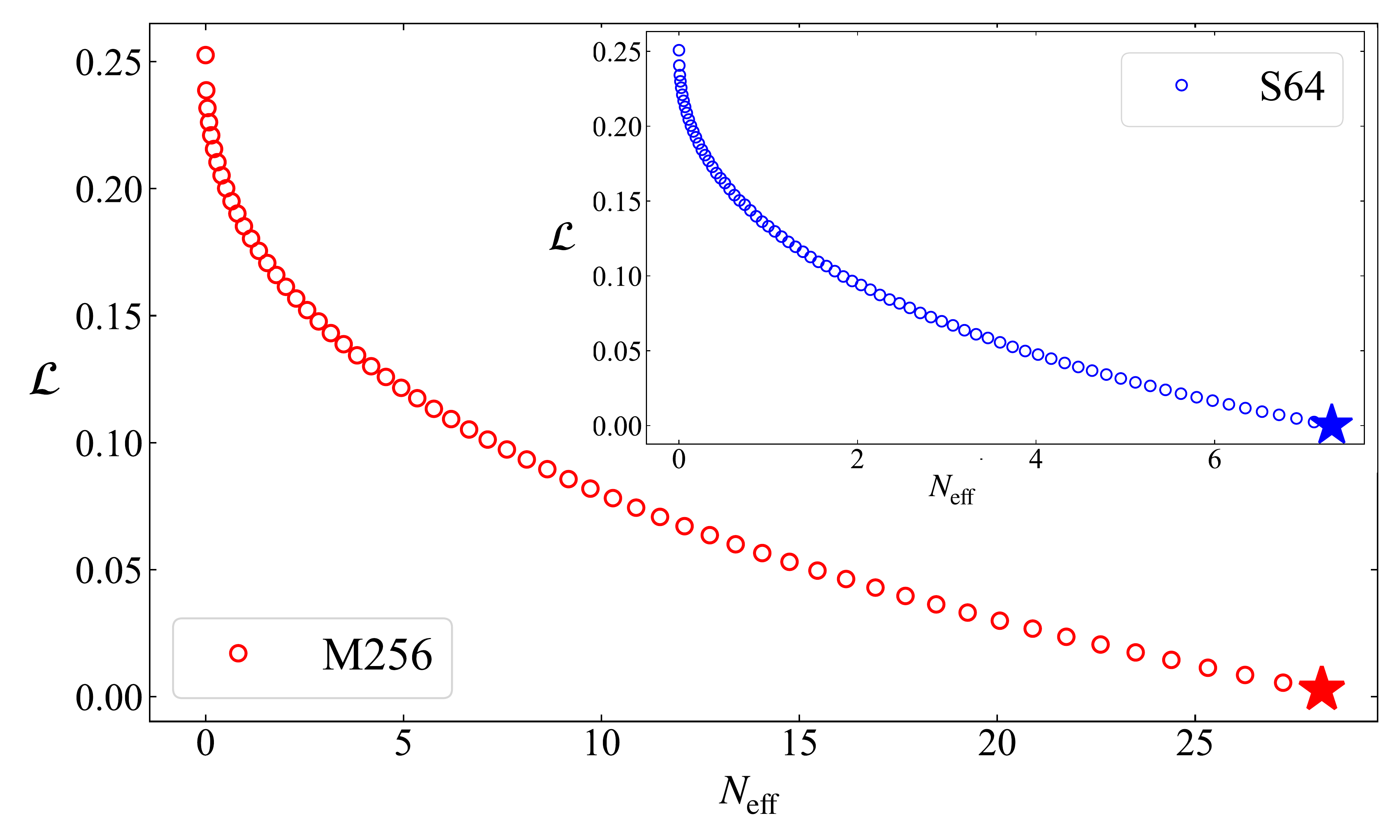}
    \caption{Evolution of the loss $\mathcal{L}$ [Eq.~\eqref{eq:loss}]
    as a function of the effective mode number $N_{\text{eff}}$ during the optimization process for S64 and M256. Convergence is achieved after approximately 3800 steps for S64 and 24000 steps for M256. For visual clarity, data points are shown at intervals of every 50 steps for S64 and every 400 steps for M256.}
    \label{fig:optimization}
\end{figure}

The primary challenge in carrying out this optimization scheme lies in the computation of the gradient $\boldsymbol{G}$ in the step of gradient evaluation.
To perform the gradient descent, we first compute the directional derivative of the loss function.
Let $\boldsymbol{X} \in \mathfrak{sp}(2N)$ be a tangent vector at $\boldsymbol{S}$. The infinitesimal evolution of $\boldsymbol{S}$ along $\boldsymbol{X}$ is given by  $\boldsymbol{S} \exp(\epsilon\boldsymbol{X})$, with $\epsilon\rightarrow{}0^{+}$ being an infinitesimal number. The directional derivative of $\mathcal{L}$ at $\boldsymbol{S}$ along $\boldsymbol{X}$ is defined as
\begin{align}
    \partial_{\boldsymbol{X}}\mathcal{L}(\boldsymbol{S}) = \lim_{\epsilon \to 0} \frac{\mathcal{L}(\boldsymbol{S} + \epsilon\boldsymbol{S}\boldsymbol{X}) - \mathcal{L}(\boldsymbol{S})}{\epsilon},
\end{align}
where we have used the first-order approximation $\boldsymbol{S} \exp(\epsilon\boldsymbol{X}) \approx \boldsymbol{S}(\mathbf{I} + \epsilon\boldsymbol{X})$. Meanwhile, the loss function is evaluated by diagonalizing the matrix $\boldsymbol{W}[\boldsymbol{S}]=\boldsymbol{\Gamma}_{\text{th}} - \frac{1}{2} \boldsymbol{S} \boldsymbol{S}^T$. Assuming the lowest eigenvalue of $\boldsymbol{W}[\boldsymbol{S}]$ is non-degenerate with corresponding eigenvector $|v_0\rangle$, the first-order perturbation theory indicates that
\begin{align}
    \partial_{\boldsymbol{X}}\mathcal{L}(\boldsymbol{S}) & = \frac{1}{2} \langle v_0 | \boldsymbol{S} (\boldsymbol{X} + \boldsymbol{X}^T) \boldsymbol{S}^T | v_0 \rangle.
\end{align}
By definition, the gradient $\boldsymbol{G}$ is the tangent vector that maximizes this directional derivative for a fixed norm, which is found by solving
\begin{align}
    \label{eq:gradient}
    \boldsymbol{G}=\arg \max_{\boldsymbol{X} \in \mathfrak{sp}(2N), \|\boldsymbol{X}\|=1} \partial_{\boldsymbol{X}}\mathcal{L}(\boldsymbol{S}),
\end{align}
where $\|\cdot\|$ is the two-form norm on the symplectic manifold. The resulting $\boldsymbol{G}$ is the gradient used in the update step.

In practice, as the optimization nears convergence, the ground state of $\boldsymbol{W}$ exhibits a high degree of degeneracy, which precludes an analytical computation of the gradient. Nevertheless, the optimization algorithm based on Eq.~\eqref{eq:gradient} still works with this approximate gradient, though its efficiency is limited.

To demonstrate the effectiveness of this method, we use it to filter the quantum channels of the mixed-state covariance matrices extracted from the S64 and M256 configurations of Jiuzhang4.0 experiment. The optimization dynamics are shown in Fig.~\ref{fig:optimization}, which plots the evolution of the loss function $\mathcal{L}(\boldsymbol{S})$ against the effective particle number $N_{\text{eff}}$. A key observation is the gradual decrease in the slope of the curve. This indicates that as the optimization proceeds, achieving a reduction in $\mathcal{L}$ requires progressively larger increases in $N_{\text{eff}}$.

\newpage


\begin{thebibliography}{65}%
\makeatletter
\providecommand \@ifxundefined [1]{%
 \@ifx{#1\undefined}
}%
\providecommand \@ifnum [1]{%
 \ifnum #1\expandafter \@firstoftwo
 \else \expandafter \@secondoftwo
 \fi
}%
\providecommand \@ifx [1]{%
 \ifx #1\expandafter \@firstoftwo
 \else \expandafter \@secondoftwo
 \fi
}%
\providecommand \natexlab [1]{#1}%
\providecommand \enquote  [1]{``#1''}%
\providecommand \bibnamefont  [1]{#1}%
\providecommand \bibfnamefont [1]{#1}%
\providecommand \citenamefont [1]{#1}%
\providecommand \href@noop [0]{\@secondoftwo}%
\providecommand \href [0]{\begingroup \@sanitize@url \@href}%
\providecommand \@href[1]{\@@startlink{#1}\@@href}%
\providecommand \@@href[1]{\endgroup#1\@@endlink}%
\providecommand \@sanitize@url [0]{\catcode `\\12\catcode `\$12\catcode
  `\&12\catcode `\#12\catcode `\^12\catcode `\_12\catcode `\%12\relax}%
\providecommand \@@startlink[1]{}%
\providecommand \@@endlink[0]{}%
\providecommand \url  [0]{\begingroup\@sanitize@url \@url }%
\providecommand \@url [1]{\endgroup\@href {#1}{\urlprefix }}%
\providecommand \urlprefix  [0]{URL }%
\providecommand \Eprint [0]{\href }%
\providecommand \doibase [0]{http://dx.doi.org/}%
\providecommand \selectlanguage [0]{\@gobble}%
\providecommand \bibinfo  [0]{\@secondoftwo}%
\providecommand \bibfield  [0]{\@secondoftwo}%
\providecommand \translation [1]{[#1]}%
\providecommand \BibitemOpen [0]{}%
\providecommand \bibitemStop [0]{}%
\providecommand \bibitemNoStop [0]{.\EOS\space}%
\providecommand \EOS [0]{\spacefactor3000\relax}%
\providecommand \BibitemShut  [1]{\csname bibitem#1\endcsname}%
\let\auto@bib@innerbib\@empty
\bibitem [{\citenamefont {Aaronson}\ and\ \citenamefont
  {Arkhipov}(2011)}]{Aaronson2011}%
  \BibitemOpen
  \bibfield  {author} {\bibinfo {author} {\bibfnamefont {S.}~\bibnamefont
  {Aaronson}}\ and\ \bibinfo {author} {\bibfnamefont {A.}~\bibnamefont
  {Arkhipov}},\ }in\ \href {\doibase 10.4086/toc.2013.v009a004} {\emph
  {\bibinfo {booktitle} {Proceedings of the forty-third annual ACM symposium on
  Theory of computing}}}\ (\bibinfo {year} {2011})\ pp.\ \bibinfo {pages}
  {333--342}\BibitemShut {NoStop}%
\bibitem [{\citenamefont {Broome}\ \emph {et~al.}(2013)\citenamefont {Broome},
  \citenamefont {Fedrizzi}, \citenamefont {Rahimi-Keshari}, \citenamefont
  {Dove}, \citenamefont {Aaronson}, \citenamefont {Ralph},\ and\ \citenamefont
  {White}}]{Broome2013}%
  \BibitemOpen
  \bibfield  {author} {\bibinfo {author} {\bibfnamefont {M.~A.}\ \bibnamefont
  {Broome}}, \bibinfo {author} {\bibfnamefont {A.}~\bibnamefont {Fedrizzi}},
  \bibinfo {author} {\bibfnamefont {S.}~\bibnamefont {Rahimi-Keshari}},
  \bibinfo {author} {\bibfnamefont {J.}~\bibnamefont {Dove}}, \bibinfo {author}
  {\bibfnamefont {S.}~\bibnamefont {Aaronson}}, \bibinfo {author}
  {\bibfnamefont {T.~C.}\ \bibnamefont {Ralph}}, \ and\ \bibinfo {author}
  {\bibfnamefont {A.~G.}\ \bibnamefont {White}},\ }\href {\doibase
  10.1126/science.1231440} {\bibfield  {journal} {\bibinfo  {journal}
  {Science}\ }\textbf {\bibinfo {volume} {339}},\ \bibinfo {pages} {794}
  (\bibinfo {year} {2013})}\BibitemShut {NoStop}%
\bibitem [{\citenamefont {Tillmann}\ \emph {et~al.}(2013)\citenamefont
  {Tillmann}, \citenamefont {Daki{\'c}}, \citenamefont {Heilmann},
  \citenamefont {Nolte}, \citenamefont {Szameit},\ and\ \citenamefont
  {Walther}}]{Tillmann2013}%
  \BibitemOpen
  \bibfield  {author} {\bibinfo {author} {\bibfnamefont {M.}~\bibnamefont
  {Tillmann}}, \bibinfo {author} {\bibfnamefont {B.}~\bibnamefont {Daki{\'c}}},
  \bibinfo {author} {\bibfnamefont {R.}~\bibnamefont {Heilmann}}, \bibinfo
  {author} {\bibfnamefont {S.}~\bibnamefont {Nolte}}, \bibinfo {author}
  {\bibfnamefont {A.}~\bibnamefont {Szameit}}, \ and\ \bibinfo {author}
  {\bibfnamefont {P.}~\bibnamefont {Walther}},\ }\href {\doibase
  https://doi.org/10.1038/nphoton.2013.102} {\bibfield  {journal} {\bibinfo
  {journal} {Nat. Photon.}\ }\textbf {\bibinfo {volume} {7}},\ \bibinfo {pages}
  {540} (\bibinfo {year} {2013})}\BibitemShut {NoStop}%
\bibitem [{\citenamefont {Spring}\ \emph {et~al.}(2013)\citenamefont {Spring},
  \citenamefont {Metcalf}, \citenamefont {Humphreys}, \citenamefont
  {Kolthammer}, \citenamefont {Jin}, \citenamefont {Barbieri}, \citenamefont
  {Datta}, \citenamefont {Thomas-Peter}, \citenamefont {Langford},
  \citenamefont {Kundys} \emph {et~al.}}]{Spring2013}%
  \BibitemOpen
  \bibfield  {author} {\bibinfo {author} {\bibfnamefont {J.~B.}\ \bibnamefont
  {Spring}}, \bibinfo {author} {\bibfnamefont {B.~J.}\ \bibnamefont {Metcalf}},
  \bibinfo {author} {\bibfnamefont {P.~C.}\ \bibnamefont {Humphreys}}, \bibinfo
  {author} {\bibfnamefont {W.~S.}\ \bibnamefont {Kolthammer}}, \bibinfo
  {author} {\bibfnamefont {X.-M.}\ \bibnamefont {Jin}}, \bibinfo {author}
  {\bibfnamefont {M.}~\bibnamefont {Barbieri}}, \bibinfo {author}
  {\bibfnamefont {A.}~\bibnamefont {Datta}}, \bibinfo {author} {\bibfnamefont
  {N.}~\bibnamefont {Thomas-Peter}}, \bibinfo {author} {\bibfnamefont {N.~K.}\
  \bibnamefont {Langford}}, \bibinfo {author} {\bibfnamefont {D.}~\bibnamefont
  {Kundys}},  \emph {et~al.},\ }\href {\doibase 10.1126/science.1231692}
  {\bibfield  {journal} {\bibinfo  {journal} {Science}\ }\textbf {\bibinfo
  {volume} {339}},\ \bibinfo {pages} {798} (\bibinfo {year}
  {2013})}\BibitemShut {NoStop}%
\bibitem [{\citenamefont {Crespi}\ \emph {et~al.}(2013)\citenamefont {Crespi},
  \citenamefont {Osellame}, \citenamefont {Ramponi}, \citenamefont {Brod},
  \citenamefont {Galvao}, \citenamefont {Spagnolo}, \citenamefont {Vitelli},
  \citenamefont {Maiorino}, \citenamefont {Mataloni},\ and\ \citenamefont
  {Sciarrino}}]{Crespi2013}%
  \BibitemOpen
  \bibfield  {author} {\bibinfo {author} {\bibfnamefont {A.}~\bibnamefont
  {Crespi}}, \bibinfo {author} {\bibfnamefont {R.}~\bibnamefont {Osellame}},
  \bibinfo {author} {\bibfnamefont {R.}~\bibnamefont {Ramponi}}, \bibinfo
  {author} {\bibfnamefont {D.~J.}\ \bibnamefont {Brod}}, \bibinfo {author}
  {\bibfnamefont {E.~F.}\ \bibnamefont {Galvao}}, \bibinfo {author}
  {\bibfnamefont {N.}~\bibnamefont {Spagnolo}}, \bibinfo {author}
  {\bibfnamefont {C.}~\bibnamefont {Vitelli}}, \bibinfo {author} {\bibfnamefont
  {E.}~\bibnamefont {Maiorino}}, \bibinfo {author} {\bibfnamefont
  {P.}~\bibnamefont {Mataloni}}, \ and\ \bibinfo {author} {\bibfnamefont
  {F.}~\bibnamefont {Sciarrino}},\ }\href {\doibase
  https://doi.org/10.1038/nphoton.2013.112} {\bibfield  {journal} {\bibinfo
  {journal} {Nat. Photon.}\ }\textbf {\bibinfo {volume} {7}},\ \bibinfo {pages}
  {545} (\bibinfo {year} {2013})}\BibitemShut {NoStop}%
\bibitem [{\citenamefont {Wang}\ \emph {et~al.}(2017)\citenamefont {Wang},
  \citenamefont {He}, \citenamefont {Li}, \citenamefont {Su}, \citenamefont
  {Li}, \citenamefont {Huang}, \citenamefont {Ding}, \citenamefont {Chen},
  \citenamefont {Liu}, \citenamefont {Qin}, \citenamefont {Li}, \citenamefont
  {He}, \citenamefont {Schneider}, \citenamefont {Kamp}, \citenamefont {Peng},
  \citenamefont {H\"ofling}, \citenamefont {Lu},\ and\ \citenamefont
  {Pan}}]{WangH2017}%
  \BibitemOpen
  \bibfield  {author} {\bibinfo {author} {\bibfnamefont {H.}~\bibnamefont
  {Wang}}, \bibinfo {author} {\bibfnamefont {Y.}~\bibnamefont {He}}, \bibinfo
  {author} {\bibfnamefont {Y.-H.}\ \bibnamefont {Li}}, \bibinfo {author}
  {\bibfnamefont {Z.-E.}\ \bibnamefont {Su}}, \bibinfo {author} {\bibfnamefont
  {B.}~\bibnamefont {Li}}, \bibinfo {author} {\bibfnamefont {H.-L.}\
  \bibnamefont {Huang}}, \bibinfo {author} {\bibfnamefont {X.}~\bibnamefont
  {Ding}}, \bibinfo {author} {\bibfnamefont {M.-C.}\ \bibnamefont {Chen}},
  \bibinfo {author} {\bibfnamefont {C.}~\bibnamefont {Liu}}, \bibinfo {author}
  {\bibfnamefont {J.}~\bibnamefont {Qin}}, \bibinfo {author} {\bibfnamefont
  {J.-P.}\ \bibnamefont {Li}}, \bibinfo {author} {\bibfnamefont {Y.-M.}\
  \bibnamefont {He}}, \bibinfo {author} {\bibfnamefont {C.}~\bibnamefont
  {Schneider}}, \bibinfo {author} {\bibfnamefont {M.}~\bibnamefont {Kamp}},
  \bibinfo {author} {\bibfnamefont {C.-Z.}\ \bibnamefont {Peng}}, \bibinfo
  {author} {\bibfnamefont {S.}~\bibnamefont {H\"ofling}}, \bibinfo {author}
  {\bibfnamefont {C.-Y.}\ \bibnamefont {Lu}}, \ and\ \bibinfo {author}
  {\bibfnamefont {J.-W.}\ \bibnamefont {Pan}},\ }\href {\doibase
  https://doi.org/10.1038/nphoton.2017.63} {\bibfield  {journal} {\bibinfo
  {journal} {Nat. Photon.}\ }\textbf {\bibinfo {volume} {11}},\ \bibinfo
  {pages} {361} (\bibinfo {year} {2017})}\BibitemShut {NoStop}%
\bibitem [{\citenamefont {He}\ \emph {et~al.}(2017)\citenamefont {He},
  \citenamefont {Ding}, \citenamefont {Su}, \citenamefont {Huang},
  \citenamefont {Qin}, \citenamefont {Wang}, \citenamefont {Unsleber},
  \citenamefont {Chen}, \citenamefont {Wang}, \citenamefont {He}, \citenamefont
  {Wang}, \citenamefont {Zhang}, \citenamefont {Chen}, \citenamefont
  {Schneider}, \citenamefont {Kamp}, \citenamefont {You}, \citenamefont {Wang},
  \citenamefont {H\"ofling}, \citenamefont {Lu},\ and\ \citenamefont
  {Pan}}]{HeY2017}%
  \BibitemOpen
  \bibfield  {author} {\bibinfo {author} {\bibfnamefont {Y.}~\bibnamefont
  {He}}, \bibinfo {author} {\bibfnamefont {X.}~\bibnamefont {Ding}}, \bibinfo
  {author} {\bibfnamefont {Z.-E.}\ \bibnamefont {Su}}, \bibinfo {author}
  {\bibfnamefont {H.-L.}\ \bibnamefont {Huang}}, \bibinfo {author}
  {\bibfnamefont {J.}~\bibnamefont {Qin}}, \bibinfo {author} {\bibfnamefont
  {C.}~\bibnamefont {Wang}}, \bibinfo {author} {\bibfnamefont {S.}~\bibnamefont
  {Unsleber}}, \bibinfo {author} {\bibfnamefont {C.}~\bibnamefont {Chen}},
  \bibinfo {author} {\bibfnamefont {H.}~\bibnamefont {Wang}}, \bibinfo {author}
  {\bibfnamefont {Y.-M.}\ \bibnamefont {He}}, \bibinfo {author} {\bibfnamefont
  {X.-L.}\ \bibnamefont {Wang}}, \bibinfo {author} {\bibfnamefont {W.-J.}\
  \bibnamefont {Zhang}}, \bibinfo {author} {\bibfnamefont {S.-J.}\ \bibnamefont
  {Chen}}, \bibinfo {author} {\bibfnamefont {C.}~\bibnamefont {Schneider}},
  \bibinfo {author} {\bibfnamefont {M.}~\bibnamefont {Kamp}}, \bibinfo {author}
  {\bibfnamefont {L.-X.}\ \bibnamefont {You}}, \bibinfo {author} {\bibfnamefont
  {Z.}~\bibnamefont {Wang}}, \bibinfo {author} {\bibfnamefont {S.}~\bibnamefont
  {H\"ofling}}, \bibinfo {author} {\bibfnamefont {C.-Y.}\ \bibnamefont {Lu}}, \
  and\ \bibinfo {author} {\bibfnamefont {J.-W.}\ \bibnamefont {Pan}},\ }\href
  {\doibase 10.1103/PhysRevLett.118.190501} {\bibfield  {journal} {\bibinfo
  {journal} {Phys. Rev. Lett.}\ }\textbf {\bibinfo {volume} {118}},\ \bibinfo
  {pages} {190501} (\bibinfo {year} {2017})}\BibitemShut {NoStop}%
\bibitem [{\citenamefont {Loredo}\ \emph {et~al.}(2017)\citenamefont {Loredo},
  \citenamefont {Broome}, \citenamefont {Hilaire}, \citenamefont {Gazzano},
  \citenamefont {Sagnes}, \citenamefont {Lemaitre}, \citenamefont {Almeida},
  \citenamefont {Senellart},\ and\ \citenamefont {White}}]{Loredo2017}%
  \BibitemOpen
  \bibfield  {author} {\bibinfo {author} {\bibfnamefont {J.~C.}\ \bibnamefont
  {Loredo}}, \bibinfo {author} {\bibfnamefont {M.~A.}\ \bibnamefont {Broome}},
  \bibinfo {author} {\bibfnamefont {P.}~\bibnamefont {Hilaire}}, \bibinfo
  {author} {\bibfnamefont {O.}~\bibnamefont {Gazzano}}, \bibinfo {author}
  {\bibfnamefont {I.}~\bibnamefont {Sagnes}}, \bibinfo {author} {\bibfnamefont
  {A.}~\bibnamefont {Lemaitre}}, \bibinfo {author} {\bibfnamefont {M.~P.}\
  \bibnamefont {Almeida}}, \bibinfo {author} {\bibfnamefont {P.}~\bibnamefont
  {Senellart}}, \ and\ \bibinfo {author} {\bibfnamefont {A.~G.}\ \bibnamefont
  {White}},\ }\href {\doibase 10.1103/PhysRevLett.118.130503} {\bibfield
  {journal} {\bibinfo  {journal} {Phys. Rev. Lett.}\ }\textbf {\bibinfo
  {volume} {118}},\ \bibinfo {pages} {130503} (\bibinfo {year}
  {2017})}\BibitemShut {NoStop}%
\bibitem [{\citenamefont {Lund}\ \emph {et~al.}(2014)\citenamefont {Lund},
  \citenamefont {Laing}, \citenamefont {Rahimi-Keshari}, \citenamefont
  {Rudolph}, \citenamefont {O'Brien},\ and\ \citenamefont {Ralph}}]{Lund2014}%
  \BibitemOpen
  \bibfield  {author} {\bibinfo {author} {\bibfnamefont {A.~P.}\ \bibnamefont
  {Lund}}, \bibinfo {author} {\bibfnamefont {A.}~\bibnamefont {Laing}},
  \bibinfo {author} {\bibfnamefont {S.}~\bibnamefont {Rahimi-Keshari}},
  \bibinfo {author} {\bibfnamefont {T.}~\bibnamefont {Rudolph}}, \bibinfo
  {author} {\bibfnamefont {J.~L.}\ \bibnamefont {O'Brien}}, \ and\ \bibinfo
  {author} {\bibfnamefont {T.~C.}\ \bibnamefont {Ralph}},\ }\href {\doibase
  10.1103/PhysRevLett.113.100502} {\bibfield  {journal} {\bibinfo  {journal}
  {Phys. Rev. Lett.}\ }\textbf {\bibinfo {volume} {113}},\ \bibinfo {pages}
  {100502} (\bibinfo {year} {2014})}\BibitemShut {NoStop}%
\bibitem [{\citenamefont {Zhong}\ \emph {et~al.}(2019)\citenamefont {Zhong},
  \citenamefont {Peng}, \citenamefont {Li}, \citenamefont {Hu}, \citenamefont
  {Li}, \citenamefont {Qin}, \citenamefont {Wu}, \citenamefont {Zhang},
  \citenamefont {Li}, \citenamefont {Zhang}, \citenamefont {Wang},
  \citenamefont {You}, \citenamefont {Jiang}, \citenamefont {Li}, \citenamefont
  {Liu}, \citenamefont {Dowling}, \citenamefont {Lu},\ and\ \citenamefont
  {Pan}}]{ZhongHS2019}%
  \BibitemOpen
  \bibfield  {author} {\bibinfo {author} {\bibfnamefont {H.-S.}\ \bibnamefont
  {Zhong}}, \bibinfo {author} {\bibfnamefont {L.-C.}\ \bibnamefont {Peng}},
  \bibinfo {author} {\bibfnamefont {Y.}~\bibnamefont {Li}}, \bibinfo {author}
  {\bibfnamefont {Y.}~\bibnamefont {Hu}}, \bibinfo {author} {\bibfnamefont
  {W.}~\bibnamefont {Li}}, \bibinfo {author} {\bibfnamefont {J.}~\bibnamefont
  {Qin}}, \bibinfo {author} {\bibfnamefont {D.}~\bibnamefont {Wu}}, \bibinfo
  {author} {\bibfnamefont {W.}~\bibnamefont {Zhang}}, \bibinfo {author}
  {\bibfnamefont {H.}~\bibnamefont {Li}}, \bibinfo {author} {\bibfnamefont
  {L.}~\bibnamefont {Zhang}}, \bibinfo {author} {\bibfnamefont
  {Z.}~\bibnamefont {Wang}}, \bibinfo {author} {\bibfnamefont {L.}~\bibnamefont
  {You}}, \bibinfo {author} {\bibfnamefont {X.}~\bibnamefont {Jiang}}, \bibinfo
  {author} {\bibfnamefont {L.}~\bibnamefont {Li}}, \bibinfo {author}
  {\bibfnamefont {N.-L.}\ \bibnamefont {Liu}}, \bibinfo {author} {\bibfnamefont
  {J.~P.}\ \bibnamefont {Dowling}}, \bibinfo {author} {\bibfnamefont {C.-Y.}\
  \bibnamefont {Lu}}, \ and\ \bibinfo {author} {\bibfnamefont {J.-W.}\
  \bibnamefont {Pan}},\ }\href {\doibase
  https://doi.org/10.1016/j.scib.2019.04.007} {\bibfield  {journal} {\bibinfo
  {journal} {Sci. Bull.}\ }\textbf {\bibinfo {volume} {64}},\ \bibinfo {pages}
  {511} (\bibinfo {year} {2019})}\BibitemShut {NoStop}%
\bibitem [{\citenamefont {Zhong}\ \emph {et~al.}(2020)\citenamefont {Zhong},
  \citenamefont {Wang}, \citenamefont {Deng}, \citenamefont {Chen},
  \citenamefont {Peng}, \citenamefont {Luo}, \citenamefont {Qin}, \citenamefont
  {Wu}, \citenamefont {Ding}, \citenamefont {Hu}, \citenamefont {Hu},
  \citenamefont {Yang}, \citenamefont {Zhang}, \citenamefont {Li},
  \citenamefont {Li}, \citenamefont {Jiang}, \citenamefont {Gan}, \citenamefont
  {Yang}, \citenamefont {You}, \citenamefont {Wang}, \citenamefont {Li},
  \citenamefont {Liu}, \citenamefont {Lu},\ and\ \citenamefont
  {Pan}}]{ZhongHS2020}%
  \BibitemOpen
  \bibfield  {author} {\bibinfo {author} {\bibfnamefont {H.-S.}\ \bibnamefont
  {Zhong}}, \bibinfo {author} {\bibfnamefont {H.}~\bibnamefont {Wang}},
  \bibinfo {author} {\bibfnamefont {Y.-H.}\ \bibnamefont {Deng}}, \bibinfo
  {author} {\bibfnamefont {M.-C.}\ \bibnamefont {Chen}}, \bibinfo {author}
  {\bibfnamefont {L.-C.}\ \bibnamefont {Peng}}, \bibinfo {author}
  {\bibfnamefont {Y.-H.}\ \bibnamefont {Luo}}, \bibinfo {author} {\bibfnamefont
  {J.}~\bibnamefont {Qin}}, \bibinfo {author} {\bibfnamefont {D.}~\bibnamefont
  {Wu}}, \bibinfo {author} {\bibfnamefont {X.}~\bibnamefont {Ding}}, \bibinfo
  {author} {\bibfnamefont {Y.}~\bibnamefont {Hu}}, \bibinfo {author}
  {\bibfnamefont {P.}~\bibnamefont {Hu}}, \bibinfo {author} {\bibfnamefont
  {X.-Y.}\ \bibnamefont {Yang}}, \bibinfo {author} {\bibfnamefont {W.-J.}\
  \bibnamefont {Zhang}}, \bibinfo {author} {\bibfnamefont {H.}~\bibnamefont
  {Li}}, \bibinfo {author} {\bibfnamefont {Y.}~\bibnamefont {Li}}, \bibinfo
  {author} {\bibfnamefont {X.}~\bibnamefont {Jiang}}, \bibinfo {author}
  {\bibfnamefont {L.}~\bibnamefont {Gan}}, \bibinfo {author} {\bibfnamefont
  {G.}~\bibnamefont {Yang}}, \bibinfo {author} {\bibfnamefont {L.}~\bibnamefont
  {You}}, \bibinfo {author} {\bibfnamefont {Z.}~\bibnamefont {Wang}}, \bibinfo
  {author} {\bibfnamefont {L.}~\bibnamefont {Li}}, \bibinfo {author}
  {\bibfnamefont {N.-L.}\ \bibnamefont {Liu}}, \bibinfo {author} {\bibfnamefont
  {C.-Y.}\ \bibnamefont {Lu}}, \ and\ \bibinfo {author} {\bibfnamefont {J.-W.}\
  \bibnamefont {Pan}},\ }\href {\doibase 10.1126/science.abe8770} {\bibfield
  {journal} {\bibinfo  {journal} {Science}\ }\textbf {\bibinfo {volume}
  {370}},\ \bibinfo {pages} {1460} (\bibinfo {year} {2020})}\BibitemShut
  {NoStop}%
\bibitem [{\citenamefont {Zhong}\ \emph {et~al.}(2021)\citenamefont {Zhong},
  \citenamefont {Deng}, \citenamefont {Qin}, \citenamefont {Wang},
  \citenamefont {Chen}, \citenamefont {Peng}, \citenamefont {Luo},
  \citenamefont {Wu}, \citenamefont {Gong}, \citenamefont {Su}, \citenamefont
  {Hu}, \citenamefont {Hu}, \citenamefont {Yang}, \citenamefont {Zhang},
  \citenamefont {Li}, \citenamefont {Li}, \citenamefont {Jiang}, \citenamefont
  {Gan}, \citenamefont {Yang}, \citenamefont {You}, \citenamefont {Wang},
  \citenamefont {Li}, \citenamefont {Liu}, \citenamefont {Renema},
  \citenamefont {Lu},\ and\ \citenamefont {Pan}}]{Zhong2021}%
  \BibitemOpen
  \bibfield  {author} {\bibinfo {author} {\bibfnamefont {H.-S.}\ \bibnamefont
  {Zhong}}, \bibinfo {author} {\bibfnamefont {Y.-H.}\ \bibnamefont {Deng}},
  \bibinfo {author} {\bibfnamefont {J.}~\bibnamefont {Qin}}, \bibinfo {author}
  {\bibfnamefont {H.}~\bibnamefont {Wang}}, \bibinfo {author} {\bibfnamefont
  {M.-C.}\ \bibnamefont {Chen}}, \bibinfo {author} {\bibfnamefont {L.-C.}\
  \bibnamefont {Peng}}, \bibinfo {author} {\bibfnamefont {Y.-H.}\ \bibnamefont
  {Luo}}, \bibinfo {author} {\bibfnamefont {D.}~\bibnamefont {Wu}}, \bibinfo
  {author} {\bibfnamefont {S.-Q.}\ \bibnamefont {Gong}}, \bibinfo {author}
  {\bibfnamefont {H.}~\bibnamefont {Su}}, \bibinfo {author} {\bibfnamefont
  {Y.}~\bibnamefont {Hu}}, \bibinfo {author} {\bibfnamefont {P.}~\bibnamefont
  {Hu}}, \bibinfo {author} {\bibfnamefont {X.-Y.}\ \bibnamefont {Yang}},
  \bibinfo {author} {\bibfnamefont {W.-J.}\ \bibnamefont {Zhang}}, \bibinfo
  {author} {\bibfnamefont {H.}~\bibnamefont {Li}}, \bibinfo {author}
  {\bibfnamefont {Y.}~\bibnamefont {Li}}, \bibinfo {author} {\bibfnamefont
  {X.}~\bibnamefont {Jiang}}, \bibinfo {author} {\bibfnamefont
  {L.}~\bibnamefont {Gan}}, \bibinfo {author} {\bibfnamefont {G.}~\bibnamefont
  {Yang}}, \bibinfo {author} {\bibfnamefont {L.}~\bibnamefont {You}}, \bibinfo
  {author} {\bibfnamefont {Z.}~\bibnamefont {Wang}}, \bibinfo {author}
  {\bibfnamefont {L.}~\bibnamefont {Li}}, \bibinfo {author} {\bibfnamefont
  {N.-L.}\ \bibnamefont {Liu}}, \bibinfo {author} {\bibfnamefont {J.~J.}\
  \bibnamefont {Renema}}, \bibinfo {author} {\bibfnamefont {C.-Y.}\
  \bibnamefont {Lu}}, \ and\ \bibinfo {author} {\bibfnamefont {J.-W.}\
  \bibnamefont {Pan}},\ }\href {\doibase 10.1103/PhysRevLett.127.180502}
  {\bibfield  {journal} {\bibinfo  {journal} {Phys. Rev. Lett.}\ }\textbf
  {\bibinfo {volume} {127}},\ \bibinfo {pages} {180502} (\bibinfo {year}
  {2021})}\BibitemShut {NoStop}%
\bibitem [{\citenamefont {Madsen}\ \emph {et~al.}(2022)\citenamefont {Madsen},
  \citenamefont {Laudenbach}, \citenamefont {Askarani}, \citenamefont
  {Rortais}, \citenamefont {Vincent}, \citenamefont {Bulmer}, \citenamefont
  {Miatto}, \citenamefont {Neuhaus}, \citenamefont {Helt}, \citenamefont
  {Collins} \emph {et~al.}}]{Madsen2022}%
  \BibitemOpen
  \bibfield  {author} {\bibinfo {author} {\bibfnamefont {L.~S.}\ \bibnamefont
  {Madsen}}, \bibinfo {author} {\bibfnamefont {F.}~\bibnamefont {Laudenbach}},
  \bibinfo {author} {\bibfnamefont {M.~F.}\ \bibnamefont {Askarani}}, \bibinfo
  {author} {\bibfnamefont {F.}~\bibnamefont {Rortais}}, \bibinfo {author}
  {\bibfnamefont {T.}~\bibnamefont {Vincent}}, \bibinfo {author} {\bibfnamefont
  {J.~F.}\ \bibnamefont {Bulmer}}, \bibinfo {author} {\bibfnamefont {F.~M.}\
  \bibnamefont {Miatto}}, \bibinfo {author} {\bibfnamefont {L.}~\bibnamefont
  {Neuhaus}}, \bibinfo {author} {\bibfnamefont {L.~G.}\ \bibnamefont {Helt}},
  \bibinfo {author} {\bibfnamefont {M.~J.}\ \bibnamefont {Collins}},  \emph
  {et~al.},\ }\href {\doibase https://doi.org/10.1038/s41586-022-04725-x}
  {\bibfield  {journal} {\bibinfo  {journal} {Nature}\ }\textbf {\bibinfo
  {volume} {606}},\ \bibinfo {pages} {75} (\bibinfo {year} {2022})}\BibitemShut
  {NoStop}%
\bibitem [{\citenamefont {Deng}\ \emph {et~al.}(2023)\citenamefont {Deng},
  \citenamefont {Gu}, \citenamefont {Liu}, \citenamefont {Gong}, \citenamefont
  {Su}, \citenamefont {Zhang}, \citenamefont {Tang}, \citenamefont {Jia},
  \citenamefont {Xu}, \citenamefont {Chen}, \citenamefont {Qin}, \citenamefont
  {Peng}, \citenamefont {Yan}, \citenamefont {Hu}, \citenamefont {Huang},
  \citenamefont {Li}, \citenamefont {Li}, \citenamefont {Chen}, \citenamefont
  {Jiang}, \citenamefont {Gan}, \citenamefont {Yang}, \citenamefont {You},
  \citenamefont {Li}, \citenamefont {Zhong}, \citenamefont {Wang},
  \citenamefont {Liu}, \citenamefont {Renema}, \citenamefont {Lu},\ and\
  \citenamefont {Pan}}]{Deng2023}%
  \BibitemOpen
  \bibfield  {author} {\bibinfo {author} {\bibfnamefont {Y.-H.}\ \bibnamefont
  {Deng}}, \bibinfo {author} {\bibfnamefont {Y.-C.}\ \bibnamefont {Gu}},
  \bibinfo {author} {\bibfnamefont {H.-L.}\ \bibnamefont {Liu}}, \bibinfo
  {author} {\bibfnamefont {S.-Q.}\ \bibnamefont {Gong}}, \bibinfo {author}
  {\bibfnamefont {H.}~\bibnamefont {Su}}, \bibinfo {author} {\bibfnamefont
  {Z.-J.}\ \bibnamefont {Zhang}}, \bibinfo {author} {\bibfnamefont {H.-Y.}\
  \bibnamefont {Tang}}, \bibinfo {author} {\bibfnamefont {M.-H.}\ \bibnamefont
  {Jia}}, \bibinfo {author} {\bibfnamefont {J.-M.}\ \bibnamefont {Xu}},
  \bibinfo {author} {\bibfnamefont {M.-C.}\ \bibnamefont {Chen}}, \bibinfo
  {author} {\bibfnamefont {J.}~\bibnamefont {Qin}}, \bibinfo {author}
  {\bibfnamefont {L.-C.}\ \bibnamefont {Peng}}, \bibinfo {author}
  {\bibfnamefont {J.}~\bibnamefont {Yan}}, \bibinfo {author} {\bibfnamefont
  {Y.}~\bibnamefont {Hu}}, \bibinfo {author} {\bibfnamefont {J.}~\bibnamefont
  {Huang}}, \bibinfo {author} {\bibfnamefont {H.}~\bibnamefont {Li}}, \bibinfo
  {author} {\bibfnamefont {Y.}~\bibnamefont {Li}}, \bibinfo {author}
  {\bibfnamefont {Y.}~\bibnamefont {Chen}}, \bibinfo {author} {\bibfnamefont
  {X.}~\bibnamefont {Jiang}}, \bibinfo {author} {\bibfnamefont
  {L.}~\bibnamefont {Gan}}, \bibinfo {author} {\bibfnamefont {G.}~\bibnamefont
  {Yang}}, \bibinfo {author} {\bibfnamefont {L.}~\bibnamefont {You}}, \bibinfo
  {author} {\bibfnamefont {L.}~\bibnamefont {Li}}, \bibinfo {author}
  {\bibfnamefont {H.-S.}\ \bibnamefont {Zhong}}, \bibinfo {author}
  {\bibfnamefont {H.}~\bibnamefont {Wang}}, \bibinfo {author} {\bibfnamefont
  {N.-L.}\ \bibnamefont {Liu}}, \bibinfo {author} {\bibfnamefont {J.~J.}\
  \bibnamefont {Renema}}, \bibinfo {author} {\bibfnamefont {C.-Y.}\
  \bibnamefont {Lu}}, \ and\ \bibinfo {author} {\bibfnamefont {J.-W.}\
  \bibnamefont {Pan}},\ }\href {\doibase 10.1103/PhysRevLett.131.150601}
  {\bibfield  {journal} {\bibinfo  {journal} {Phys. Rev. Lett.}\ }\textbf
  {\bibinfo {volume} {131}},\ \bibinfo {pages} {150601} (\bibinfo {year}
  {2023})}\BibitemShut {NoStop}%
\bibitem [{\citenamefont {Hamilton}\ \emph {et~al.}(2017)\citenamefont
  {Hamilton}, \citenamefont {Kruse}, \citenamefont {Sansoni}, \citenamefont
  {Barkhofen}, \citenamefont {Silberhorn},\ and\ \citenamefont
  {Jex}}]{Hamilton2017}%
  \BibitemOpen
  \bibfield  {author} {\bibinfo {author} {\bibfnamefont {C.~S.}\ \bibnamefont
  {Hamilton}}, \bibinfo {author} {\bibfnamefont {R.}~\bibnamefont {Kruse}},
  \bibinfo {author} {\bibfnamefont {L.}~\bibnamefont {Sansoni}}, \bibinfo
  {author} {\bibfnamefont {S.}~\bibnamefont {Barkhofen}}, \bibinfo {author}
  {\bibfnamefont {C.}~\bibnamefont {Silberhorn}}, \ and\ \bibinfo {author}
  {\bibfnamefont {I.}~\bibnamefont {Jex}},\ }\href {\doibase
  10.1103/PhysRevLett.119.170501} {\bibfield  {journal} {\bibinfo  {journal}
  {Phys. Rev. Lett.}\ }\textbf {\bibinfo {volume} {119}},\ \bibinfo {pages}
  {170501} (\bibinfo {year} {2017})}\BibitemShut {NoStop}%
\bibitem [{\citenamefont {Valiant}(1979)}]{Valiant1979}%
  \BibitemOpen
  \bibfield  {author} {\bibinfo {author} {\bibfnamefont {L.}~\bibnamefont
  {Valiant}},\ }\href {\doibase https://doi.org/10.1016/0304-3975(79)90044-6}
  {\bibfield  {journal} {\bibinfo  {journal} {Theor. Comput. Sci.}\ }\textbf
  {\bibinfo {volume} {8}},\ \bibinfo {pages} {189} (\bibinfo {year}
  {1979})}\BibitemShut {NoStop}%
\bibitem [{\citenamefont {Kalai}\ and\ \citenamefont
  {Kindler}(2014)}]{Kalai2014}%
  \BibitemOpen
  \bibfield  {author} {\bibinfo {author} {\bibfnamefont {G.}~\bibnamefont
  {Kalai}}\ and\ \bibinfo {author} {\bibfnamefont {G.}~\bibnamefont
  {Kindler}},\ }\href {https://arxiv.org/abs/1409.3093} {\bibfield  {journal}
  {\bibinfo  {journal} {arXiv:1409.3093}\ } (\bibinfo {year}
  {2014})}\BibitemShut {NoStop}%
\bibitem [{\citenamefont {Rahimi-Keshari}\ \emph {et~al.}(2015)\citenamefont
  {Rahimi-Keshari}, \citenamefont {Lund},\ and\ \citenamefont
  {Ralph}}]{Rahimi-Keshari2015}%
  \BibitemOpen
  \bibfield  {author} {\bibinfo {author} {\bibfnamefont {S.}~\bibnamefont
  {Rahimi-Keshari}}, \bibinfo {author} {\bibfnamefont {A.~P.}\ \bibnamefont
  {Lund}}, \ and\ \bibinfo {author} {\bibfnamefont {T.~C.}\ \bibnamefont
  {Ralph}},\ }\href {\doibase 10.1103/PhysRevLett.114.060501} {\bibfield
  {journal} {\bibinfo  {journal} {Phys. Rev. Lett.}\ }\textbf {\bibinfo
  {volume} {114}},\ \bibinfo {pages} {060501} (\bibinfo {year}
  {2015})}\BibitemShut {NoStop}%
\bibitem [{\citenamefont {Oszmaniec}\ and\ \citenamefont
  {Brod}(2018)}]{Oszmaniec2018}%
  \BibitemOpen
  \bibfield  {author} {\bibinfo {author} {\bibfnamefont {M.}~\bibnamefont
  {Oszmaniec}}\ and\ \bibinfo {author} {\bibfnamefont {D.~J.}\ \bibnamefont
  {Brod}},\ }\href {\doibase 10.1088/1367-2630/aadfa8} {\bibfield  {journal}
  {\bibinfo  {journal} {New J. Phys.}\ }\textbf {\bibinfo {volume} {20}},\
  \bibinfo {pages} {092002} (\bibinfo {year} {2018})}\BibitemShut {NoStop}%
\bibitem [{\citenamefont {Qi}\ \emph {et~al.}(2020)\citenamefont {Qi},
  \citenamefont {Brod}, \citenamefont {Quesada},\ and\ \citenamefont
  {Garc\'{\i}a-Patr\'on}}]{Qi2020}%
  \BibitemOpen
  \bibfield  {author} {\bibinfo {author} {\bibfnamefont {H.}~\bibnamefont
  {Qi}}, \bibinfo {author} {\bibfnamefont {D.~J.}\ \bibnamefont {Brod}},
  \bibinfo {author} {\bibfnamefont {N.}~\bibnamefont {Quesada}}, \ and\
  \bibinfo {author} {\bibfnamefont {R.}~\bibnamefont {Garc\'{\i}a-Patr\'on}},\
  }\href {\doibase 10.1103/PhysRevLett.124.100502} {\bibfield  {journal}
  {\bibinfo  {journal} {Phys. Rev. Lett.}\ }\textbf {\bibinfo {volume} {124}},\
  \bibinfo {pages} {100502} (\bibinfo {year} {2020})}\BibitemShut {NoStop}%
\bibitem [{\citenamefont {Aharonov}\ \emph {et~al.}(2023)\citenamefont
  {Aharonov}, \citenamefont {Gao}, \citenamefont {Landau}, \citenamefont
  {Liu},\ and\ \citenamefont {Vazirani}}]{Aharonov2023}%
  \BibitemOpen
  \bibfield  {author} {\bibinfo {author} {\bibfnamefont {D.}~\bibnamefont
  {Aharonov}}, \bibinfo {author} {\bibfnamefont {X.}~\bibnamefont {Gao}},
  \bibinfo {author} {\bibfnamefont {Z.}~\bibnamefont {Landau}}, \bibinfo
  {author} {\bibfnamefont {Y.}~\bibnamefont {Liu}}, \ and\ \bibinfo {author}
  {\bibfnamefont {U.}~\bibnamefont {Vazirani}},\ }in\ \href {\doibase
  https://doi.org/1Schuch0.1145/3564246.3585234} {\emph {\bibinfo {booktitle}
  {Proceedings of the 55th Annual ACM Symposium on Theory of Computing}}}\
  (\bibinfo {year} {2023})\ pp.\ \bibinfo {pages} {945--957}\BibitemShut
  {NoStop}%
\bibitem [{\citenamefont {Oh}\ \emph {et~al.}(2023)\citenamefont {Oh},
  \citenamefont {Jiang},\ and\ \citenamefont {Fefferman}}]{Oh2023}%
  \BibitemOpen
  \bibfield  {author} {\bibinfo {author} {\bibfnamefont {C.}~\bibnamefont
  {Oh}}, \bibinfo {author} {\bibfnamefont {L.}~\bibnamefont {Jiang}}, \ and\
  \bibinfo {author} {\bibfnamefont {B.}~\bibnamefont {Fefferman}},\ }\href
  {https://arxiv.org/abs/2301.11532} {\bibfield  {journal} {\bibinfo  {journal}
  {arXiv:2301.11532}\ } (\bibinfo {year} {2023})}\BibitemShut {NoStop}%
\bibitem [{\citenamefont {Oh}\ \emph {et~al.}(2024)\citenamefont {Oh},
  \citenamefont {Liu}, \citenamefont {Alexeev}, \citenamefont {Fefferman},\
  and\ \citenamefont {Jiang}}]{Oh2024}%
  \BibitemOpen
  \bibfield  {author} {\bibinfo {author} {\bibfnamefont {C.}~\bibnamefont
  {Oh}}, \bibinfo {author} {\bibfnamefont {M.}~\bibnamefont {Liu}}, \bibinfo
  {author} {\bibfnamefont {Y.}~\bibnamefont {Alexeev}}, \bibinfo {author}
  {\bibfnamefont {B.}~\bibnamefont {Fefferman}}, \ and\ \bibinfo {author}
  {\bibfnamefont {L.}~\bibnamefont {Jiang}},\ }\href {\doibase
  10.1038/s41567-024-02535-8} {\bibfield  {journal} {\bibinfo  {journal} {Nat.
  Phys.}\ }\textbf {\bibinfo {volume} {20}},\ \bibinfo {pages} {1461} (\bibinfo
  {year} {2024})}\BibitemShut {NoStop}%
\bibitem [{\citenamefont {Auerbach}(2012)}]{Auerbach-Book}%
  \BibitemOpen
  \bibfield  {author} {\bibinfo {author} {\bibfnamefont {A.}~\bibnamefont
  {Auerbach}},\ }\href {\doibase 10.1007/978-1-4612-0869-3} {\emph {\bibinfo
  {title} {Interacting electrons and quantum magnetism}}}\ (\bibinfo
  {publisher} {Springer New York},\ \bibinfo {year} {2012})\BibitemShut
  {NoStop}%
\bibitem [{\citenamefont {Arovas}\ and\ \citenamefont
  {Auerbach}(1988)}]{Arovas1988}%
  \BibitemOpen
  \bibfield  {author} {\bibinfo {author} {\bibfnamefont {D.~P.}\ \bibnamefont
  {Arovas}}\ and\ \bibinfo {author} {\bibfnamefont {A.}~\bibnamefont
  {Auerbach}},\ }\href {\doibase 10.1103/PhysRevB.38.316} {\bibfield  {journal}
  {\bibinfo  {journal} {Phys. Rev. B}\ }\textbf {\bibinfo {volume} {38}},\
  \bibinfo {pages} {316} (\bibinfo {year} {1988})}\BibitemShut {NoStop}%
\bibitem [{\citenamefont {Sachdev}(1992)}]{Sachdev1992}%
  \BibitemOpen
  \bibfield  {author} {\bibinfo {author} {\bibfnamefont {S.}~\bibnamefont
  {Sachdev}},\ }\href {\doibase 10.1103/PhysRevB.45.12377} {\bibfield
  {journal} {\bibinfo  {journal} {Phys. Rev. B}\ }\textbf {\bibinfo {volume}
  {45}},\ \bibinfo {pages} {12377} (\bibinfo {year} {1992})}\BibitemShut
  {NoStop}%
\bibitem [{\citenamefont {Wang}\ and\ \citenamefont
  {Vishwanath}(2006)}]{WangF2006}%
  \BibitemOpen
  \bibfield  {author} {\bibinfo {author} {\bibfnamefont {F.}~\bibnamefont
  {Wang}}\ and\ \bibinfo {author} {\bibfnamefont {A.}~\bibnamefont
  {Vishwanath}},\ }\href {\doibase 10.1103/PhysRevB.74.174423} {\bibfield
  {journal} {\bibinfo  {journal} {Phys. Rev. B}\ }\textbf {\bibinfo {volume}
  {74}},\ \bibinfo {pages} {174423} (\bibinfo {year} {2006})}\BibitemShut
  {NoStop}%
\bibitem [{\citenamefont {Keeling}\ \emph {et~al.}(2025)\citenamefont
  {Keeling}, \citenamefont {Stoudenmire}, \citenamefont {Bañuls},\ and\
  \citenamefont {Reichman}}]{Keeling2025}%
  \BibitemOpen
  \bibfield  {author} {\bibinfo {author} {\bibfnamefont {J.}~\bibnamefont
  {Keeling}}, \bibinfo {author} {\bibfnamefont {E.~M.}\ \bibnamefont
  {Stoudenmire}}, \bibinfo {author} {\bibfnamefont {M.-C.}\ \bibnamefont
  {Bañuls}}, \ and\ \bibinfo {author} {\bibfnamefont {D.~R.}\ \bibnamefont
  {Reichman}},\ }\href {https://arxiv.org/abs/2509.07661} {\bibfield  {journal}
  {\bibinfo  {journal} {arXiv:2509.07661}\ } (\bibinfo {year}
  {2025})}\BibitemShut {NoStop}%
\bibitem [{\citenamefont {Link}\ \emph {et~al.}(2024)\citenamefont {Link},
  \citenamefont {Tu},\ and\ \citenamefont {Strunz}}]{Link2024}%
  \BibitemOpen
  \bibfield  {author} {\bibinfo {author} {\bibfnamefont {V.}~\bibnamefont
  {Link}}, \bibinfo {author} {\bibfnamefont {H.-H.}\ \bibnamefont {Tu}}, \ and\
  \bibinfo {author} {\bibfnamefont {W.~T.}\ \bibnamefont {Strunz}},\ }\href
  {\doibase 10.1103/PhysRevLett.132.200403} {\bibfield  {journal} {\bibinfo
  {journal} {Phys. Rev. Lett.}\ }\textbf {\bibinfo {volume} {132}},\ \bibinfo
  {pages} {200403} (\bibinfo {year} {2024})}\BibitemShut {NoStop}%
\bibitem [{\citenamefont {Chen}\ and\ \citenamefont {Liu}(2025)}]{ChenC2025}%
  \BibitemOpen
  \bibfield  {author} {\bibinfo {author} {\bibfnamefont {C.}~\bibnamefont
  {Chen}}\ and\ \bibinfo {author} {\bibfnamefont {R.-B.}\ \bibnamefont {Liu}},\
  }\href {https://arxiv.org/abs/2509.00424} {\bibfield  {journal} {\bibinfo
  {journal} {arXiv:2509.00424}\ } (\bibinfo {year} {2025})}\BibitemShut
  {NoStop}%
\bibitem [{\citenamefont {Chanda}\ \emph {et~al.}(2020)\citenamefont {Chanda},
  \citenamefont {Zakrzewski}, \citenamefont {Lewenstein},\ and\ \citenamefont
  {Tagliacozzo}}]{Chanda2020}%
  \BibitemOpen
  \bibfield  {author} {\bibinfo {author} {\bibfnamefont {T.}~\bibnamefont
  {Chanda}}, \bibinfo {author} {\bibfnamefont {J.}~\bibnamefont {Zakrzewski}},
  \bibinfo {author} {\bibfnamefont {M.}~\bibnamefont {Lewenstein}}, \ and\
  \bibinfo {author} {\bibfnamefont {L.}~\bibnamefont {Tagliacozzo}},\ }\href
  {\doibase 10.1103/PhysRevLett.124.180602} {\bibfield  {journal} {\bibinfo
  {journal} {Phys. Rev. Lett.}\ }\textbf {\bibinfo {volume} {124}},\ \bibinfo
  {pages} {180602} (\bibinfo {year} {2020})}\BibitemShut {NoStop}%
\bibitem [{\citenamefont {Jerrum}\ \emph {et~al.}(2004)\citenamefont {Jerrum},
  \citenamefont {Sinclair},\ and\ \citenamefont {Vigoda}}]{Jerrum2004}%
  \BibitemOpen
  \bibfield  {author} {\bibinfo {author} {\bibfnamefont {M.}~\bibnamefont
  {Jerrum}}, \bibinfo {author} {\bibfnamefont {A.}~\bibnamefont {Sinclair}}, \
  and\ \bibinfo {author} {\bibfnamefont {E.}~\bibnamefont {Vigoda}},\ }\href
  {\doibase 10.1145/1008731.1008738} {\bibfield  {journal} {\bibinfo  {journal}
  {J. ACM}\ }\textbf {\bibinfo {volume} {51}},\ \bibinfo {pages} {671}
  (\bibinfo {year} {2004})}\BibitemShut {NoStop}%
\bibitem [{\citenamefont {Aaronson}\ and\ \citenamefont
  {Hance}(2012)}]{Aaronson2012}%
  \BibitemOpen
  \bibfield  {author} {\bibinfo {author} {\bibfnamefont {S.}~\bibnamefont
  {Aaronson}}\ and\ \bibinfo {author} {\bibfnamefont {T.}~\bibnamefont
  {Hance}},\ }\href {https://arxiv.org/abs/1212.0025} {\bibfield  {journal}
  {\bibinfo  {journal} {arXiv:1212.0025}\ } (\bibinfo {year}
  {2012})}\BibitemShut {NoStop}%
\bibitem [{\citenamefont {Lim}\ and\ \citenamefont {Oh}(2025)}]{LimY2025}%
  \BibitemOpen
  \bibfield  {author} {\bibinfo {author} {\bibfnamefont {Y.}~\bibnamefont
  {Lim}}\ and\ \bibinfo {author} {\bibfnamefont {C.}~\bibnamefont {Oh}},\
  }\href {https://arxiv.org/abs/2502.12882} {\bibfield  {journal} {\bibinfo
  {journal} {arXiv:2502.12882}\ } (\bibinfo {year} {2025})}\BibitemShut
  {NoStop}%
\bibitem [{\citenamefont {Quesada}\ \emph {et~al.}(2022)\citenamefont
  {Quesada}, \citenamefont {Chadwick}, \citenamefont {Bell}, \citenamefont
  {Arrazola}, \citenamefont {Vincent}, \citenamefont {Qi},\ and\ \citenamefont
  {Garc\'{\i}a-Patr\'on}}]{Quesada2022}%
  \BibitemOpen
  \bibfield  {author} {\bibinfo {author} {\bibfnamefont {N.}~\bibnamefont
  {Quesada}}, \bibinfo {author} {\bibfnamefont {R.~S.}\ \bibnamefont
  {Chadwick}}, \bibinfo {author} {\bibfnamefont {B.~A.}\ \bibnamefont {Bell}},
  \bibinfo {author} {\bibfnamefont {J.~M.}\ \bibnamefont {Arrazola}}, \bibinfo
  {author} {\bibfnamefont {T.}~\bibnamefont {Vincent}}, \bibinfo {author}
  {\bibfnamefont {H.}~\bibnamefont {Qi}}, \ and\ \bibinfo {author}
  {\bibfnamefont {R.}~\bibnamefont {Garc\'{\i}a-Patr\'on}},\ }\href {\doibase
  10.1103/PRXQuantum.3.010306} {\bibfield  {journal} {\bibinfo  {journal} {PRX
  Quantum}\ }\textbf {\bibinfo {volume} {3}},\ \bibinfo {pages} {010306}
  (\bibinfo {year} {2022})}\BibitemShut {NoStop}%
\bibitem [{\citenamefont {Bulmer}\ \emph {et~al.}(2022)\citenamefont {Bulmer},
  \citenamefont {Bell}, \citenamefont {Chadwick}, \citenamefont {Jones},
  \citenamefont {Moise}, \citenamefont {Rigazzi}, \citenamefont {Thorbecke},
  \citenamefont {Haus}, \citenamefont {Vaerenbergh}, \citenamefont {Patel},
  \citenamefont {Walmsley},\ and\ \citenamefont {Laing}}]{Blumer2022}%
  \BibitemOpen
  \bibfield  {author} {\bibinfo {author} {\bibfnamefont {J.~F.~F.}\
  \bibnamefont {Bulmer}}, \bibinfo {author} {\bibfnamefont {B.~A.}\
  \bibnamefont {Bell}}, \bibinfo {author} {\bibfnamefont {R.~S.}\ \bibnamefont
  {Chadwick}}, \bibinfo {author} {\bibfnamefont {A.~E.}\ \bibnamefont {Jones}},
  \bibinfo {author} {\bibfnamefont {D.}~\bibnamefont {Moise}}, \bibinfo
  {author} {\bibfnamefont {A.}~\bibnamefont {Rigazzi}}, \bibinfo {author}
  {\bibfnamefont {J.}~\bibnamefont {Thorbecke}}, \bibinfo {author}
  {\bibfnamefont {U.-U.}\ \bibnamefont {Haus}}, \bibinfo {author}
  {\bibfnamefont {T.~V.}\ \bibnamefont {Vaerenbergh}}, \bibinfo {author}
  {\bibfnamefont {R.~B.}\ \bibnamefont {Patel}}, \bibinfo {author}
  {\bibfnamefont {I.~A.}\ \bibnamefont {Walmsley}}, \ and\ \bibinfo {author}
  {\bibfnamefont {A.}~\bibnamefont {Laing}},\ }\href {\doibase
  10.1126/sciadv.abl9236} {\bibfield  {journal} {\bibinfo  {journal} {Sci.
  Adv.}\ }\textbf {\bibinfo {volume} {8}},\ \bibinfo {pages} {eabl9236}
  (\bibinfo {year} {2022})}\BibitemShut {NoStop}%
\bibitem [{\citenamefont {Dodd}\ \emph {et~al.}(2025)\citenamefont {Dodd},
  \citenamefont {Mart{\'\i}nez-Cifuentes}, \citenamefont {Brown}, \citenamefont
  {Quesada},\ and\ \citenamefont {Garc{\'\i}a-Patr{\'o}n}}]{Dodd2025}%
  \BibitemOpen
  \bibfield  {author} {\bibinfo {author} {\bibfnamefont {T.}~\bibnamefont
  {Dodd}}, \bibinfo {author} {\bibfnamefont {J.}~\bibnamefont
  {Mart{\'\i}nez-Cifuentes}}, \bibinfo {author} {\bibfnamefont {O.~T.}\
  \bibnamefont {Brown}}, \bibinfo {author} {\bibfnamefont {N.}~\bibnamefont
  {Quesada}}, \ and\ \bibinfo {author} {\bibfnamefont {R.}~\bibnamefont
  {Garc{\'\i}a-Patr{\'o}n}},\ }\href {https://arxiv.org/abs/2511.14923}
  {\bibfield  {journal} {\bibinfo  {journal} {arXiv:2511.14923}\ } (\bibinfo
  {year} {2025})}\BibitemShut {NoStop}%
\bibitem [{\citenamefont {\"Ostlund}\ and\ \citenamefont
  {Rommer}(1995)}]{Ostlund1995}%
  \BibitemOpen
  \bibfield  {author} {\bibinfo {author} {\bibfnamefont {S.}~\bibnamefont
  {\"Ostlund}}\ and\ \bibinfo {author} {\bibfnamefont {S.}~\bibnamefont
  {Rommer}},\ }\href {\doibase 10.1103/PhysRevLett.75.3537} {\bibfield
  {journal} {\bibinfo  {journal} {Phys. Rev. Lett.}\ }\textbf {\bibinfo
  {volume} {75}},\ \bibinfo {pages} {3537} (\bibinfo {year}
  {1995})}\BibitemShut {NoStop}%
\bibitem [{\citenamefont {Vidal}(2003)}]{Vidal2003b}%
  \BibitemOpen
  \bibfield  {author} {\bibinfo {author} {\bibfnamefont {G.}~\bibnamefont
  {Vidal}},\ }\href {\doibase 10.1103/PhysRevLett.91.147902} {\bibfield
  {journal} {\bibinfo  {journal} {Phys. Rev. Lett.}\ }\textbf {\bibinfo
  {volume} {91}},\ \bibinfo {pages} {147902} (\bibinfo {year}
  {2003})}\BibitemShut {NoStop}%
\bibitem [{\citenamefont {Perez-Garcia}\ \emph {et~al.}(2007)\citenamefont
  {Perez-Garcia}, \citenamefont {Verstraete}, \citenamefont {Wolf},\ and\
  \citenamefont {Cirac}}]{Perez-Garcia2007}%
  \BibitemOpen
  \bibfield  {author} {\bibinfo {author} {\bibfnamefont {D.}~\bibnamefont
  {Perez-Garcia}}, \bibinfo {author} {\bibfnamefont {F.}~\bibnamefont
  {Verstraete}}, \bibinfo {author} {\bibfnamefont {M.~M.}\ \bibnamefont
  {Wolf}}, \ and\ \bibinfo {author} {\bibfnamefont {J.~I.}\ \bibnamefont
  {Cirac}},\ }\href {\doibase 10.26421/QIC7.5-6-1} {\bibfield  {journal}
  {\bibinfo  {journal} {Quantum Inf. and Comp.}\ }\textbf {\bibinfo {volume}
  {7}},\ \bibinfo {pages} {401} (\bibinfo {year} {2007})}\BibitemShut {NoStop}%
\bibitem [{\citenamefont {Verstraete}\ \emph {et~al.}(2008)\citenamefont
  {Verstraete}, \citenamefont {Murg},\ and\ \citenamefont
  {Cirac}}]{Verstraete2008}%
  \BibitemOpen
  \bibfield  {author} {\bibinfo {author} {\bibfnamefont {F.}~\bibnamefont
  {Verstraete}}, \bibinfo {author} {\bibfnamefont {V.}~\bibnamefont {Murg}}, \
  and\ \bibinfo {author} {\bibfnamefont {J.~I.}\ \bibnamefont {Cirac}},\ }\href
  {\doibase 10.1080/14789940801912366} {\bibfield  {journal} {\bibinfo
  {journal} {Adv. Phys.}\ }\textbf {\bibinfo {volume} {57}},\ \bibinfo {pages}
  {143} (\bibinfo {year} {2008})}\BibitemShut {NoStop}%
\bibitem [{\citenamefont {Schollw{\"o}ck}(2011)}]{Schollwoeck2011}%
  \BibitemOpen
  \bibfield  {author} {\bibinfo {author} {\bibfnamefont {U.}~\bibnamefont
  {Schollw{\"o}ck}},\ }\href {\doibase 10.1016/j.aop.2010.09.012} {\bibfield
  {journal} {\bibinfo  {journal} {Ann. Phys.}\ }\textbf {\bibinfo {volume}
  {326}},\ \bibinfo {pages} {96 } (\bibinfo {year} {2011})}\BibitemShut
  {NoStop}%
\bibitem [{\citenamefont {Cirac}\ \emph {et~al.}(2021)\citenamefont {Cirac},
  \citenamefont {P\'erez-Garc\'{\i}a}, \citenamefont {Schuch},\ and\
  \citenamefont {Verstraete}}]{Cirac2021}%
  \BibitemOpen
  \bibfield  {author} {\bibinfo {author} {\bibfnamefont {J.~I.}\ \bibnamefont
  {Cirac}}, \bibinfo {author} {\bibfnamefont {D.}~\bibnamefont
  {P\'erez-Garc\'{\i}a}}, \bibinfo {author} {\bibfnamefont {N.}~\bibnamefont
  {Schuch}}, \ and\ \bibinfo {author} {\bibfnamefont {F.}~\bibnamefont
  {Verstraete}},\ }\href {\doibase 10.1103/RevModPhys.93.045003} {\bibfield
  {journal} {\bibinfo  {journal} {Rev. Mod. Phys.}\ }\textbf {\bibinfo {volume}
  {93}},\ \bibinfo {pages} {045003} (\bibinfo {year} {2021})}\BibitemShut
  {NoStop}%
\bibitem [{\citenamefont {Xiang}(2023)}]{XiangT-Book}%
  \BibitemOpen
  \bibfield  {author} {\bibinfo {author} {\bibfnamefont {T.}~\bibnamefont
  {Xiang}},\ }\href {\doibase 10.1017/9781009398671.002} {\emph {\bibinfo
  {title} {Density Matrix and Tensor Network Renormalization}}}\ (\bibinfo
  {publisher} {Cambridge University Press},\ \bibinfo {year}
  {2023})\BibitemShut {NoStop}%
\bibitem [{\citenamefont {Fishman}\ and\ \citenamefont
  {White}(2015)}]{Fishman2015}%
  \BibitemOpen
  \bibfield  {author} {\bibinfo {author} {\bibfnamefont {M.~T.}\ \bibnamefont
  {Fishman}}\ and\ \bibinfo {author} {\bibfnamefont {S.~R.}\ \bibnamefont
  {White}},\ }\href {\doibase 10.1103/PhysRevB.92.075132} {\bibfield  {journal}
  {\bibinfo  {journal} {Phys. Rev. B}\ }\textbf {\bibinfo {volume} {92}},\
  \bibinfo {pages} {075132} (\bibinfo {year} {2015})}\BibitemShut {NoStop}%
\bibitem [{\citenamefont {Schuch}\ and\ \citenamefont
  {Bauer}(2019)}]{Schuch2019}%
  \BibitemOpen
  \bibfield  {author} {\bibinfo {author} {\bibfnamefont {N.}~\bibnamefont
  {Schuch}}\ and\ \bibinfo {author} {\bibfnamefont {B.}~\bibnamefont {Bauer}},\
  }\href {\doibase 10.1103/PhysRevB.100.245121} {\bibfield  {journal} {\bibinfo
   {journal} {Phys. Rev. B}\ }\textbf {\bibinfo {volume} {100}},\ \bibinfo
  {pages} {245121} (\bibinfo {year} {2019})}\BibitemShut {NoStop}%
\bibitem [{\citenamefont {Wu}\ \emph {et~al.}(2020)\citenamefont {Wu},
  \citenamefont {Wang},\ and\ \citenamefont {Tu}}]{WuYH2020}%
  \BibitemOpen
  \bibfield  {author} {\bibinfo {author} {\bibfnamefont {Y.-H.}\ \bibnamefont
  {Wu}}, \bibinfo {author} {\bibfnamefont {L.}~\bibnamefont {Wang}}, \ and\
  \bibinfo {author} {\bibfnamefont {H.-H.}\ \bibnamefont {Tu}},\ }\href
  {\doibase 10.1103/PhysRevLett.124.246401} {\bibfield  {journal} {\bibinfo
  {journal} {Phys. Rev. Lett.}\ }\textbf {\bibinfo {volume} {124}},\ \bibinfo
  {pages} {246401} (\bibinfo {year} {2020})}\BibitemShut {NoStop}%
\bibitem [{\citenamefont {Jin}\ \emph {et~al.}(2020)\citenamefont {Jin},
  \citenamefont {Tu},\ and\ \citenamefont {Zhou}}]{JinHK2020}%
  \BibitemOpen
  \bibfield  {author} {\bibinfo {author} {\bibfnamefont {H.-K.}\ \bibnamefont
  {Jin}}, \bibinfo {author} {\bibfnamefont {H.-H.}\ \bibnamefont {Tu}}, \ and\
  \bibinfo {author} {\bibfnamefont {Y.}~\bibnamefont {Zhou}},\ }\href {\doibase
  10.1103/PhysRevB.101.165135} {\bibfield  {journal} {\bibinfo  {journal}
  {Phys. Rev. B}\ }\textbf {\bibinfo {volume} {101}},\ \bibinfo {pages}
  {165135} (\bibinfo {year} {2020})}\BibitemShut {NoStop}%
\bibitem [{\citenamefont {Aghaei}\ \emph {et~al.}(2020)\citenamefont {Aghaei},
  \citenamefont {Bauer}, \citenamefont {Shtengel},\ and\ \citenamefont
  {Mishmash}}]{Aghaei2020}%
  \BibitemOpen
  \bibfield  {author} {\bibinfo {author} {\bibfnamefont {A.~M.}\ \bibnamefont
  {Aghaei}}, \bibinfo {author} {\bibfnamefont {B.}~\bibnamefont {Bauer}},
  \bibinfo {author} {\bibfnamefont {K.}~\bibnamefont {Shtengel}}, \ and\
  \bibinfo {author} {\bibfnamefont {R.~V.}\ \bibnamefont {Mishmash}},\ }\href
  {https://arxiv.org/abs/2009.12435} {\bibfield  {journal} {\bibinfo  {journal}
  {arXiv:2009.12435}\ } (\bibinfo {year} {2020})}\BibitemShut {NoStop}%
\bibitem [{\citenamefont {Petrica}\ \emph {et~al.}(2021)\citenamefont
  {Petrica}, \citenamefont {Zheng}, \citenamefont {Chan},\ and\ \citenamefont
  {Clark}}]{Petrica2021}%
  \BibitemOpen
  \bibfield  {author} {\bibinfo {author} {\bibfnamefont {G.}~\bibnamefont
  {Petrica}}, \bibinfo {author} {\bibfnamefont {B.-X.}\ \bibnamefont {Zheng}},
  \bibinfo {author} {\bibfnamefont {G.~K.-L.}\ \bibnamefont {Chan}}, \ and\
  \bibinfo {author} {\bibfnamefont {B.~K.}\ \bibnamefont {Clark}},\ }\href
  {\doibase 10.1103/PhysRevB.103.125161} {\bibfield  {journal} {\bibinfo
  {journal} {Phys. Rev. B}\ }\textbf {\bibinfo {volume} {103}},\ \bibinfo
  {pages} {125161} (\bibinfo {year} {2021})}\BibitemShut {NoStop}%
\bibitem [{\citenamefont {N\"u\ss{}eler}\ \emph {et~al.}(2021)\citenamefont
  {N\"u\ss{}eler}, \citenamefont {Dhand}, \citenamefont {Huelga},\ and\
  \citenamefont {Plenio}}]{Nuesseler2021}%
  \BibitemOpen
  \bibfield  {author} {\bibinfo {author} {\bibfnamefont {A.}~\bibnamefont
  {N\"u\ss{}eler}}, \bibinfo {author} {\bibfnamefont {I.}~\bibnamefont
  {Dhand}}, \bibinfo {author} {\bibfnamefont {S.~F.}\ \bibnamefont {Huelga}}, \
  and\ \bibinfo {author} {\bibfnamefont {M.~B.}\ \bibnamefont {Plenio}},\
  }\href {\doibase 10.1103/PhysRevA.104.012415} {\bibfield  {journal} {\bibinfo
   {journal} {Phys. Rev. A}\ }\textbf {\bibinfo {volume} {104}},\ \bibinfo
  {pages} {012415} (\bibinfo {year} {2021})}\BibitemShut {NoStop}%
\bibitem [{\citenamefont {Jin}\ \emph {et~al.}(2022)\citenamefont {Jin},
  \citenamefont {Sun}, \citenamefont {Zhou},\ and\ \citenamefont
  {Tu}}]{JinHK2022a}%
  \BibitemOpen
  \bibfield  {author} {\bibinfo {author} {\bibfnamefont {H.-K.}\ \bibnamefont
  {Jin}}, \bibinfo {author} {\bibfnamefont {R.-Y.}\ \bibnamefont {Sun}},
  \bibinfo {author} {\bibfnamefont {Y.}~\bibnamefont {Zhou}}, \ and\ \bibinfo
  {author} {\bibfnamefont {H.-H.}\ \bibnamefont {Tu}},\ }\href {\doibase
  10.1103/PhysRevB.105.L081101} {\bibfield  {journal} {\bibinfo  {journal}
  {Phys. Rev. B}\ }\textbf {\bibinfo {volume} {105}},\ \bibinfo {pages}
  {L081101} (\bibinfo {year} {2022})}\BibitemShut {NoStop}%
\bibitem [{\citenamefont {Liu}\ \emph {et~al.}(2025{\natexlab{a}})\citenamefont
  {Liu}, \citenamefont {Wu}, \citenamefont {Tu},\ and\ \citenamefont
  {Xiang}}]{LiuT2025a}%
  \BibitemOpen
  \bibfield  {author} {\bibinfo {author} {\bibfnamefont {T.}~\bibnamefont
  {Liu}}, \bibinfo {author} {\bibfnamefont {Y.-H.}\ \bibnamefont {Wu}},
  \bibinfo {author} {\bibfnamefont {H.-H.}\ \bibnamefont {Tu}}, \ and\ \bibinfo
  {author} {\bibfnamefont {T.}~\bibnamefont {Xiang}},\ }\href {\doibase
  10.1088/2058-9565/addae0} {\bibfield  {journal} {\bibinfo  {journal} {Quantum
  Sci. Technol.}\ }\textbf {\bibinfo {volume} {10}},\ \bibinfo {pages} {035033}
  (\bibinfo {year} {2025}{\natexlab{a}})}\BibitemShut {NoStop}%
\bibitem [{\citenamefont {Li}\ \emph {et~al.}(2025)\citenamefont {Li},
  \citenamefont {Zhang},\ and\ \citenamefont {Po}}]{LiKL2025}%
  \BibitemOpen
  \bibfield  {author} {\bibinfo {author} {\bibfnamefont {K.}~\bibnamefont
  {Li}}, \bibinfo {author} {\bibfnamefont {Y.-B.}\ \bibnamefont {Zhang}}, \
  and\ \bibinfo {author} {\bibfnamefont {H.~C.}\ \bibnamefont {Po}},\ }\href
  {\doibase 10.1103/PhysRevResearch.7.013182} {\bibfield  {journal} {\bibinfo
  {journal} {Phys. Rev. Res.}\ }\textbf {\bibinfo {volume} {7}},\ \bibinfo
  {pages} {013182} (\bibinfo {year} {2025})}\BibitemShut {NoStop}%
\bibitem [{\citenamefont {Jin}\ \emph {et~al.}(2025)\citenamefont {Jin},
  \citenamefont {Sun}, \citenamefont {Tu},\ and\ \citenamefont
  {Zhou}}]{JinHK2025}%
  \BibitemOpen
  \bibfield  {author} {\bibinfo {author} {\bibfnamefont {H.-K.}\ \bibnamefont
  {Jin}}, \bibinfo {author} {\bibfnamefont {R.-Y.}\ \bibnamefont {Sun}},
  \bibinfo {author} {\bibfnamefont {H.-H.}\ \bibnamefont {Tu}}, \ and\ \bibinfo
  {author} {\bibfnamefont {Y.}~\bibnamefont {Zhou}},\ }\href
  {https://doi.org/10.1007/s43673-025-00156-8} {\bibfield  {journal} {\bibinfo
  {journal} {AAPPS Bull.}\ }\textbf {\bibinfo {volume} {35}},\ \bibinfo {pages}
  {16} (\bibinfo {year} {2025})}\BibitemShut {NoStop}%
\bibitem [{\citenamefont {Hille}\ and\ \citenamefont
  {Szabó}(2025)}]{Hille2025}%
  \BibitemOpen
  \bibfield  {author} {\bibinfo {author} {\bibfnamefont {S.~H.}\ \bibnamefont
  {Hille}}\ and\ \bibinfo {author} {\bibfnamefont {A.}~\bibnamefont {Szabó}},\
  }\href {https://arxiv.org/abs/2510.05227} {\bibfield  {journal} {\bibinfo
  {journal} {arXiv:2510.05227}\ } (\bibinfo {year} {2025})}\BibitemShut
  {NoStop}%
\bibitem [{Note1()}]{Note1}%
  \BibitemOpen
  \bibinfo {note} {Local displacements are omitted because they can always be
  removed locally and do not affect entanglement properties.}\BibitemShut
  {Stop}%
\bibitem [{Note2()}]{Note2}%
  \BibitemOpen
  \bibinfo {note} {Strictly speaking, the GSVD of a pure BGS takes the form
  $|\phi \rangle = \langle I_m|\left (|\phi _L\rangle \otimes |\phi _R\rangle
  \right )$, where $|\phi _L\rangle = \DOTSB \prod@ \slimits@
  _{q=1}^{n_e}\protect \mathrm {exp}\left [b^{\dagger }_{L,q}l^{\dagger
  }_{q}\right ]|0\rangle _{b_L,l}$ and $|\phi _R\rangle = \DOTSB \prod@
  \slimits@ _{q=1}^{n_e}\protect \mathrm {exp}\left [b^{\dagger
  }_{R,q}r^{\dagger }_{q}\right ]|0\rangle _{b_R,r}$ are Gaussian
  ``isometries'', and the bond state $|I_m\rangle = \DOTSB \prod@ \slimits@
  _{q=1}^{n_e}\protect \mathrm {exp}\left [\Lambda _q l^{\dagger
  }_{q}r^{\dagger }_{q}\right ]|0\rangle _{l,r}$ carries the Gaussian
  ``singular values'' $\Lambda _q$. For convenience of the ensuing algorithm,
  we absorb the factors $\Lambda _q$ into $|\phi _R\rangle $ and take
  $|I_m\rangle $ to be a maximally entangled state of virtual modes mediating
  the entanglement across the cut.}\BibitemShut {Stop}%
\bibitem [{\citenamefont {Schuch}\ \emph {et~al.}(2008)\citenamefont {Schuch},
  \citenamefont {Cirac},\ and\ \citenamefont {Wolf}}]{Schuch2008}%
  \BibitemOpen
  \bibfield  {author} {\bibinfo {author} {\bibfnamefont {N.}~\bibnamefont
  {Schuch}}, \bibinfo {author} {\bibfnamefont {J.~I.}\ \bibnamefont {Cirac}}, \
  and\ \bibinfo {author} {\bibfnamefont {M.~M.}\ \bibnamefont {Wolf}},\ }in\
  \href {https://arxiv.org/abs/1201.3945} {\emph {\bibinfo {booktitle} {Quantum
  Information and Many Body Quantum Systems}}},\ Vol.~\bibinfo {volume} {5},\
  \bibinfo {editor} {edited by\ \bibinfo {editor} {\bibfnamefont
  {M.}~\bibnamefont {Ericsson}}\ and\ \bibinfo {editor} {\bibfnamefont
  {S.}~\bibnamefont {Montangero}}}\ (\bibinfo  {publisher} {Edizioni della
  Normale},\ \bibinfo {address} {Pisa},\ \bibinfo {year} {2008})\ pp.\ \bibinfo
  {pages} {129--142},\ \Eprint {http://arxiv.org/abs/quant-ph/0509166}
  {arXiv:quant-ph/0509166} \BibitemShut {NoStop}%
\bibitem [{lap()}]{laptop}%
  \BibitemOpen
  \href@noop {} {}\bibinfo {note} {Apple MacBook Air (M4, 2025)}\BibitemShut
  {NoStop}%
\bibitem [{\citenamefont {Liu}\ \emph {et~al.}(2025{\natexlab{b}})\citenamefont
  {Liu}, \citenamefont {Su}, \citenamefont {Gong}, \citenamefont {Gu},
  \citenamefont {Tang}, \citenamefont {Jia}, \citenamefont {Wei}, \citenamefont
  {Song}, \citenamefont {Wang}, \citenamefont {Zheng}, \citenamefont {Chen},
  \citenamefont {Li}, \citenamefont {Ren}, \citenamefont {Zhu}, \citenamefont
  {Wang}, \citenamefont {Chen}, \citenamefont {Liu}, \citenamefont {Song},
  \citenamefont {Yang}, \citenamefont {Chen}, \citenamefont {An}, \citenamefont
  {Zhang}, \citenamefont {Gan}, \citenamefont {Yang}, \citenamefont {Xu},
  \citenamefont {He}, \citenamefont {Wang}, \citenamefont {Zhong},
  \citenamefont {Chen}, \citenamefont {Jiang}, \citenamefont {Li},
  \citenamefont {Liu}, \citenamefont {Deng}, \citenamefont {Su}, \citenamefont
  {Zhang}, \citenamefont {Lu},\ and\ \citenamefont {Pan}}]{LiuHL2025}%
  \BibitemOpen
  \bibfield  {author} {\bibinfo {author} {\bibfnamefont {H.-L.}\ \bibnamefont
  {Liu}}, \bibinfo {author} {\bibfnamefont {H.}~\bibnamefont {Su}}, \bibinfo
  {author} {\bibfnamefont {S.-Q.}\ \bibnamefont {Gong}}, \bibinfo {author}
  {\bibfnamefont {Y.-C.}\ \bibnamefont {Gu}}, \bibinfo {author} {\bibfnamefont
  {H.-Y.}\ \bibnamefont {Tang}}, \bibinfo {author} {\bibfnamefont {M.-H.}\
  \bibnamefont {Jia}}, \bibinfo {author} {\bibfnamefont {Q.}~\bibnamefont
  {Wei}}, \bibinfo {author} {\bibfnamefont {Y.}~\bibnamefont {Song}}, \bibinfo
  {author} {\bibfnamefont {D.}~\bibnamefont {Wang}}, \bibinfo {author}
  {\bibfnamefont {M.}~\bibnamefont {Zheng}}, \bibinfo {author} {\bibfnamefont
  {F.}~\bibnamefont {Chen}}, \bibinfo {author} {\bibfnamefont {L.}~\bibnamefont
  {Li}}, \bibinfo {author} {\bibfnamefont {S.}~\bibnamefont {Ren}}, \bibinfo
  {author} {\bibfnamefont {X.}~\bibnamefont {Zhu}}, \bibinfo {author}
  {\bibfnamefont {M.}~\bibnamefont {Wang}}, \bibinfo {author} {\bibfnamefont
  {Y.}~\bibnamefont {Chen}}, \bibinfo {author} {\bibfnamefont {Y.}~\bibnamefont
  {Liu}}, \bibinfo {author} {\bibfnamefont {L.}~\bibnamefont {Song}}, \bibinfo
  {author} {\bibfnamefont {P.}~\bibnamefont {Yang}}, \bibinfo {author}
  {\bibfnamefont {J.}~\bibnamefont {Chen}}, \bibinfo {author} {\bibfnamefont
  {H.}~\bibnamefont {An}}, \bibinfo {author} {\bibfnamefont {L.}~\bibnamefont
  {Zhang}}, \bibinfo {author} {\bibfnamefont {L.}~\bibnamefont {Gan}}, \bibinfo
  {author} {\bibfnamefont {G.}~\bibnamefont {Yang}}, \bibinfo {author}
  {\bibfnamefont {J.-M.}\ \bibnamefont {Xu}}, \bibinfo {author} {\bibfnamefont
  {Y.-M.}\ \bibnamefont {He}}, \bibinfo {author} {\bibfnamefont
  {H.}~\bibnamefont {Wang}}, \bibinfo {author} {\bibfnamefont {H.-S.}\
  \bibnamefont {Zhong}}, \bibinfo {author} {\bibfnamefont {M.-C.}\ \bibnamefont
  {Chen}}, \bibinfo {author} {\bibfnamefont {X.}~\bibnamefont {Jiang}},
  \bibinfo {author} {\bibfnamefont {L.}~\bibnamefont {Li}}, \bibinfo {author}
  {\bibfnamefont {N.-L.}\ \bibnamefont {Liu}}, \bibinfo {author} {\bibfnamefont
  {Y.-H.}\ \bibnamefont {Deng}}, \bibinfo {author} {\bibfnamefont {X.-L.}\
  \bibnamefont {Su}}, \bibinfo {author} {\bibfnamefont {Q.}~\bibnamefont
  {Zhang}}, \bibinfo {author} {\bibfnamefont {C.-Y.}\ \bibnamefont {Lu}}, \
  and\ \bibinfo {author} {\bibfnamefont {J.-W.}\ \bibnamefont {Pan}},\ }\href
  {https://arxiv.org/abs/2508.09092} {\bibfield  {journal} {\bibinfo  {journal}
  {arXiv:2508.09092}\ } (\bibinfo {year} {2025}{\natexlab{b}})}\BibitemShut
  {NoStop}%
\bibitem [{\citenamefont {Liu}(2023)}]{LiuMZ2023-Data}%
  \BibitemOpen
  \bibfield  {author} {\bibinfo {author} {\bibfnamefont {M.}~\bibnamefont
  {Liu}},\ }\href {\doibase 10.5281/zenodo.7992736} {\enquote {\bibinfo {title}
  {sss441803/bosonsupremacy: Version 0.0 [data set]},}\ } (\bibinfo {year}
  {2023})\BibitemShut {NoStop}%
\bibitem [{\citenamefont {Liu}\ \emph {et~al.}(2026)\citenamefont {Liu},
  \citenamefont {Xiong},\ and\ \citenamefont {Wang}}]{LiuK2026}%
  \BibitemOpen
  \bibfield  {author} {\bibinfo {author} {\bibfnamefont {K.}~\bibnamefont
  {Liu}}, \bibinfo {author} {\bibfnamefont {F.}~\bibnamefont {Xiong}}, \ and\
  \bibinfo {author} {\bibfnamefont {F.}~\bibnamefont {Wang}},\ }\href {\doibase
  10.1103/gckn-mc7w} {\bibfield  {journal} {\bibinfo  {journal} {Phys. Rev. B}\
  }\textbf {\bibinfo {volume} {113}},\ \bibinfo {pages} {024413} (\bibinfo
  {year} {2026})}\BibitemShut {NoStop}%
\bibitem [{\citenamefont {Liao}\ \emph {et~al.}(2019)\citenamefont {Liao},
  \citenamefont {Liu}, \citenamefont {Wang},\ and\ \citenamefont
  {Xiang}}]{LiaoHJ2019}%
  \BibitemOpen
  \bibfield  {author} {\bibinfo {author} {\bibfnamefont {H.-J.}\ \bibnamefont
  {Liao}}, \bibinfo {author} {\bibfnamefont {J.-G.}\ \bibnamefont {Liu}},
  \bibinfo {author} {\bibfnamefont {L.}~\bibnamefont {Wang}}, \ and\ \bibinfo
  {author} {\bibfnamefont {T.}~\bibnamefont {Xiang}},\ }\href {\doibase
  10.1103/PhysRevX.9.031041} {\bibfield  {journal} {\bibinfo  {journal} {Phys.
  Rev. X}\ }\textbf {\bibinfo {volume} {9}},\ \bibinfo {pages} {031041}
  (\bibinfo {year} {2019})}\BibitemShut {NoStop}%
\bibitem [{\citenamefont {Tay}\ and\ \citenamefont
  {Motrunich}(2011)}]{Tay2011}%
  \BibitemOpen
  \bibfield  {author} {\bibinfo {author} {\bibfnamefont {T.}~\bibnamefont
  {Tay}}\ and\ \bibinfo {author} {\bibfnamefont {O.~I.}\ \bibnamefont
  {Motrunich}},\ }\href {\doibase 10.1103/PhysRevB.84.020404} {\bibfield
  {journal} {\bibinfo  {journal} {Phys. Rev. B}\ }\textbf {\bibinfo {volume}
  {84}},\ \bibinfo {pages} {020404} (\bibinfo {year} {2011})}\BibitemShut
  {NoStop}%
\end{thebibliography}
%

\end{document}